\documentclass[openacc]{rsproca_new}%

\usepackage{mathptmx}
\usepackage{courier}

\usepackage{graphicx}
\usepackage{amsmath}
\usepackage{tikz}
\usepackage{tikz-3dplot}
\usetikzlibrary{shapes,calc,positioning}
\usetikzlibrary{calc,intersections}

\usepackage{amsmath, amssymb, amsthm, amscd}
\usepackage{xcolor}
\colorlet{mountainred}{red}
\usepackage{booktabs}
\usepackage{multirow}
\usepackage{array}
\usepackage[export]{adjustbox}
\usepackage{tabularray}
\UseTblrLibrary{booktabs}
\newlength{\paramheight}
\newtheorem{theorem}{Theorem}

\newtheorem{corollary}{Corollary}

\usepackage{subfig}

\usepackage{amsmath,amsfonts,amsxtra}
\usepackage{amssymb,amstext,amsbsy} %
\usepackage{latexsym} %
\usepackage{textcomp} %
\usepackage{nicefrac} %
\newcounter{rchremark}

\newcounter{aliceremark}

\newcounter{danielremark}

\newcounter{homamremark}

\newcounter{sascharemark}

\newcommand{\xlo}{x^{\ulcorner}}
\newcommand{\xlu}{x^{\llcorner}}
\newcommand{\xro}{x^{\urcorner}}
\newcommand{\xru}{x^{\lrcorner}}
\newcommand{\ylo}{y^{\ulcorner}}
\newcommand{\ylu}{y^{\llcorner}}
\newcommand{\yro}{y^{\urcorner}}
\newcommand{\yru}{y^{\lrcorner}}
\newcommand{\zlo}{z^{\ulcorner}}
\newcommand{\zlu}{z^{\llcorner}}
\newcommand{\zro}{z^{\urcorner}}
\newcommand{\zru}{z^{\lrcorner}}
\newcommand{\Ulo}{U^{\ulcorner}}
\newcommand{\Ulu}{U^{\llcorner}}
\newcommand{\Uro}{U^{\urcorner}}
\newcommand{\Uru}{U^{\lrcorner}}
\newcommand{\Vl}{V^{\vdash}}
\newcommand{\Vr}{V^{\dashv}}
\newcommand{\Tro}{T_{\urcorner}}
\newcommand{\Tru}{T_{\lrcorner}}
\newcommand{\tro}{t_{\urcorner}}

\newcommand{\Qro}{Q_{\urcorner}}

\newcommand{\lena}{a}
\newcommand{\lenb}{b}
\newcommand{\lenc}{c}
\newcommand{\lend}{d}
\newcommand{\sqrlend}{D}
\newcommand{\sqrlene}{E}

\newcommand{\lenv}{v}
\newcommand{\lenl}{\ell_1}
\newcommand{\lenlt}{\ell_2}
\newcommand{\foldanglebeta}{\beta}
\newcommand{\foldanglegamma}{\gamma}
\newcommand{\dihedral}{\phi}

\newcommand{\textsave}[1]{}

\titlehead{Research}

\begin{document}

\title{Closed-form solution for semi-symmetric folding of waterbomb tessellations}

\author{%
S. Stüttgen$^{1}$, H.\ Spartali$^{2}$, A.C.\ Niemeyer$^{3}$, R.\ Chudoba$^{4}$ and D.\ Robertz$^{1}$ 
}
\address{$^{1}$Chair for Algebra and Number Theory, RWTH Aachen University, Pontdriesch 14/16, 52062 Aachen, Germany\\
$^{2}$Kempen Krause Ingenieure GmbH, Ritterstraße 20, 52072 Aachen, Germany\\
$^{3}$Chair for Algebra and Representation Theory, RWTH Aachen University, Pontdriesch 10-16, 52062 Aachen, Germany\\
$^{4}$Institute of Structural Concrete, RWTH Aachen University, Mies-van-der-Rohe Str. 1, 52072 Aachen, Germany}

\subject{origami, structural engineering, kinematics}

\keywords{
waterbomb origami tessellation, folding kinematics, analytical solution, shells, crease pattern
}

\corres{Sascha Stüttgen\\
\email{sascha.stuettgen@rwth-aachen.de}}

\begin{abstract}
Rigid-origami folding has versatile applications in the fields of architecture, engineering, medicine and aerospace. 
Given the design parameters of a single waterbomb cell, the global geometry of the origami tessellated shell  at an intermediate state of folding is directly determined.
Even though existing numerical methods of rigid-origami can be used to describe geometric forms, algebraic solutions of the waterbomb folding kinematics and tessellations 
reduce the algorithmic complexity and determine the range of admissible configurations exactly.
The existing analytical solution for a symmetrically folded waterbomb cell and tessellation is extended to the case of a cell folded semi-symmetrically, i.e. a cell with a 180° rotation symmetry. The algebraic derivation provides the solution for the entire folding process. The resulting functional relationship between the design parameters and the 3D cell geometry of the folded waterbomb cell enables efficient studies of achievable shell geometries.
This development facilitates a deeper understanding of the correspondence between 
the geometric parameters of a single waterbomb cell and the statically relevant 
geometric properties, i.e.\ the shell curvature and the effective cross-sectional height. The systematic description of this correspondence
provides the basis for innovative design concepts using modern cementitious composites, e.g.\ 
textile-reinforced concrete, to produce thin-walled, high-performance, lightweight shells.
\absbreak
\end{abstract}

\rsbreak

\section{Introduction}

Rigid-origami kinematics, renowned for enabling deployability and packability of shapes, have been utilized across various technical fields such as robotics \cite{origami_robots_2025}, engineering \cite{non_flat_2024, engineering_origami_review_2021, Liu_Miura_2023}, medicine \cite{stents_2025}, meta-materials \cite{jamalimehr_rigidly_2022}, aerospace engineering \cite{Yue_Aerospace_2023} and many more. Mathematical descriptions of rigid folding kinematics have recently been provided for various folding patterns such as hybrid patterns combining the Miura-Ori and Eggbox folding patterns \cite{hybrid_origami_2024} or radially expanding Bloom patterns \cite{bloom_2025}. 

\begin{figure}[h]
\centering
\includegraphics[width=\textwidth]{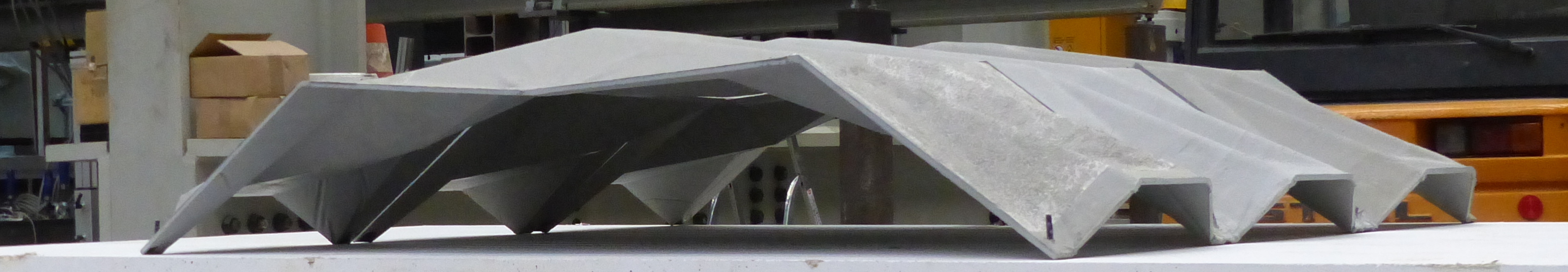}
\caption{Thin-walled textile-reinforced concrete element composed of symmetric waterbomb cells and produced by the fold-in-fresh principle \cite{spartali_iass2022}.}
\label{fig:wb_demonstrator}
\end{figure}

A motivating example is the fold-in-fresh production of thin-walled textile-reinforced concrete elements, as shown in Figure~\ref{fig:wb_demonstrator} \cite{spartali_iass2022}. Thin layers of fine-grained concrete and carbon textile fabric are laminated on a flat, foldable formwork. While the concrete is fresh, the formwork folds the laminate into its target configuration; the element is removed after hardening. For this process, it is not sufficient to prescribe only the final shape. The formwork must follow compatible vertex trajectories throughout the folding motion. An exact algebraic description of the waterbomb kinematics therefore provides the link between the crease-pattern parameters and a geometrically reachable structural form.

The waterbomb pattern is particularly relevant to this application because its cell parameters control both the global curvature of the shell, which influences its membrane action, and the effective cross-sectional height, which contributes to its local bending stiffness. Extending the design space from symmetric to semi-symmetric waterbomb cells adds control over the orientation of the cylindrical axis and the diagonal ribs while preserving the periodicity needed for modular assembly and repetitive production. This provides a means of coordinating the global shape and rib layout with the structural requirements of a given load case. The planar configurations contained in the same family further motivate material-efficient ribbed floor and ceiling elements.

These design objectives require a rigorous description of the interplay between the crease-pattern parameters and the resulting folded geometry. While the last two decades have seen an influx of numerical models~\cite{Tachi2009,Tachi2010, origami_simulator_2018} and form-fitting approaches~\cite{nassar_eggbox_2016, zhao_computational_2021}, closed-form solutions specific to waterbomb origami kinematics provide a direct route to the admissible configurations~\cite{fonseca_symmetries_2021,farnham_rigid_2022,zhao_analysis_2021, chen_symmetric_2016, Li_Quadruple_Waterbomb_2025, Brandon_Waterbomb_2014}. Algebraic methods complement these numerical tools in two respects. First, their closed-form expressions certify the existence and exact admissible range of rigid configurations, including degenerate flat-folded and planar states. Second, once the master-cell configuration is known, a tessellation with any prescribed number of cells can be generated without iteratively solving compatibility or equilibrium equations, at a computational cost that grows linearly with the number of cells.

 The present work builds on two preceding conference contributions. In \cite{chudoba2021}, the symmetric waterbomb-cell kinematics and the corresponding tessellated shell geometries were studied. A preliminary semi-symmetric formulation, including an angle-based parametrization of the cell, the construction of compatible three-cell assemblies, and parametric studies of the resulting shell geometries, was presented in \cite{robertz_iass2022}. The present journal article consolidates and extends these results in a unified framework. In particular, it provides a new length-based algebraic parametrization of the semi-symmetric waterbomb pattern, a rigorous curvature theorem for uniform semi-symmetric tessellations with a complete proof, and a characterization of planar waterbomb patterns. It also corrects the sign in the earlier derivation of the cell-to-cell rotation angle and the sign of \(\sin\gamma\) in the symmetric-coordinate formula used for \(\beta_{\mathrm{sym}}\).

The paper is organized as follows. We present two approaches to the derivation of the semi-symmetric folding of a waterbomb base in Sec.~\ref{sec:parametrization}. With the base solutions at hand, we build on the compatibility conditions of \cite{chudoba2021, robertz_iass2022} and reformulate them in the present notation to derive the algebraic expressions for the tessellations in Sec.~\ref{sec:tessellating_the_waterbomb_cell}. These closed-form solutions allow us to perform parametric studies in Sec.~\ref{sec:folded_geometries}, investigating the relationship between the cell parameters and the resulting ribbed shell geometries. We explore both symmetric and semi-symmetric folding cases. 
To demonstrate the practical implications of our findings, we provide examples of direct applications in terms of quantified curvature and rib height for varied cell parameters. We examine constrained solutions that yield slab elements with zero curvature, as well as vault geometries with ribs arranged along a circular curve. These applications offer valuable design support for structural elements.

\begin{table*}[t]
\centering
\small
\caption{Nomenclature used throughout the paper.}
\label{tab:nomenclature}
\begin{tblr}{colspec={l l X[l]}, row{1}={font=\bfseries}}
\toprule
Group & Symbol & Meaning \\
\midrule
Cell geometry & \(a,b,c\) & edge lengths of the flat base cell (Figure~\ref{fig:pattern_semisym}(a)) \\
 & \(d\) & diagonal length, \(d=\sqrt{(a-c)^2+b^2}\) \\
 & \(\alpha_1,\alpha_2\) & planar sector angles of the flat cell \\
 & \(\eta=b/a,\ \zeta=c/a\) & dimensionless cell-shape parameters \\
Vertices & \(O\); \(U^{\ulcorner},U^{\urcorner},U^{\llcorner},U^{\lrcorner}\); \(V^{\vdash},V^{\dashv}\) & center, corner, and middle vertices of a cell \\
Folding & \(\gamma,\beta\) & angle-based folding parameters (Figure~\ref{fig:pattern_semisym}(b)) \\
 & \(\beta_{\mathrm{sym}},\Delta\beta\) & symmetric value of \(\beta\) and its semi-symmetric deviation \\
 & \(v,w\) & length-based folding parameters \\
 & \(\ell_1,\ell_2\) & auxiliary distances used in the length-based parametrization \\
 & \(\phi_1,\ldots,\phi_4\) & dihedral angles (Figure~\ref{fig:pattern_semisym}(d)) \\
Tessellation & \(\theta,\delta\) & cell-to-cell rotation angle and translation in the symmetric case \\
 & \(T_{\urcorner},T_{\lrcorner}\) & commuting neighbor-cell isometries \\
 & \(Q_{\urcorner},t_{\urcorner}\) & rotation and translation parts of \(T_{\urcorner}\) \\
Global shape & \(A,C,z_C,\tau\) & cylinder axis, cross-section center, its height, and axis-direction angle \\
 & \(r,\kappa=1/r\) & cylinder radius and curvature magnitude \\
 & \(z^{\dashv}=c\cos\gamma\) & effective cross-sectional height at the middle vertex \\
\bottomrule
\end{tblr}
\end{table*}

\section{Development of parametric waterbomb cell}\label{sec:parametrization} 

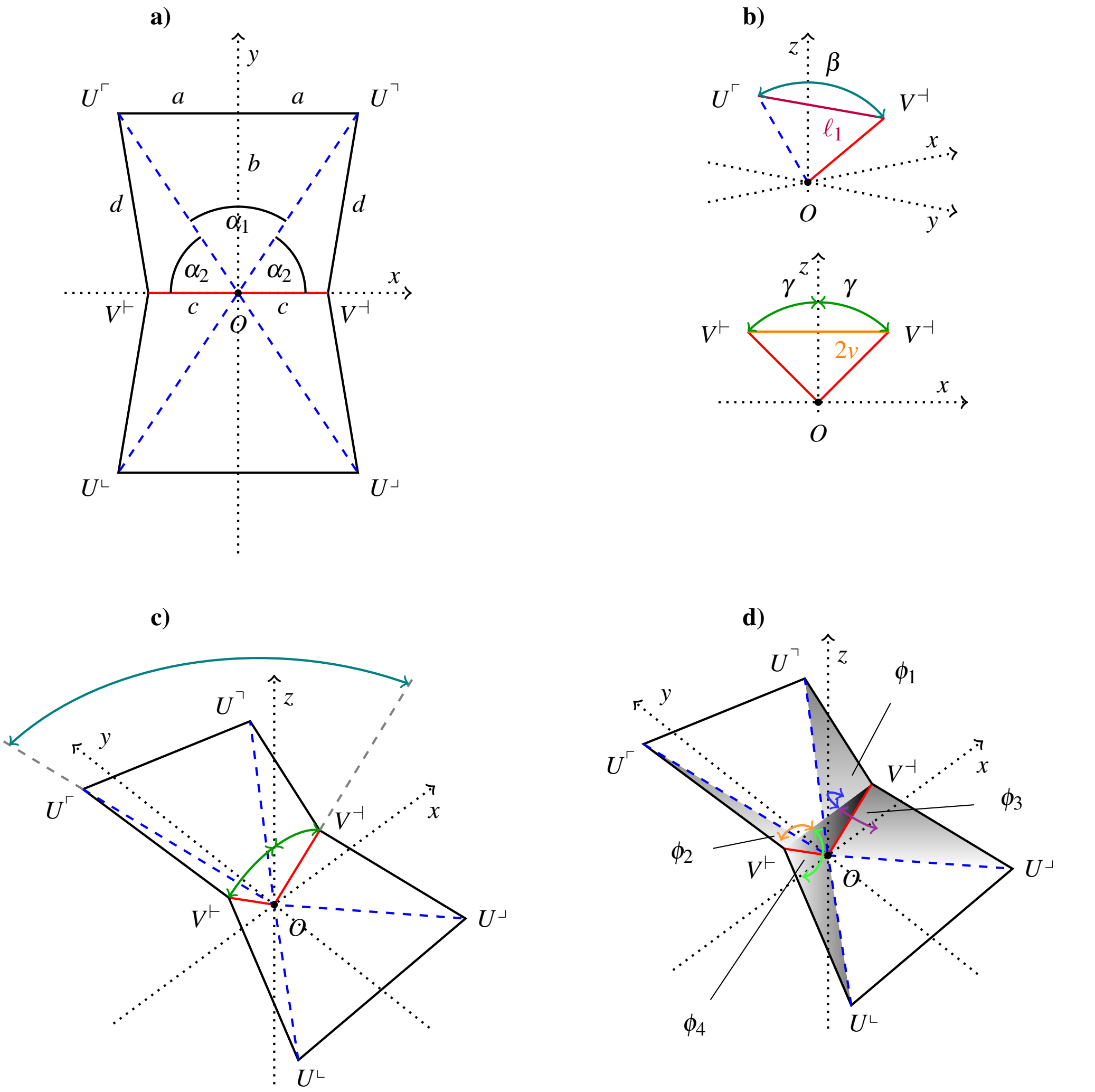
\begin{figure*}[h]
\centering
\begin{minipage}[t]{0.45\textwidth}
\textbf{a)} \\ \centering
\begin{tikzpicture}[thick,scale=1.5]
\def\lengtha{1}
\def\lengthb{1.5}
\def\lengthc{0.75}
\def\radius{sqrt(\lengtha^2+\lengthb^2)}
\coordinate (OO) at (0,0);
\coordinate (Ulo) at (-\lengtha,\lengthb);
\coordinate (Uro) at (\lengtha,\lengthb);
\coordinate (Ulu) at (-\lengtha,-\lengthb);
\coordinate (Uru) at (\lengtha,-\lengthb);
\coordinate (Vl) at (-\lengthc,0);
\coordinate (Vr) at (\lengthc,0);
\coordinate (Mo) at (0,\lengthb);

\draw[dotted,->] (-1.45*\lengtha,0) -- node [pos=0.95, above] {$x$} (1.45*\lengtha,0);
\draw[dotted,->] (0,-1.45*\lengthb) -- node [pos=0.95, right] {$y$} node [pos=0.75, right] {$b$} (0,1.45*\lengthb);

\draw[solid] (Vl) -- (Ulu) -- (Uru) -- (Vr) -- node [midway, right] {$d$} (Uro) -- node [midway, above] {$a$} (Mo) -- node [midway, above] {$a$} (Ulo) -- node [midway, left] {$d$} cycle;
\draw[solid,mountainred] (Vl) -- node [midway, below, black] {$c$} (OO) -- node [midway, below, black] {$c$} (Vr);
\draw[dashed,blue] (Ulo) -- (Uru);
\draw[dashed,blue] (Uro) -- (Ulu);

\draw[solid] (OO) + (0.75*\lengthc,0) arc (0:acos(\lengtha / sqrt(\lengtha^2+\lengthb^2)):0.75*\lengthc)
  node [pos=0.3, below, left] {{\small $\alpha_2$}};
\draw[solid] (OO) + (-0.75*\lengthc,0) arc (180:180-acos(\lengtha / sqrt(\lengtha^2+\lengthb^2)):0.75*\lengthc)
  node [pos=0.3, below, right] {{\small $\alpha_2$}};
\draw[solid] (OO) + (-0.4*\lengtha, 0.4*\lengthb) arc (180-acos(\lengtha / sqrt(\lengtha*\lengtha+\lengthb*\lengthb)):acos(\lengtha / sqrt(\lengtha*\lengtha+\lengthb*\lengthb)):0.4*\radius)
  node [midway, below] {{\small $\alpha_1$}};

\node[circle, fill=black, inner sep=0.325mm] at (OO) {};
\node[below=0.5em of OO] {$O$};
\node[above left=-1pt and -1pt of Ulo] {$\Ulo$};
\node[above right=-1pt and 1pt of Uro] {$\Uro$};
\node[below left=-1pt and -1pt of Ulu] {$\Ulu$};
\node[below right=-1pt and 1pt of Uru] {$\Uru$};
\node[below left=-1pt and 1pt of Vl] {$\Vl$};
\node[below right=-1pt and 1pt of Vr] {$\Vr$};
\end{tikzpicture}
\end{minipage} \hfill
\begin{minipage}[t]{0.45\textwidth}
\textbf{b)} \\ \centering
\raisebox{7em}{\begin{tabular}{c}
\begin{tikzpicture}[thick,scale=1.25]
\coordinate (OO) at (0,0);
\coordinate (Ulo) at (120:1);
\coordinate (Vr) at (40:1);

\draw[dotted,->] (-1,-0.2) -- node [pos=0.9, above] {$x$} (1.5,0.3);
\draw[dotted,->] (-1,0.2) -- node [pos=0.9, below] {$y$} (1.5,-0.3);
\draw[dotted,->] (0,-0.1) -- node [pos=0.9, left] {$z$} (0,1.5);

\draw[dashed,blue] (OO) -- (Ulo);
\draw[solid,mountainred] (OO) -- (Vr);
\draw[solid,purple] (Ulo) -- node [pos=0.6, below] {$\lenl$} (Vr);

\draw[solid,->,teal] (0,1) arc (90:40:1) node [pos=0.3, above, black] {$\foldanglebeta$};
\draw[solid,->,teal] (0,1) arc (90:120:1) node [pos=0.4, below, black] {};

\node[circle, fill=black, inner sep=0.325mm] at (OO) {};
\node[below=0.5em of OO] {$O$};
\node[left=0.25em of Ulo] {$\Ulo$};
\node[above right=-1pt and 2pt of Vr] {$\Vr$};
\end{tikzpicture}\\
\begin{tikzpicture}[thick,scale=1.25]
\coordinate (OO) at (0,0);
\coordinate (Vl) at (135:1);
\coordinate (Vr) at (45:1);

\draw[dotted,->] (-1,0) -- node [pos=0.9, above] {$x$} (1.5,0);
\draw[dotted,->] (0,-0.1) -- node [pos=0.9, left] {$z$} (0,1.5);

\draw[solid,mountainred] (OO) -- (Vl);
\draw[solid,mountainred] (OO) -- (Vr);

\draw[solid,orange] (Vl) -- node [pos=0.7, below] {$2 \lenv$} (Vr);

\draw[solid,<->,green!60!black] (0,1) arc (90:45:1) node [pos=0.4, above, black] {$\foldanglegamma$};
\draw[solid,<->,green!60!black] (0,1) arc (90:135:1) node [pos=0.4, above, black] {$\foldanglegamma$};

\node[circle, fill=black, inner sep=0.325mm] at (OO) {};
\node[below=0.5em of OO] {$O$};
\node[left=0.25em of Vl] {$\Vl$};
\node[right=0.25em of Vr] {$\Vr$};
\end{tikzpicture}
\end{tabular}}
\end{minipage} \\[2em]
\begin{minipage}[t]{0.45\textwidth}
\textbf{c)} \\ \centering
\tdplotsetmaincoords{40}{-44}
\begin{tikzpicture}[tdplot_main_coords,thick,scale=2.8]
\coordinate (P11) at (0., 0., 0.);
\coordinate (P12) at (0.2828427124, 0., 0.2828427124);
\coordinate (P13) at (-0.2828427124, 0., 0.2828427124);
\coordinate (P14) at (0.6097893972, 0.7865660925, 0.0973173838);
\coordinate (P15) at (-0.3535533905, 0.8660254040, 0.3535533905);
\coordinate (P16) at (-0.6097893972, -0.7865660925, 0.0973173838);
\coordinate (P17) at (0.3535533905, -0.8660254040, 0.3535533905);
\coordinate (extP12) at (3.1*0.2828427124, 3.1*0., 3.1*0.2828427124);
\coordinate (extP15) at (1.45*-0.3535533905, 1.45*0.8660254040, 1.45*0.3535533905);

\coordinate (xAxis1) at (-1.0, 0., 0.);
\coordinate (xAxis2) at (1.0, 0., 0.);
\coordinate (yAxis1) at (0., -1.0, 0.);
\coordinate (yAxis2) at (0., 1.3, 0.);
\coordinate (zAxis1) at (0., 0., -1.25);
\coordinate (zAxis2) at (0., 0., 1.6);

\draw[solid,mountainred] (P13) -- (P11) -- (P12);
\draw[solid] (P12) -- (P14) -- (P15) -- (P13);
\draw[solid] (P12) -- (P17) -- (P16) -- (P13);
\draw[dashed, blue] (P15) -- (P11) -- (P14);
\draw[dashed, blue] (P16) -- (P11) -- (P17);

\draw[dotted,->] (xAxis1) -- node [pos=0.95, below right] {$x$} (xAxis2);
\draw[dotted,->] (yAxis1) -- node [pos=0.95, above right] {$y$} (yAxis2);
\draw[dotted,->] (zAxis1) -- node [pos=0.9, above right] {$z$} (zAxis2);

\draw[dashed,gray] (P12) -- (extP12);
\draw[dashed,gray] (P15) -- (extP15);

\node (OO) [below right=0.25em of P11] {$O$};
\node (VR) [above right=-3pt and 2pt of P12] {$\Vr$};
\node (VL) [below left=-1pt and -1pt of P13] {$\Vl$};
\node (ULO) [below left=-1pt and -1pt of P15] {$\Ulo$};
\node (ULU) [below right=0pt and -4pt of P16] {$\Ulu$};
\node (URO) [above left=-1pt and -2pt of P14] {$\Uro$};
\node (URU) [right=1pt of P17] {$\Uru$};

\node[circle, fill=black, inner sep=0.325mm] at (P11) {};

\tdplotsetrotatedcoords{90}{90}{0}
\tdplotdrawarc[tdplot_rotated_coords,thick,<->,green!60!black]{(0,0,0)}{0.4}{180}{135}{}{}
\tdplotdrawarc[tdplot_rotated_coords,thick,<->,green!60!black]{(0,0,0)}{0.4}{180}{225}{}{}

\tdplotsetrotatedcoords{0}{4}{22.2077}
\tdplotdrawarc[tdplot_rotated_coords,thick,<->,teal]{(0,0,0)}{1.5}{0}{76}{}{}

\end{tikzpicture}
\end{minipage} \hfill
\begin{minipage}[t]{0.45\textwidth}
\textbf{d)} \\ \centering
\tdplotsetmaincoords{40}{-44}
\begin{tikzpicture}[tdplot_main_coords,thick,scale=2.7]
\coordinate (P11) at (0., 0., 0.);
\coordinate (P12) at (0.2828427124, 0., 0.2828427124);
\coordinate (P13) at (-0.2828427124, 0., 0.2828427124);
\coordinate (P14) at (0.6097893972, 0.7865660925, 0.0973173838);
\coordinate (P15) at (-0.3535533905, 0.8660254040, 0.3535533905);
\coordinate (P16) at (-0.6097893972, -0.7865660925, 0.0973173838);
\coordinate (P17) at (0.3535533905, -0.8660254040, 0.3535533905);

\coordinate (xAxis1) at (-1.0, 0., 0.);
\coordinate (xAxis2) at (1.0, 0., 0.);
\coordinate (yAxis1) at (0., -1.0, 0.);
\coordinate (yAxis2) at (0., 1.3, 0.);
\coordinate (zAxis1) at (0., 0., -1.25);
\coordinate (zAxis2) at (0., 0., 1.6);

\shade (P11) -- (P12) -- (P14) -- cycle;
\shade (P11) -- (P13) -- (P15) -- cycle;
\shade (P11) -- (P12) -- (P17) -- cycle;
\shade [left color=white,right color=gray] (P11) -- (P13) -- (P16) -- cycle;
\shade [left color=white,right color=black] (P13) -- (P11) -- (P12) -- cycle;

\draw[solid,mountainred] (P13) -- (P11) -- (P12);
\draw[solid] (P12) -- (P14) -- (P15) -- (P13);
\draw[solid] (P12) -- (P17) -- (P16) -- (P13);
\draw[dashed, blue] (P15) -- (P11) -- (P14);
\draw[dashed, blue] (P16) -- (P11) -- (P17);

\draw[dotted,->] (xAxis1) -- node [pos=0.95, below right] {$x$} (xAxis2);
\draw[dotted,->] (yAxis1) -- node [pos=0.95, above right] {$y$} (yAxis2);
\draw[dotted,->] (zAxis1) -- node [pos=0.9, above right] {$z$} (zAxis2);

\node (OO) [below right=0.25em of P11] {$O$};
\node (VR) [above right=-3pt and 2pt of P12] {$\Vr$};
\node (VL) [below left=-1pt and -1pt of P13] {$\Vl$};
\node (ULO) [below left=-1pt and -1pt of P15] {$\Ulo$};
\node (ULU) [below right=0pt and -4pt of P16] {$\Ulu$};
\node (URO) [above left=-1pt and -2pt of P14] {$\Uro$};
\node (URU) [right=1pt of P17] {$\Uru$};

\node[circle, fill=black, inner sep=0.325mm] at (P11) {};

\tdplotsetrotatedcoords{0}{45}{0}
\tdplotdrawarc[tdplot_rotated_coords,thick,<->,blue!80!white]{(0,0,0.25)}{0.15}{60}{180}{}{}
\tdplotdrawarc[tdplot_rotated_coords,thick,<->,violet!80!white]{(0,0,0.25)}{0.15}{180}{270}{}{}
\tdplotsetrotatedcoords{0}{-45}{0}
\tdplotdrawarc[tdplot_rotated_coords,thick,<->,green!80!white]{(0,0,0.25)}{0.12}{-140}{0}{}{}
\tdplotdrawarc[tdplot_rotated_coords,thick,<->,orange!80!white]{(0,0,0.25)}{0.12}{0}{100}{}{}

\draw[solid,thin] (-0.5,0.25,0.25) -- (-0.25,0.1,0.26);
\node at (-0.65,0.3,0.3) {$\dihedral_2$};
\draw[solid,thin] (-1,-0.25,0) -- (-0.25,-0.1,0.26);
\node at (-1.1,-0.25,-0.1) {$\dihedral_4$};
\draw[solid,thin] (0.3,0.15,0.2) -- (0.5,0.1,0.6);
\node at (0.65,0.15,0.65) {$\dihedral_1$};
\draw[solid,thin] (0.25,0,0.1) -- (0.7,-0.25,0);
\node at (0.85,-0.35,0) {$\dihedral_3$};
\end{tikzpicture}
\end{minipage}
\caption{Waterbomb cell configuration: a) base cell dimensions, b) folding parameters; angles $\beta$ and $\gamma$, lengths $\ell_1$ and $v$, c) perspective view of folded cell with folding parameters indicated, d) dihedral angles of faces. Reproduced and enhanced from \cite{robertz_iass2022}.}
\label{fig:pattern_semisym}
\end{figure*}

Consider a rigidly folded waterbomb cell as given in 
Figure~\ref{fig:pattern_semisym}.
Two approaches to describe the kinematics of the semi-symmetric folding of the waterbomb cell are derived. The first approach, referred to as the \emph{length-based solution}, yields a square root expression for the intermediate folding state positions. This expression is determined by two parameters $v$ and $w$ that control the distance between certain vertices. On the other hand, the second approach, referred to as the \emph{angle-based solution}, involves trigonometric expressions based on the spherical angles defined within the cell (cf. \cite{chudoba2021, robertz_iass2022}).

We employ a Cartesian coordinate system with origin $O$ to describe the waterbomb base cell.
Figure~\ref{fig:pattern_semisym} shows the flat base cell on the left, indicating edge lengths $\lena$, $\lenb$, $\lenc$, $\lend = \sqrt{(\lena-\lenc)^2+\lenb^2}$,
planar angles $\alpha_1$, $\alpha_2$ as well as the vertex labels $\Ulo$, $\Uro$, $\Vr$, $\Uru$, $\Ulu$, $\Vl$. The mountain fold lines are on the $x$-axis (red line)
and the valley folds on the lines
$y=\frac{b}{a}x$ and $y=-\frac{b}{a}x$ (dashed blue lines), see Figure~\ref{fig:pattern_semisym}(a).
Moreover,  Figure~\ref{fig:pattern_semisym}(b)  introduces two lengths $\lenl$ and $\lenv$ and two folding angles $\foldanglegamma$, $\foldanglebeta$ and 
Figure~\ref{fig:pattern_semisym}(c) 
shows a folded base cell.
For later use we note
\begin{equation}\label{eq:alpha1}
\cos(\alpha_1) \, = \, \frac{\lenb^2-\lena^2}{\lena^2+\lenb^2}\,, \qquad
\sin(\alpha_1) \, = \, \frac{2 \, \lena \, \lenb}{\lena^2+\lenb^2}\,,
\end{equation}
\begin{equation}\label{eq:alpha2}
\cos(\alpha_2) \, = \, \frac{\lena}{\sqrt{\lena^2+\lenb^2}}\,, \qquad
\sin(\alpha_2) \, = \, \frac{\lenb}{\sqrt{\lena^2+\lenb^2}}\,.
\end{equation}
\begin{equation*}
v = c \sin(\foldanglegamma)\,, \quad \sqrt{c^2-v^2}=c\cos(\foldanglegamma)\,,
\quad
(\lenl)^2 = a^2+b^2+c^2-2c\sqrt{a^2+b^2}\cos(\foldanglebeta).
\end{equation*}
Moreover, we introduce the notation
$$\begin{array}{rclrcl}
\Ulo & = & [\xlo,\ylo,\zlo], & \Uro & = & [\xro,\yro,\zro],\\[0.2em]
\Ulu & = & [\xlu,\ylu,\zlu], & \Uru & = & [\xru,\yru,\zru],\\[0.2em]
\Vl & = & [-\lenc\sin(\foldanglegamma), 0, \lenc\cos(\foldanglegamma)] & \Vr & = & [\lenc\sin(\foldanglegamma), 0, \lenc\cos(\foldanglegamma)]\\[0.2em]
& = & [-v, 0, \sqrt{\lenc^2 - v^2}], & & = & [v, 0, \sqrt{\lenc^2 - v^2}].
\end{array}
$$
In \cite{chudoba2021, robertz_iass2022} we derived equations $E_1$, \ldots, $E_5$
that are reproduced in Table~\ref{tab:wbequations}
for the upper half of the cell and which hold correspondingly for the lower half when
$[\xru, \yru, \zru]$ and $[\xlu, \ylu, \zlu]$ are substituted
for $[\xro, \yro, \zro]$ and $[\xlo, \ylo, \zlo]$, respectively.
 The reference frame is fixed by placing \(O\) at the origin and \(\Vl,\Vr\) in the \(xz\)-plane, symmetrically with respect to the \(z\)-axis, as shown in Figure~\ref{fig:pattern_semisym}. These conditions remove the whole-body translations and rotations from the cell kinematics.

As the base cell is folded rigidly, angles between adjacent valley and mountain fold lines
($E_1,E_2$)
and
angles between adjacent valley fold lines ($E_5$)
are preserved as well as 
distances of the corners to the origin are preserved ($E_3,E_4$).

We say the cell is \emph{semi-symmetrically folded}
if $[\xro, \yro, \zro] =[-\xlu, -\ylu, \zlu]$ and $[\xlo, \ylo, \zlo] =  [-\xru, -\yru, \zru]$ and
\emph{symmetrically folded} if, in addition,
$[\xro, \yro, \zro]=[\xru, -\yru, \zru].$
 A generic waterbomb cell without symmetry assumptions has three internal degrees of freedom. Theorem~\ref{the:main} gives an explicit parametrization of the semi-symmetric waterbomb cell by two independent parameters, either $(v,w)$ or $(\gamma,\beta)$. In the length-based formulation, $v$ and $w$ are geometric quantities recoverable from the configuration, and their admissible set has a two-dimensional interior. Thus, on each non-degenerate algebraic branch, the semi-symmetric configuration family is two-dimensional. The additional reflection symmetry is given by $w=4av$, so Corollary~\ref{cor:symm_param} identifies the symmetric configurations as a one-dimensional subfamily. The one-parameter families used later are selected folding paths within the two-dimensional semi-symmetric family.

We also introduce the dimensionless parameters $\eta = b/a$ and $\zeta=c/a$ to eliminate the geometric magnitudes in the theorem below.

\begin{theorem}\label{the:main}
The kinematics of a single semi-symmetric waterbomb cell
with edge lengths $\lena$, $\lenb$, $\lenc$
and corner vertices
can be parametrized in two ways:
\begin{enumerate}
\item[(a)]  In terms of two  parameters $(\lenv,w)$ where
$\lenv \in (\max\{0, \frac{\lena^2-\lenb^2}{\lena^2+\lenb^2}c
\},\lenc]$ denotes the distance of $V^\dashv$ to the $z$-axis and
$w= ||U^{\urcorner}-V^\vdash||^2-
||U^{\urcorner}-V^\dashv||^2$ 
as follows:
\[
\begin{array}{rcl}
\hspace{-4em}
[\xro, \yro, \zro] \!\! & \!\! = \!\! & \!\!\!\!
\frac{1}{4vk}\left[wk,\, \sqrt{16(a^2+b^2)v^2k^2 -w^2k^2 -(w-4ac)^2v^2},\,  (4ac-w)v
\right]
\!\! %
,\\[0.5em]
\hspace{-2em}
[\xlo, \ylo, \zlo] \!\! & \!\! = \!\! & \!\!\!\!
\frac{1}{4vk}\left[-uk,\, \sqrt{16(a^2+b^2)v^2k^2 -u^2k^2 -(u-4ac)^2v^2},\,  (4ac-u)v
\right],
\end{array}\]
where  $k=\sqrt{c^2-v^2}$ and  
$u = u_1/(4b^2c^2+4ac w-w^2)$ with
\begin{align*}
  u_1 &= \left(8 k^{2} a^{2}-4 b^{2} c^{2}+8 b^{2} v^{2}\right) w +32 a^{3} c \,v^{2}-2 a c \,w^{2}\\
&\pm  2 \sqrt{\left(4 a c -w \right) \left(4 a^{3} c -4 a \,b^{2} c -a^{2} w -b^{2} w \right) \left(16 (a^{2} +b^2)v^{4}-16 b^{2} c^{2} v^{2}-8 a c \,v^{2} w +c^{2} w^{2}\right)}
\end{align*}
and
$w \in \left[\frac{4 v}{c}\left(a v -b k\right),\frac{4v}{c} \left(a v +b k\right)
\right]$.
Moreover, the cell is symmetric when 
$w= 4av$.\\
\item[(b)]  In terms of two angles $\foldanglegamma$, $\foldanglebeta$
as follows:
\[
\begin{array}{rclcl}
\hspace{-2em}
[\xro, \yro, \zro] \!\! & \!\! = \!\! & \!\!
[\lena \sin(\foldanglegamma) - \lenb \cos(\dihedral_1) \cos(\foldanglegamma), \,
\lenb \sin(\dihedral_1), \,
\lena \cos(\foldanglegamma) + \lenb \cos(\dihedral_1) \sin(\foldanglegamma)]
\!\! & \!\! = \!\! & \!\!
[-\xlu, -\ylu, \zlu]\,,\\[0.5em]
\hspace{-2em}
[\xlo, \ylo, \zlo] \!\! & \!\! = \!\! & \!\!
[-\lena \sin(\foldanglegamma) + \lenb \cos(\dihedral_2) \cos(\foldanglegamma), \,
\lenb \sin(\dihedral_2), \,
\lena \cos(\foldanglegamma) + \lenb \cos(\dihedral_2) \sin(\foldanglegamma)]
\!\! & \!\! = \!\! & \!\!
[-\xru, -\yru, \zru]\,,
\end{array}
\]
where (cf.\ also \eqref{eq:cospsi1}--\eqref{eq:sinpsi6})
\[
\begin{array}{rcl}
\cos(\dihedral_1) \!\! & \!\! = \!\! & \!\! \displaystyle
\frac{(\lenb^2-\lena^2) - \lena \, \sqrt{\lena^2+\lenb^2} \cos(\foldanglebeta)}{\lenb \, \sqrt{\lena^2+\lenb^2} \sin(\foldanglebeta)} \,
\frac{\lena - \sqrt{\lena^2+\lenb^2} \cos(\foldanglebeta) \cos(2 \foldanglegamma)}{\sqrt{\lena^2+\lenb^2} \sin(\foldanglebeta) \sin(2 \foldanglegamma)} \, -\\[1.5em]
& & \displaystyle
\frac{\sqrt{\, \lena^2 \, (3 \, \lenb^2 - \lena^2) + 2 \, \lena \, (\lenb^2 - \lena^2) \sqrt{\lena^2+\lenb^2} \cos(\foldanglebeta) - (\lena^2+\lenb^2)^2 \cos^2(\foldanglebeta) \,}}{\lenb \, \sqrt{\lena^2+\lenb^2} \sin(\foldanglebeta)} \, \times\\[1.5em]
& & \displaystyle
\frac{\sqrt{\, \lenb^2 + 2 \, \lena \, \sqrt{\lena^2+\lenb^2} \cos(\foldanglebeta) \cos(2 \foldanglegamma) - (\lena^2+\lenb^2) \, (\cos^2(\foldanglebeta) + \cos^2(2 \foldanglegamma)) \,}}{\sqrt{\lena^2+\lenb^2} \, \sin(\foldanglebeta) \sin(2 \foldanglegamma)}\,,
\end{array}
\]
\[
\begin{array}{rcl}
\sin(\dihedral_1) \!\! & \!\! = \!\! & \!\! \displaystyle
\frac{\sqrt{\, \lena^2 \, (3 \, \lenb^2 - \lena^2) + 2 \, \lena \, (\lenb^2 - \lena^2) \sqrt{\lena^2+\lenb^2} \cos(\foldanglebeta) - (\lena^2+\lenb^2)^2 \cos^2(\foldanglebeta) \,}}{\lenb \, \sqrt{\lena^2+\lenb^2} \sin(\foldanglebeta)} \, \times\\[1.5em]
\!\! & \!\! \!\! & \!\! \displaystyle
\frac{\lena - \sqrt{\lena^2+\lenb^2} \cos(\foldanglebeta) \cos(2 \foldanglegamma)}{\sqrt{\lena^2+\lenb^2} \sin(\foldanglebeta) \sin(2 \foldanglegamma)} +
\frac{(\lenb^2-\lena^2) - \lena \, \sqrt{\lena^2+\lenb^2} \cos(\foldanglebeta)}{\lenb \, \sqrt{\lena^2+\lenb^2} \sin(\foldanglebeta)} \, \times\\[1.5em]
\!\! & \!\! \!\! & \!\! \displaystyle
\frac{\sqrt{\, \lenb^2 + 2 \, \lena \, \sqrt{\lena^2+\lenb^2} \cos(\foldanglebeta) \cos(2 \foldanglegamma) - (\lena^2+\lenb^2) \, (\cos^2(\foldanglebeta) + \cos^2(2 \foldanglegamma)) \,}}{\sqrt{\lena^2+\lenb^2} \, \sin(\foldanglebeta) \sin(2 \foldanglegamma)}\,,
\end{array}
\]
\[
\begin{array}{rcl}
\cos(\dihedral_2) \!\! & \!\! = \!\! & \!\! \displaystyle
\frac{\sqrt{\lena^2+\lenb^2} \cos(\foldanglebeta) - \lena \cos(2 \foldanglegamma)}{\lenb \sin(2 \foldanglegamma)}\,,\\[1.5em]
\sin(\dihedral_2) \!\! & \!\! = \!\! & \!\! \displaystyle
\frac{\sqrt{\, \lenb^2 + 2 \, \lena \, \sqrt{\lena^2+\lenb^2} \cos(\foldanglebeta) \cos(2 \foldanglegamma) - (\lena^2+\lenb^2) \, (\cos^2(\foldanglebeta) + \cos^2(2 \foldanglegamma)) \,}}{\lenb \, \sin(2 \foldanglegamma)}\,.
\end{array}
\]
\end{enumerate}
\end{theorem}

\begin{corollary}\label{cor:symm_param}
The kinematics of a single symmetric
waterbomb cell
with edge lengths $a$, $b$, $c$
can be parametrized in two ways:
\begin{enumerate}
    \item In terms of a parameter 
     $\lenv \in [\max\{0, \frac{\lena^2-\lenb^2}{\lena^2+\lenb^2}c
\},\lenc]$  by \[
\begin{array}{rcl}
[\xro, \yro, \zro] & = &
\left[\lena, \,\sqrt{\lenb^2 -\lena^2 \frac{c-v}{c+v}},\,
\lena \sqrt{\frac{c-v}{c+v}}\,\right].
\end{array}\]
\item In terms of an angle \(\gamma \in [0, \pi/2)\) by
\[
    \begin{array}{rcl}
        [\xro, \yro, \zro] & = &
        \left[\lena, \,\frac{\sqrt{b^2 \cos^2(\gamma) - a^2(1 - \sin(\gamma))^2}}{\cos(\gamma)},\,
        \frac{a(1 - \sin(\gamma))}{\cos(\gamma)}\,\right].
    \end{array}
\]
\end{enumerate}
\end{corollary}

These symbolic parametrizations are derived in the following sections starting with the specification of common geometric parameters and the criteria of rigid foldability of a cell followed by the length- and angle-based solutions.

\begin{table}
  \centering
\begin{tabular}{llcl}
  \toprule
  $E_1$: &   $c \sin(\gamma)\xro + c \cos(\gamma) \zro - ac $ &  $=$& $0$\\
  $E_2$: &  $-c \sin(\gamma) \xlo + c \cos(\gamma) \zlo - ac $&  $=$& $0$\\
\midrule
$E_3$: & $(\xro)^2 + {(\yro)}^2 + (\zro)^2 - (a^2+b^2) $&  $=$& $0$\\
$E_4$: & $(\xlo)^2 + {(\ylo)}^2 + (\zlo)^2 - (a^2+b^2)$&  $=$& $0$\\
\midrule
$E_5$: & $\xro\xlo + \yro\ylo + \zro\zlo + (a^2-b^2)$&  $=$& $0$\\
  \bottomrule
\end{tabular}
\caption{Equations for the semi-symmetric waterbomb base cell, reproduced from \cite[Table 2]{chudoba2021}.}\label{tab:wbequations}
\end{table}

\subsection{Length-based parametrization of a semi-symmetric waterbomb cell}\label{subsec:parametrizationlengths}

We now prove Theorem~\ref{the:main}(a). To model the kinematics of the waterbomb base cell,
we introduce   the lengths $\lenv$ and $\lenl$, where 
$\lenl := || U^\ulcorner - V^\dashv||$. 
We start by determining the range for $v$. When $v=c$, the cell is flat. To find the lower bound for $v$ we imagine the cell
to be completely folded in the $xz$-plane  (ignoring rigidity for the moment), that is when $U^{\urcorner}=[a,0,b]=U^{\lrcorner}.$ 
Solving $
||U^{\urcorner}-V^\dashv||^2 = d^2 = (a-c)^2 + b^2$ for $v$
yields $v = \frac{a^2-b^2}{a^2+b^2}c$ or $v=c.$ The rigid folding
condition then enforces
$v\in [\max\{0, \frac{a^2-b^2}{a^2+b^2}c
\},c]$. Note that $v=0$ is only possible when $a\le b$. Temporarily we
also introduce $\lenlt :=
|| U^\urcorner - V^\vdash||.$ The parameters $u$ and $w$ are now
chosen such that $u = \lenl^2-d^2$ and $w = \lenlt^2-d^2$. Thus
$w= \lenlt^2 -d^2 = ||U^{\urcorner}-V^\vdash||^2-
||U^{\urcorner}-V^\dashv||^2$ 
and $u= \ell_1^2-d^2 =||U^{\ulcorner}-V^\dashv||^2 -
||U^{\ulcorner}-V^\vdash||^2$.
This 
yields
$(x^\urcorner -v)^2 - (x^\urcorner + v)^2
= d^2 -\ell_2^2=-w$
and $(x^\ulcorner +v)^2 - (x^\ulcorner - v)^2
= d^2 -\ell_1^2=-u.$ Suppose first, $v>0.$
Then, 
\begin{equation}\label{eq:Xurlcorner}
x^\urcorner = \frac{w}{4v}\quad \textrm{and}\quad
x^\ulcorner = -\frac{u}{4v}, \quad \textrm{if\ } v >0.
\end{equation}
Noting that $v=c\sin(\gamma)$ and $\sqrt{c^2-v^2} = c\cos(\gamma),$
Equations $(E_1)$ and $(E_2)$ immediately yield
\begin{equation}\label{eq:Zurlcorner}
z^\urcorner = \frac{4ac -w}{4\sqrt{c^2-v^2}}\quad \textrm{and} \quad
z^\ulcorner = \frac{4ac -u}{4\sqrt{c^2-v^2}}, \quad \textrm{if\ } v < c,
\end{equation}
with  $z^\urcorner = z^\ulcorner= 0$ if $v=c$.
Finally, Equations $(E_3)$ and $(E_4)$ yield
\begin{equation}\label{eq:Yurlcorner}
y^\urcorner=\frac{1}{4}\sqrt{16 (a^{2}+b^{2})-\frac{w^{2}}{v^{2}}-\frac{16 \left(a c -\frac{w}{4}\right)^{2}}{c^{2}-v^{2}}}
\quad\textrm{and}\quad
y^\ulcorner=\frac{1}{4}\sqrt{16 (a^{2}+b^{2})-\frac{u^{2}}{v^{2}}-\frac{16 \left(a c -\frac{u}{4}\right)^{2}}{c^{2}-v^{2}}},
\end{equation}
where $y^\ulcorner=y^\urcorner=b$ when $v=c.$
If $v=0$, then $z^\urcorner = z^\ulcorner =a $ by Equations $(E_1), (E_2)$ and $||U^\urcorner||=||U^\ulcorner||=d$. Thus 
$(x^\urcorner)^2+(y^\urcorner)^2 = (x^\ulcorner)^2+(y^\ulcorner)^2= d^2-a^2$ describes a circle.
Thus we have obtained 
expressions for $U^\urcorner$ and $U^\ulcorner$ in terms of
the design parameters $a,b,c$ and three parameters $u,v, w.$

We first investigate the situation in which the cell is folded symmetrically. In this case $x^\urcorner=-x^\ulcorner=a$,  $y^\urcorner=y^\ulcorner$ and $z^\urcorner=z^\ulcorner.$ In particular, the kinematics of the cell is described by the single parameter $v\in [\max\{0, \frac{a^2-b^2}{a^2+b^2}c
\},c].$ In this case Equation~\eqref{eq:Xurlcorner} gives
$w^{sym}:=w = u= 4av.$

For the description of the semi-symmetric waterbomb pattern
Equation~$(E_5)$ yields an additional quadratic equation relating $u$ to $v$ and $w.$
Solving this equation for $u$ yields the stated expression for $u$.

The next aim is to determine, for a fixed 
$v\in [\max\{0, \frac{a^2-b^2}{a^2+b^2}c\},c]$, the range of $w.$ Note that $w$ is maximal or minimal if and only if $\lenlt$ is. As $\lenlt$ is minimal   when $y^\ulcorner=y^\urcorner=0 $  and maximal when
the cell is flat, whence $w=4ac+b^2$. Solving Equation~$(E_3)$ %
for $w$ by inserting Equation~\eqref{eq:Zurlcorner}.
This yields $w = \frac{4v}{c}(av\pm bk)$ and thereby determines the range for $w$. 

\subsection{Angle-based parametrization of a semi-symmetric waterbomb cell}\label{subsec:parametrizationangles}

 This subsection follows the construction of \cite[Section 3]{robertz_iass2022} and reformulates it in the notation used here. First we derive, for fixed value of $\foldanglegamma$, a new parametrization of the points $\Ulo$, $\Uro$
on the upper half of the cell, in terms of the
dihedral angles $\dihedral_1$, $\dihedral_2$ defined by the pairs of triangles $O\Vl\Vr$, $O\Vr\Uro$ and $O\Vl\Vr$, $O\Vl\Ulo$, respectively (cf. Fig.~\ref{fig:pattern_semisym}(d)).
In an analogous way we derive a parametrization of $\Ulu$, $\Uru$ on the lower half of the cell in terms of dihedral angles $\dihedral_3$, $\dihedral_4$
defined by the pairs of triangles $O\Vl\Vr$, $O\Vr\Uru$
and $O\Vl\Vr$, $O\Vl\Ulu$, respectively.

Taking the point of view that the triangle $O\Vr\Uro$ rotates around the axis through $O$ spanned by the vector $O\Vr = [\lenc\sin(\foldanglegamma), 0, \lenc\cos(\foldanglegamma)]$,
the point $\Uro$ moves on a circle with center $C_1$ and radius $r_1$,
where $C_1$ is the orthogonal projection of $\Uro$ on the previously mentioned axis of rotation and $r_1$ is the distance between $\Uro$ and $C_1$.
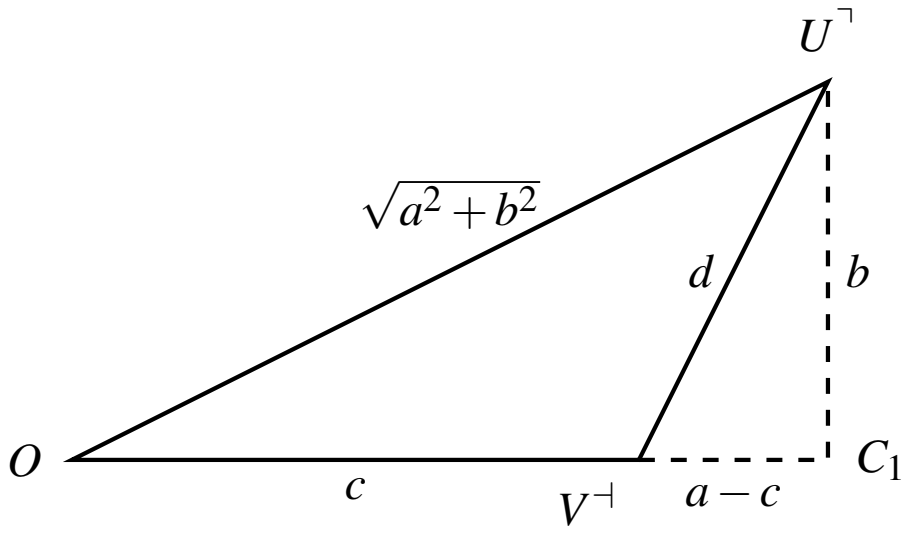
\begin{figure*}
\centering
\begin{tikzpicture}[thick,scale=1.25]
\coordinate (OO) at (0,0);
\coordinate (Uro) at (4,2);
\coordinate (Vr) at (3,0);
\coordinate (C1) at (4,0);

\draw[dashed] (Vr) -- node [midway,below] {$\lena-\lenc$} (C1) -- node [midway,right] {$\lenb$} (Uro);
\draw[solid] (OO) -- node [midway,below] {$\lenc$} (Vr) -- node [midway,left] {$\lend$} (Uro) -- node [midway,above=0.5em] {$\sqrt{\lena^2+\lenb^2}$} cycle;

\node (LO) [left=0.25em of OO] {$O$};
\node (LO) [right=0.25em of C1] {$C_1$};
\node (LVr) [below left=1pt of Vr] {$\Vr$};
\node (AUro) [above=0.25em of Uro] {$\Uro$};
\end{tikzpicture}
\caption{ Orthogonal projection of $\Uro$ on the (horizontal) axis of rotation,  reproduced from \cite{robertz_iass2022}} 
\label{fig:triangles}
\end{figure*}
Due to the geometry of the involved triangle shown in Figure~\ref{fig:triangles}
we obtain $C_1 = [\lena \sin(\foldanglegamma), 0, \lena \cos(\foldanglegamma)]$ and $r_1 = \lenb$,
and therefore,
\[
\Uro \, = \, [\xro,\yro,\zro] \, = \, C_1 + \lenb \, (\sin(\dihedral_1) \, n^{(1)}_1 + \cos(\dihedral_1) \, n^{(1)}_2)\,,
\]
where we choose the vectors $n^{(1)}_1 = [0, 1, 0]$ and $n^{(1)}_2 = [-\cos(\foldanglegamma), 0, \sin(\foldanglegamma)]$
so that $n^{(1)}_1 \perp O\Vr \perp n^{(1)}_2$ and $n^{(1)}_1 \perp n^{(1)}_2$ and $|n^{(1)}_1| = 1 = |n^{(1)}_2|$.
Then the case $\dihedral_1 = 0$ corresponds to the situation where the triangle $O\Vr\Uro$ lies on top of the triangle $O\Vl\Vr$.

In an analogous way, if the triangle $O\Vl\Ulo$ rotates around the axis through $O$ spanned by the vector $O\Vl = [-\lenc\sin(\foldanglegamma), 0, \lenc\cos(\foldanglegamma)]$,
the point $\Ulo$ moves on a circle with center $C_2$ and radius $r_2$,
where $C_2$ is the orthogonal projection of $\Ulo$ on the now considered axis of rotation and $r_2$ is the distance between $\Ulo$ and $C_2$.
Correspondingly, we obtain $C_2 = [-\lena \sin(\foldanglegamma), 0, \lena \cos(\foldanglegamma)]$ and $r_2 = \lenb$,
and therefore,
\[
\Ulo \, = \, [\xlo,\ylo,\zlo] \, = \, C_2 + \lenb \, (\sin(\dihedral_2) \, n^{(2)}_1 + \cos(\dihedral_2) \, n^{(2)}_2)\,,
\]
where we choose the vectors $n^{(2)}_1 = [0, 1, 0]$ and $n^{(2)}_2 = [\cos(\foldanglegamma), 0, \sin(\foldanglegamma)]$
so that $n^{(2)}_1 \perp O\Vl \perp n^{(2)}_2$ and $n^{(2)}_1 \perp n^{(2)}_2$ and $|n^{(2)}_1| = 1 = |n^{(2)}_2|$.
Then the case $\dihedral_2 = 0$ corresponds to the situation where the triangle $O\Vl\Ulo$ lies on top of the triangle $O\Vl\Vr$.

A similar discussion for the lower half of the waterbomb base cell
involving dihedral angles $\dihedral_3$ and $\dihedral_4$
thus leads altogether to the following parametrization of the points $\Uro$, $\Ulo$, $\Uru$, $\Ulu$:
\begin{equation}\label{eq:paramphi}
\begin{array}{rcl}
[\xro, \yro, \zro] & = &
[\lena \sin(\foldanglegamma) - \lenb \cos(\dihedral_1) \cos(\foldanglegamma), \,
\lenb \sin(\dihedral_1), \,
\lena \cos(\foldanglegamma) + \lenb \cos(\dihedral_1) \sin(\foldanglegamma)]\\[0.2em]
[\xlo, \ylo, \zlo] & = &
[-\lena \sin(\foldanglegamma) + \lenb \cos(\dihedral_2) \cos(\foldanglegamma), \,
\lenb \sin(\dihedral_2), \,
\lena \cos(\foldanglegamma) + \lenb \cos(\dihedral_2) \sin(\foldanglegamma)]\\[0.2em]
[\xru, \yru, \zru] & = &
[\lena \sin(\foldanglegamma) - \lenb \cos(\dihedral_3) \cos(\foldanglegamma), \,
-\lenb \sin(\dihedral_3), \,
\lena \cos(\foldanglegamma) + \lenb \cos(\dihedral_3) \sin(\foldanglegamma)]\\[0.2em]
[\xlu, \ylu, \zlu] & = &
[-\lena \sin(\foldanglegamma) + \lenb \cos(\dihedral_4) \cos(\foldanglegamma), \,
-\lenb \sin(\dihedral_4), \,
\lena \cos(\foldanglegamma) + \lenb \cos(\dihedral_4) \sin(\foldanglegamma)]
\end{array}
\end{equation}
Equations $(E_1)$, \ldots, $(E_4)$ for the waterbomb pattern (cf.\ Table~\ref{tab:wbequations})
are identically satisfied in $[\xro, \yro, \zro]$ and $[\xlo, \ylo, \zlo]$,
whereas equation $(E_5)$ constitutes a non-trivial relation between $\dihedral_1$, $\dihedral_2$, $\foldanglegamma$ (involving also the parameters $\lena$, $\lenb$).
The same statements hold for the lower half of the waterbomb cell, i.e.\ for the
substitution of $[\xru, \yru, \zru]$ and $[\xlu, \ylu, \zlu]$
for $[\xro, \yro, \zro]$ and $[\xlo, \ylo, \zlo]$, respectively,
in equations $E_1$, \ldots, $E_5$, where $E_5$ then correspondingly defines a relation
between $\dihedral_3$, $\dihedral_4$, $\foldanglegamma$ (depending also on $\lena$, $\lenb$).

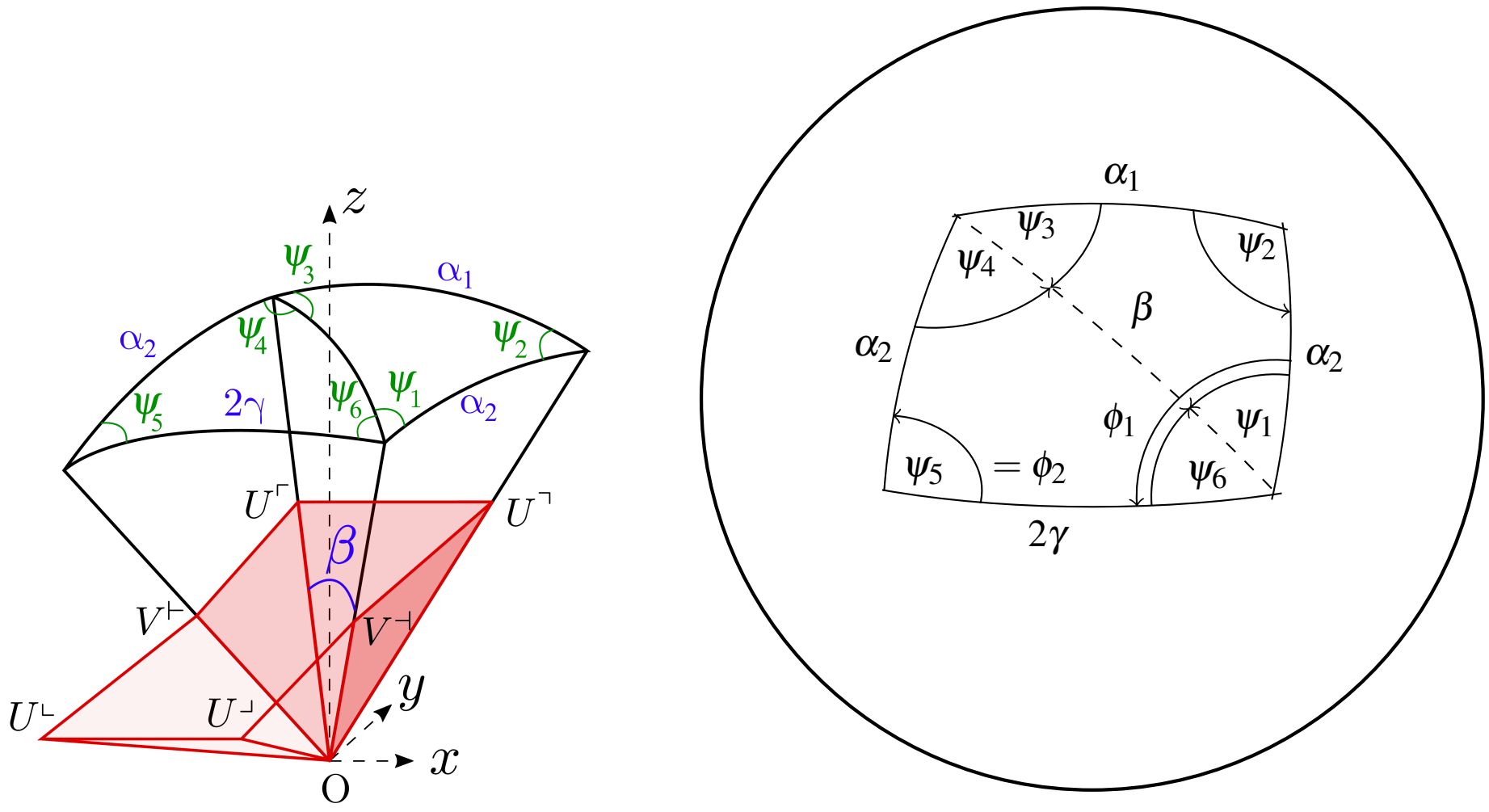
\begin{figure*}
\centering
\includegraphics[width=.36\textwidth]{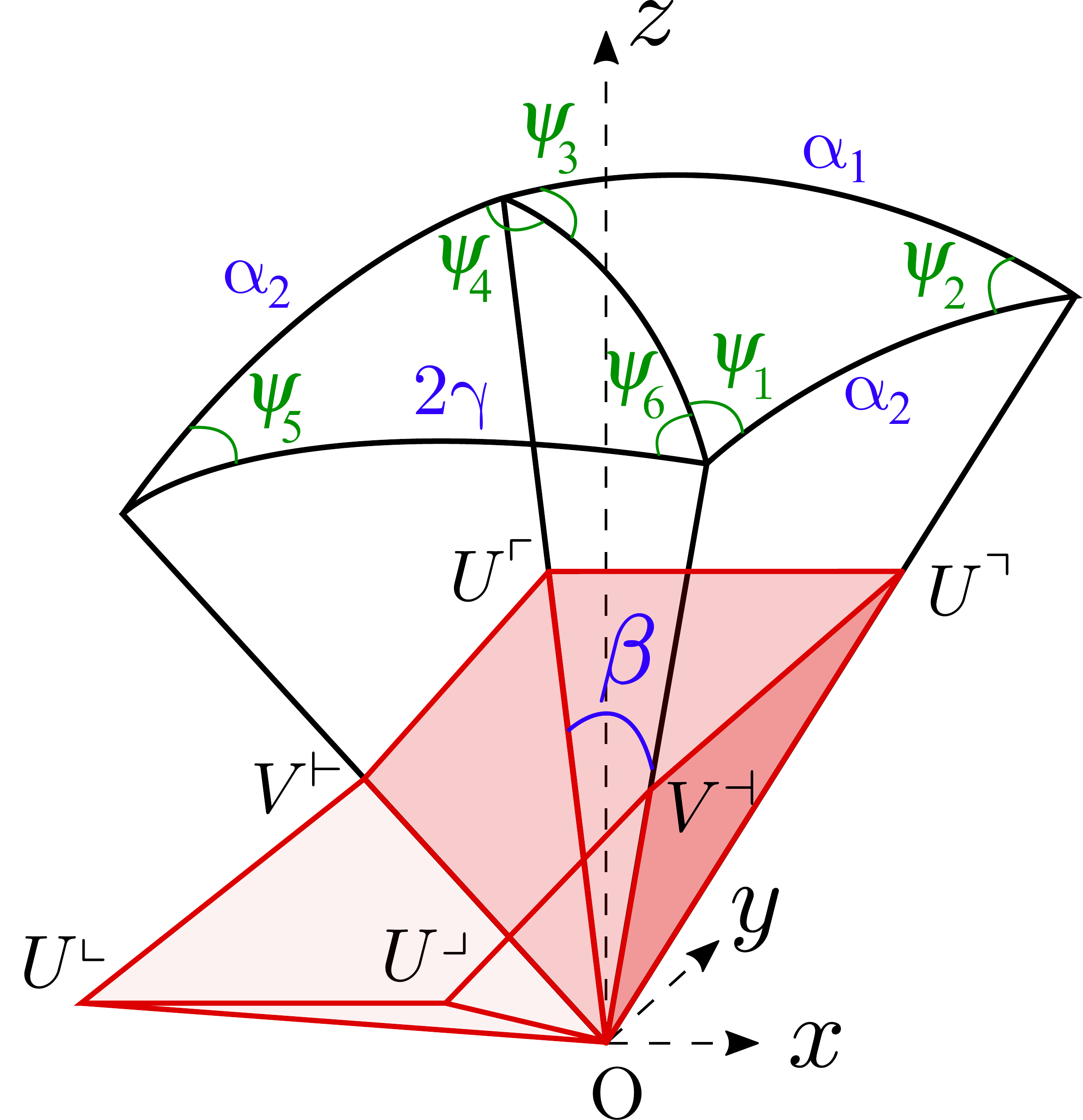}
\hspace{14pt}
\tdplotsetmaincoords{90}{90}
\begin{tikzpicture}[scale=4,tdplot_main_coords]

\tdplotsetthetaplanecoords{90}
\tdplotdrawarc[tdplot_rotated_coords,thick]{(0,0,0)}{0.8}{0}{360}{}{}

\tdplotsetrotatedcoords{60}{96}{0}
\tdplotdrawarc[tdplot_rotated_coords,name path=P1]{(0,0,0)}{0.8}{75}{117}{right}{$\alpha_2$}

\tdplotsetrotatedcoords{120}{103}{0}
\tdplotdrawarc[tdplot_rotated_coords,name path=P2]{(0,0,0)}{0.8}{76}{119}{left}{$\alpha_2$}

\tdplotsetrotatedcoords{180}{16}{0}
\tdplotdrawarc[tdplot_rotated_coords,name path=P3]{(0,0,0)}{0.8}{-29}{32.5}{below left}{$2 \foldanglegamma$}

\tdplotsetrotatedcoords{180}{-30}{0}
\tdplotdrawarc[tdplot_rotated_coords,name path=P4]{(0,0,0)}{0.8}{-30}{21}{above}{$\alpha_1$}

\tdplotsetrotatedcoords{77}{42}{0}
\tdplotdrawarc[dashed,tdplot_rotated_coords,name path=P5]{(0,0,0)}{0.8}{70}{135}{above right}{$\foldanglebeta$}

\path [name intersections={of={P1 and P3}, total=\n}]
\foreach \i in {1,...,\n}{(intersection-\i) circle [radius=0.5pt] coordinate(A\i){}};

\path [name intersections={of={P1 and P4}, total=\n}]
\foreach \i in {1,...,\n}{(intersection-\i) circle [radius=0.5pt] coordinate(B\i){}};

\path [name intersections={of={P2 and P4}, total=\n}]
\foreach \i in {1,...,\n}{(intersection-\i) circle [radius=0.5pt] coordinate(C\i){}};

\path [name intersections={of={P2 and P3}, total=\n}]
\foreach \i in {1,...,\n}{(intersection-\i) circle [radius=0.5pt] coordinate(D\i){}};

\draw (A1) node[above left=10pt and -4pt]{$\psi_1$};
\draw (A1) node[above left=-2pt and 7pt]{$\psi_6$};
\draw (B1) node[below left=-2pt and -2pt]{$\psi_2$};
\draw (C1) node[below right=5pt and -4pt]{$\psi_4$};
\draw (C1) node[below right=-4pt and 10pt]{$\psi_3$};
\draw (D1) node[above right=-2pt and 1pt]{$\psi_5$};

\tdplotsetrotatedcoords{180}{-50}{0}
\tdplotdrawarc[tdplot_rotated_coords,->]{(D1)}{0.2}{-100}{-5}{}{}

\tdplotsetrotatedcoords{27}{60}{0}
\tdplotdrawarc[tdplot_rotated_coords,->]{(B1)}{0.2}{-103}{-10}{}{}

\tdplotsetrotatedcoords{165}{130}{5}
\tdplotdrawarc[tdplot_rotated_coords,->]{(C1)}{0.3}{-101}{-55}{}{}
\tdplotdrawarc[tdplot_rotated_coords,<-]{(C1)}{0.3}{-55}{3}{}{}

\tdplotsetrotatedcoords{5}{100}{0}
\tdplotdrawarc[tdplot_rotated_coords,->]{(A1)}{0.25}{-187}{-136}{}{}
\tdplotdrawarc[tdplot_rotated_coords,<-]{(A1)}{0.25}{-136}{-85}{}{}

\tdplotsetrotatedcoords{5}{100}{0}
\tdplotdrawarc[tdplot_rotated_coords,->]{(A1)}{0.28}{-187}{-85}{}{}

\draw (A1) node[above left=10pt and 28pt]{$\dihedral_1$};
\draw (D1) node[above right=-2pt and 22pt]{$=\dihedral_2$};
\end{tikzpicture}
\caption{Arcs cut out on unit sphere, illustrating the definition of the angle $\foldanglebeta$, reproduced from \cite{robertz_iass2022}.} 
\label{fig:sphere}
\end{figure*}

In order to resolve the constraint imposed by equation $E_5$ on
$\dihedral_1$, $\dihedral_2$, $\foldanglegamma$ for the upper half of the waterbomb cell
and on $\dihedral_3$, $\dihedral_4$, $\foldanglegamma$ for the lower half, we adopt a
point of view that is common in the mathematical study of origami.
We extend the triangles $O\Uro\Ulo$, $O\Ulo\Vl$, $O\Vr\Uro$ as well as $O\Vl\Vr$
indefinitely beyond their boundary edges $\Uro\Ulo$, $\Ulo\Vl$, $\Vr\Uro$ and $\Vl\Vr$, respectively,
to convex regions that are each bounded
by two rays and we consider the intersections of these regions with a sphere
with center $O$ and radius $1$.
Then the lengths of the arcs that are cut out by these regions on the sphere
are equal to the corresponding sector angles $\alpha_1$, $\alpha_2$, $\alpha_2$
and $2\foldanglegamma$, respectively (in radians).
From the triangle $\Ulo O\Vr$ we obtain a convex region whose
sector angle we denote by $\foldanglebeta$. The corresponding arc on the unit sphere
divides the spherical quadrilateral into two spherical triangles. The angles
of these two spherical triangles will be denoted by $\psi_1$, \ldots, $\psi_6$ as indicated in Figure~\ref{fig:sphere}. Of course, for the lower half of the waterbomb
base cell an analogous configuration on the sphere will be considered.

By the spherical law of cosines,
and using \eqref{eq:alpha1} and \eqref{eq:alpha2}, we have, e.g.,
\begin{eqnarray}
\label{eq:cospsi1}
\cos(\psi_1) \! & \! = \! & \! \displaystyle
\frac{\cos(\alpha_1) - \cos(\alpha_2) \cos(\foldanglebeta)}{\sin(\alpha_2) \sin(\foldanglebeta)} \, \, \, = \, \, \,
\frac{(\lenb^2-\lena^2) - \lena \, \sqrt{\lena^2+\lenb^2} \cos(\foldanglebeta)}{\lenb \, \sqrt{\lena^2+\lenb^2} \sin(\foldanglebeta)}\,,\\[0.5em]
\label{eq:cospsi5}
\cos(\psi_5) \! & \! = \! & \! \displaystyle
\frac{\cos(\foldanglebeta) - \cos(\alpha_2) \cos(2 \foldanglegamma)}{\sin(\alpha_2) \sin(2 \foldanglegamma)} \, \, \, = \, \, \,
\frac{\sqrt{\lena^2+\lenb^2} \cos(\foldanglebeta) - \lena \cos(2 \foldanglegamma)}{\lenb \sin(2 \foldanglegamma)}\,,\\[0.5em]
\label{eq:cospsi6}
\cos(\psi_6) \! & \! = \! & \! \displaystyle
\frac{\cos(\alpha_2) - \cos(\foldanglebeta) \cos(2 \foldanglegamma)}{\sin(\foldanglebeta) \sin(2 \foldanglegamma)} \, \, \, = \, \, \,
\frac{\lena - \sqrt{\lena^2+\lenb^2} \cos(\foldanglebeta) \cos(2 \foldanglegamma)}{\sqrt{\lena^2+\lenb^2} \sin(\foldanglebeta) \sin(2 \foldanglegamma)}\,.
\end{eqnarray}

From these values we obtain the non-negative values
of $\sin(\psi_1)$, \ldots, $\sin(\psi_6)$
by using the identities $\cos^2(\psi_i) + \sin^2(\psi_i) = 1$ for $i = 1$, \ldots, $6$, namely, e.g.,
\begin{eqnarray}
\label{eq:sinpsi1}
\sin(\psi_1) \! & \! = \! & \! \displaystyle
\frac{\sqrt{\, \lena^2 \, (3 \, \lenb^2 - \lena^2) + 2 \, \lena \, (\lenb^2 - \lena^2) \sqrt{\lena^2+\lenb^2} \cos(\foldanglebeta) - (\lena^2+\lenb^2)^2 \cos^2(\foldanglebeta) \,}}{\lenb \, \sqrt{\lena^2+\lenb^2} \sin(\foldanglebeta)}\,,\\[0.5em]
\label{eq:sinpsi5}
\sin(\psi_5) \! & \! = \! & \! \displaystyle
\frac{\sqrt{\, \lenb^2 + 2 \, \lena \, \sqrt{\lena^2+\lenb^2} \cos(\foldanglebeta) \cos(2 \foldanglegamma) - (\lena^2+\lenb^2) \, (\cos^2(\foldanglebeta) + \cos^2(2 \foldanglegamma)) \,}}{\lenb \, \sin(2 \foldanglegamma)}\,,\\[0.5em]
\label{eq:sinpsi6}
\sin(\psi_6) \! & \! = \! & \! \displaystyle
\frac{\sqrt{\, \lenb^2 + 2 \, \lena \, \sqrt{\lena^2+\lenb^2} \cos(\foldanglebeta) \cos(2 \foldanglegamma) - (\lena^2+\lenb^2) \, (\cos^2(\foldanglebeta) + \cos^2(2 \foldanglegamma)) \,}}{\sqrt{\lena^2+\lenb^2} \, \sin(\foldanglebeta) \sin(2 \foldanglegamma)}\,.
\end{eqnarray}

Now, for the upper half of the waterbomb base cell, we have $\dihedral_1 = \psi_1 + \psi_6$ and $\dihedral_2 = \psi_5$, hence
\[
\cos(\dihedral_1) \, = \, \cos(\psi_1)\cos(\psi_6) - \sin(\psi_1)\sin(\psi_6)\,, \qquad
\sin(\dihedral_1) \, = \, \sin(\psi_1)\cos(\psi_6) + \cos(\psi_1)\sin(\psi_6)\,,
\]
and using these equations for substituting $\cos(\dihedral_1)$, $\sin(\dihedral_1)$,
$\cos(\dihedral_2)$, $\sin(\dihedral_2)$ in the first two equations of the
parametrization \eqref{eq:paramphi}
together with \eqref{eq:cospsi1}--\eqref{eq:sinpsi6}
yields a parametrization of $\Uro$, $\Ulo$ in terms of two angles $\foldanglegamma$, $\foldanglebeta$ only (depending also on $\lena$, $\lenb$, $\lenc$).
Taking into account the semi-symmetry of the waterbomb base cell,
the dihedral angles for its lower half
are determined by $\dihedral_3 = \dihedral_2 = \psi_5$ and $\dihedral_4 = \dihedral_1 = \psi_1 + \psi_6$.
A similar substitution thus yields a parametrization of $\Uru$, $\Ulu$
in terms of $\foldanglegamma$, $\foldanglebeta$ only (depending also on $\lena$, $\lenb$, $\lenc$). This concludes the proof of Theorem~\ref{the:main}.

While a full algebraic inversion translating the length-based parameters $(v,w)$ into the angle-based parameters $(\gamma, \beta)$ is highly complex and not strictly required for the subsequent shape-finding, the forward mapping is direct. From Equation (2.3), we have \(v=c\sin(\gamma)\). Furthermore, the parameter \(w\) is explicitly determined by substituting the angle-based coordinates of \(\Vl\) and \(\Vr\) from Theorem~\ref{the:main} into the geometric definition. This establishes a link between the two parametrizations and we proceed with the angle-based parametrization for the remainder of this study.

\section{Tessellating the waterbomb cell}
\label{sec:tessellating_the_waterbomb_cell}
Throughout this section, the master cell remains in the reference frame fixed in Section~\ref{sec:parametrization}: \(O\) is the origin and \(\Vl,\Vr\) lie in the \(xz\)-plane symmetrically about the \(z\)-axis; see Figure~\ref{fig:pattern_semisym}.
This section investigates the construction of tessellated shell geometries composed of identical copies of a waterbomb base cell, characterized by uniform design parameters and identical folding angles for all cell copies, first in the case where the base cell is fully symmetric, then in the semi-symmetric case. This focus on uniform tessellations with equal cells is driven by practical convenience and economical gains in the field of architectural design and civil engineering construction. The utilization of identical cells ensures the repeatability and edge compatibility necessary for the efficient manufacturing of precast elements. Furthermore, the resulting algebraic solution delivers exact geometric data that can serve as direct input for automated fabrication processes, such as fold-in-fresh concrete casting utilizing pneumatically controlled folding devices \cite{folding_robot_2025}.\\
Distinct from existing numerical rigid-origami models, which often rely on iterative shape-fitting or simulation to resolve the mesh geometry, this study employs a purely algebraic derivation. By mathematically describing the compatibility conditions between neighboring cells, we can generate extensive tessellations efficiently. This is achieved by utilizing the parametrization of the single master cell derived in Section \ref{sec:parametrization} and applying specific rigid motions that ensure edge consistency between neighbors. This analytical approach results in a robust and computationally efficient algorithm for the determination of large-scale tessellations whose complexity scales linearly in the number of cells.

\subsection{Folding symmetrically} \label{subsec:folding_sym}
 This subsection follows the symmetric construction of \cite[Sections 3 and 4]{chudoba2021} and recapitulates it for completeness in the unified notation of this paper. Using the kinematics of the waterbomb cell, we construct the tessellated folded shell geometry. To achieve this, compatibility
between cells must be ensured. In this subsection we assume the base cell to be reflection symmetric with respect to the $xz$-plane and the $yz$-plane. The kinematics of the master cell described in Section~\ref{sec:parametrization} based on 
symmetry assumptions, enforces that the tessellated waterbomb shell is curved only in the $yz$-plane.
This fact is exploited in the derivation of the rotation and translation operators which 
have to be applied to the master cell to obtain
the tessellated shell geometry.
The equations from Table~\ref{tab:wbequations} reduce in the symmetric case into the equations in Table~\ref{tab:wbequations_sym}. \((E_1)\) describes the condition that the angle between adjacent valley and mountain fold lines equals \(\alpha_2\), \((E_3)\) that the angle between adjacent valley fold equals \(\alpha_1\) and \((E_2)\) that the distances of the corners to the origin are preserved, throughout the entire tessellation.

\begin{table} 
    \centering
    \caption{Equations for the symmetric waterbomb cell,  reproduced from \cite{chudoba2021}.}\label{tab:wbequations_sym}
    \renewcommand{\arraystretch}{2.0} %
    \begin{tabular}{c l}
        \hline
        $\mathrm{E}_1$: & $c \sin(\gamma)\xro + c \cos(\gamma)\zro - ac = 0$ \\
        \hline
        $\mathrm{E}_2$: & $(\xro)^2 + (\yro)^2 + (\zro)^2 - (a^2 + b^2) = 0$ \\
        \hline
        $\mathrm{E}_3$: & $-(\xro)^2 + (\yro)^2 + (\zro)^2 + (a^2 - b^2) = 0$ \\
        \hline
    \end{tabular}
\end{table}

\begin{figure*}
\begin{center}
    \includegraphics[width=1\textwidth]{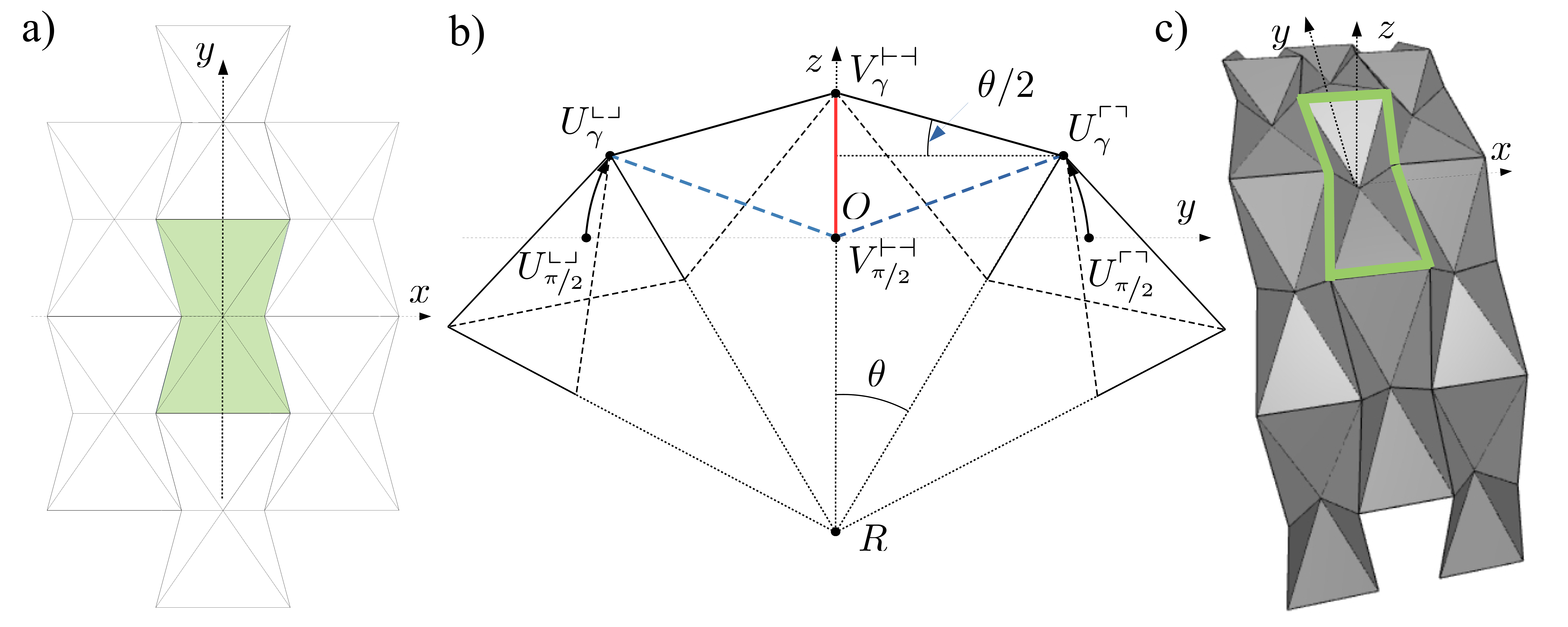}
\end{center}
\caption{ Cylindrical shell tessellation constructed as a foldable assembly of symmetric waterbomb cells,  a) Schematic top-down view of the tessellation. b) Projection of the tessellation onto the $yz$-plane, c) 3D view of cylindrical tessellation.  reproduced from \cite{chudoba2021}.}
\label{fig:generation}
\end{figure*}

The master cell has six adjacent cells (see Figure~\ref{fig:generation}(a)), two of which have their origin in the $yz$-plane. The other four cells are shifted in the $x$-direction. In the folded state, the first two cells can be obtained from 
the master cell by rotation around the $x$ axis by an angle $\pm 2\theta$. The remaining four cells are obtained from the master cell by rotation around the $x$-axis by an angle $\pm \theta$ followed by a shift $\pm \delta$.

\begin{figure}
    \centering
    \begin{tikzpicture}[
        line cap=round,
        line join=round,
        cell edge/.style={mountainred, line width=1.1pt},
        diagonal/.style={blue, dashed, line width=1.0pt},
        ang/.style={black, line width=0.8pt},
        pt/.style={circle, fill=black, inner sep=1.2pt},
        every node/.style={font=\small}
    ]

    \def\a{1.25}
    \def\b{1.90}
    \pgfmathsetmacro{\th}{atan(\b/\a)}

    \coordinate (O)   at (-\a,0);
    \coordinate (VrP) at (0,0);

    \coordinate (LLa) at (-2*\a,-\b);
    \coordinate (LTa) at (-2*\a,\b);
    \coordinate (RTa) at (0,\b);
    \coordinate (RBa) at (0,-\b);

    \coordinate (URbl) at (0,0);
    \coordinate (URbr) at (2*\a,0);
    \coordinate (URtr) at (2*\a,2*\b);
    \coordinate (URtl) at (0,2*\b);

    \coordinate (LRtl) at (0,0);
    \coordinate (LRtr) at (2*\a,0);
    \coordinate (LRbr) at (2*\a,-2*\b);
    \coordinate (LRbl) at (0,-2*\b);

    \coordinate (Cup) at (\a,\b);
    \coordinate (Clo) at (\a,-\b);

    \draw[cell edge] (LLa) rectangle (RTa);
    \draw[cell edge] (-2*\a,0) -- (VrP);

    \draw[cell edge] (URbl) rectangle (URtr);
    \draw[cell edge] (0,\b) -- (2*\a,\b);

    \draw[cell edge] (LRbl) rectangle (LRtr);
    \draw[cell edge] (0,-\b) -- (2*\a,-\b);

    \draw[diagonal] (LTa) -- (RBa);
    \draw[diagonal] (LLa) -- (RTa);

    \draw[diagonal] (URbl) -- (URtr);
    \draw[diagonal] (URtl) -- (URbr);

    \draw[diagonal] (LRtl) -- (LRbr);
    \draw[diagonal] (LRbl) -- (LRtr);

    \def\rbig{0.78}
    \def\rsmall{0.58}
    \def\rlabbig{0.52}
    \def\rlabsmall{0.39}

    \draw[ang]
      ($(O)+({180-\th}:\rbig)$)
      arc[start angle={180-\th}, end angle=\th, radius=\rbig];
    \node at ($(O)+(90:\rlabbig)$) {$\alpha_1$};

    \draw[ang]
      ($(O)+({180+\th}:\rbig)$)
      arc[start angle={180+\th}, end angle={360-\th}, radius=\rbig];
    \node at ($(O)+(270:\rlabbig)$) {$\alpha_1$};

    \draw[ang]
      ($(O)+(0:\rsmall)$)
      arc[start angle=0, end angle=\th, radius=\rsmall];
    \node at ($(O)+({0.5*\th}:\rlabsmall)$) {$\alpha_2$};

    \draw[ang]
      ($(O)+({180-\th}:\rsmall)$)
      arc[start angle={180-\th}, end angle=180, radius=\rsmall];
    \node at ($(O)+({180-0.5*\th}:\rlabsmall)$) {$\alpha_2$};

    \draw[ang]
      ($(O)+(180:\rsmall)$)
      arc[start angle=180, end angle={180+\th}, radius=\rsmall];
    \node at ($(O)+({180+0.5*\th}:\rlabsmall)$) {$\alpha_2$};

    \draw[ang]
      ($(O)+({360-\th}:\rsmall)$)
      arc[start angle={360-\th}, end angle=360, radius=\rsmall];
    \node at ($(O)+({360-0.5*\th}:\rlabsmall)$) {$\alpha_2$};

    \def\rjalpha{0.52}
    \def\rjalphaLabel{0.35}

    \def\rjhalf{0.82}
    \def\rjhalfLabel{0.56}

    \draw[ang]
      ($(VrP)+(\th:\rjhalf)$)
      arc[start angle=\th, end angle=90, radius=\rjhalf];
    \node[font=\scriptsize]
      at ($(VrP)+({0.5*(\th+90)}:\rjhalfLabel)$)
      {$\frac{\alpha_1}{2}$};

    \draw[ang]
      ($(VrP)+(0:\rjalpha)$)
      arc[start angle=0, end angle=\th, radius=\rjalpha];
    \node[font=\scriptsize]
      at ($(VrP)+({0.5*\th}:\rjalphaLabel)$)
      {$\alpha_2$};

    \draw[ang]
      ($(VrP)+({360-\th}:\rjalpha)$)
      arc[start angle={360-\th}, end angle=360, radius=\rjalpha];
    \node[font=\scriptsize]
      at ($(VrP)+({360-0.5*\th}:\rjalphaLabel)$)
      {$\alpha_2$};

    \draw[ang]
      ($(VrP)+(270:\rjhalf)$)
      arc[start angle=270, end angle={360-\th}, radius=\rjhalf];
    \node[font=\scriptsize]
      at ($(VrP)+({0.5*(270+360-\th)}:\rjhalfLabel)$)
      {$\frac{\alpha_1}{2}$};

    \def\s{0.22}
    
    \draw[ang]
      ($(VrP)+(-\s,0)$)
      -- ($(VrP)+(-\s,\s)$)
      -- ($(VrP)+(0,\s)$);
    
    \draw[ang]
      ($(VrP)+(-\s,0)$)
      -- ($(VrP)+(-\s,-\s)$)
      -- ($(VrP)+(0,-\s)$);

    \def\Olabelheight{0.88}

    \coordinate (Oelbow) at
      ($(O)+({-\a/\b*\Olabelheight},\Olabelheight)$);
    
    \node[left=6pt of Oelbow] (labO) {$O$};

    \draw[-]
      (labO.east)
      -- (Oelbow)
      -- (O);

    \node at ($(VrP)+(-0.26,-0.70)$) (labVr) {$\Vr$};
    \draw[-]
      (labVr.north)
      -- ($(VrP)$);
    
    \end{tikzpicture}
    \caption{ Angles around vertices in the symmetric waterbomb pattern where \(a = c\). Mountain edges are colored red, valley edges colored dashed blue.}
    \label{fig:Kawasaki}
\end{figure}
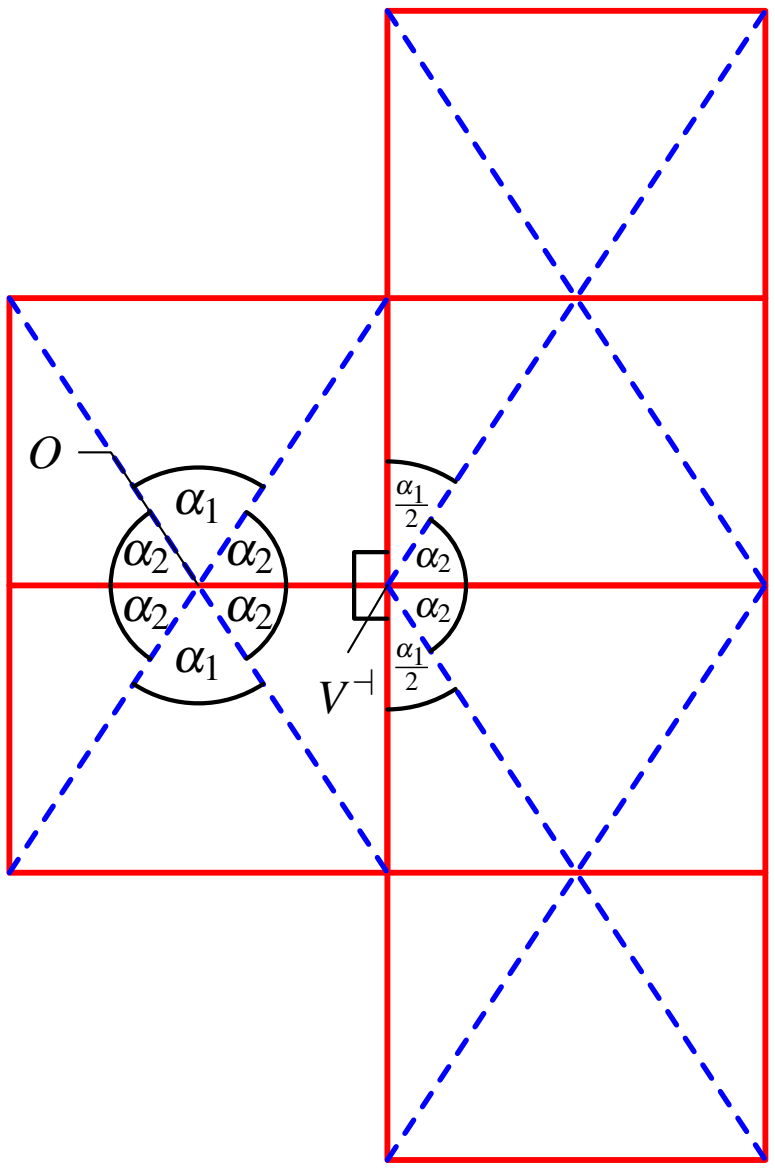

To define the rotation operator, the angle $\theta$
needs to be expressed as a function of the 
design parameters $a,b,c$ and folding angle $\gamma$. Observe that $\theta$ is the angle between the planes
$P_1$ containing $V_\gamma^\dashv, U_\gamma^\urcorner, U_\gamma^\ulcorner$ and
$P_2$ containing $V_\gamma^\dashv, U_\gamma^\lrcorner, U_\gamma^\llcorner$.
It can be determined as follows.
Let 
\begin{align*}
    N_1 & = (U^\urcorner_\gamma - U^\ulcorner_\gamma) \times (U_\gamma^\urcorner - V_\gamma^\dashv)=[0,2a(-\zro+c\cos(\gamma)),2a\yro] \text{ and} \\
    N_2 & = (U^\lrcorner_\gamma - U^\llcorner_\gamma) \times (V_\gamma^\dashv - U_\gamma^\lrcorner)=[0,2a(\zro-c\cos(\gamma)),2a\yro]
\end{align*}
denote %
vectors perpendicular to $P_1$ and $P_2$, respectively.  Note the correction of the sign for \(N_2\) compared to \cite{chudoba2021} leading to the correct angle \(\theta\) below.
Thus the equations for the planes are $P_i: (X-V_\gamma^\dashv)N_i=0.$\\
Subtracting $(E_3)$ from $(E_2)$ in Table~\ref{tab:wbequations_sym} yields $x^\urcorner$ and Equation $(E_1)$ immediately yields $z^\urcorner$, namely
\begin{align*}
    x^\urcorner &= a, \\
    z^\urcorner &= \frac{a(1 - \sin(\gamma))}{\cos(\gamma)}.
\end{align*}
Adding $E_2$ and $E_3$ yields $(y^\urcorner)^2 = b^2 - (z^\urcorner)^2$, thus the positive solution reads
\[
    y^\urcorner = \frac{\sqrt{b^2 \cos^2(\gamma) - a^2(1 - \sin(\gamma))^2}}{\cos(\gamma)}.
\]
Using this, the angle between $P_1$ and $P_2$ can be determined as follows
\begin{align}
\cos(\theta)=  \frac{N_1\cdot N_2}{|N_1||N_2|} &=
  -\frac{(\zro-c\cos(\gamma))^2 -{(\yro)}^2 }{
    {(\zro-c\cos(\gamma))^2 +(\yro)}^2
  } \nonumber \\
  &=  -\frac{c^2\cos^2(\gamma)-2ac(1-\sin(\gamma))
    +2\frac{a^2(1-\sin(\gamma))^2}{\cos^2(\gamma)} -b^2
  }{c^2\cos^2(\gamma)-2ac(1-\sin(\gamma))+b^2}.
\label{eq:cos_theta}
\end{align}
Having derived the angle \(\theta\), the global curved geometry characteristics, i.e. curvature and radius, can be determined, see Figure~\ref{fig:generation}(b) and Section~\ref{sec:global_geometry_sym}.
In the maximally folded state, i.e. \(\Vl = \Vr\), this implies that
\[\cos(\theta)=  \frac{c^2-2ac
    +2a^2-b^2
}{c^2-2ac+b^2} \quad \text{if } \gamma = 0.\]
Note that for $a=c$ this implies that in the maximally folded state all points
$V_{0}^\vdash, V_{0}^\dashv, U_{0}^\ulcorner, U_{0}^\urcorner, U_{0}^\llcorner$ and
$U_{0}^\lrcorner$ lie in one plane, i.e. the algebraic model reaches a coplanar configuration.  The standard single-vertex flat-foldability conditions clarify when this coplanar state also respects the prescribed mountain--valley assignment \cite{DemaineORourke2007}. For \(a=c\), Kawasaki's alternating-angle condition holds at both vertex classes: the alternating sum is \(\alpha_2-\alpha_1+\alpha_2-\alpha_2+\alpha_1-\alpha_2=0\) at \(O\), and \(\alpha_2-\alpha_1/2+\pi/2-\pi/2+\alpha_1/2-\alpha_2=0\) at \(\Vr\); see Figure~\ref{fig:Kawasaki}. Maekawa's condition is also satisfied, since four valley and two mountain folds meet at \(O\), whereas two valley and four mountain folds meet at \(\Vr\), so the number of mountain folds and valley folds differs by two in both cases.

The Big-Little-Big lemma adds a restriction that is not detected by coplanarity alone. At \(O\), the two sectors of size \(\alpha_1\) are each bounded by valley folds. Consequently, if \(\alpha_1<\alpha_2\), these sectors are strict local minima bounded by equal fold assignments, in conflict with the lemma. Since \(\alpha_1=\pi-2\alpha_2\) and \(\tan\alpha_2=b/a\), the mountain--valley assignment shown in Figure~\ref{fig:Kawasaki} is locally compatible with this condition precisely when \(b/a\leq\sqrt{3}\). At \(\Vr\), every potentially minimal sector is bounded by one mountain and one valley fold, so no further restriction arises. Thus, the algebraic derivation certifies a rigid path to a coplanar tessellation; a non-self-intersecting physical flat fold additionally requires the local mountain--valley and layer-ordering conditions just described.

Since \(\xro = a\), the value of the shift $\delta$ parallel to the $x$-axis is obtained as
\begin{align*}
    \delta = a + c \sin(\gamma).
\end{align*}
With the help of the parameters $\theta$ and $\delta$
defined as functions of the cell design parameters $a,b,c$ and folding parameter $\gamma$
an arbitrarily large tessellation can be
computed exactly without numerical approximation.
In contrast to existing numerical rigid origami models
\cite{Tachi2009,Tachi2010}, in the derived algebraic approach
the positions of each cell within a shell tessellation are determined independently of all other cells, solely from the position and kinematic folding state of the master cell. 
The algebraic model has been implemented using the support of computer algebra systems Maple \cite{MAPLE} and Sympy \cite{SymPyOfficial}.

\subsection{Folding semi-symmetrically}
\label{sec:semisymmetric_waterbomb_tessellation}
\begin{figure}
\centering
    \includegraphics[width=0.9\textwidth]{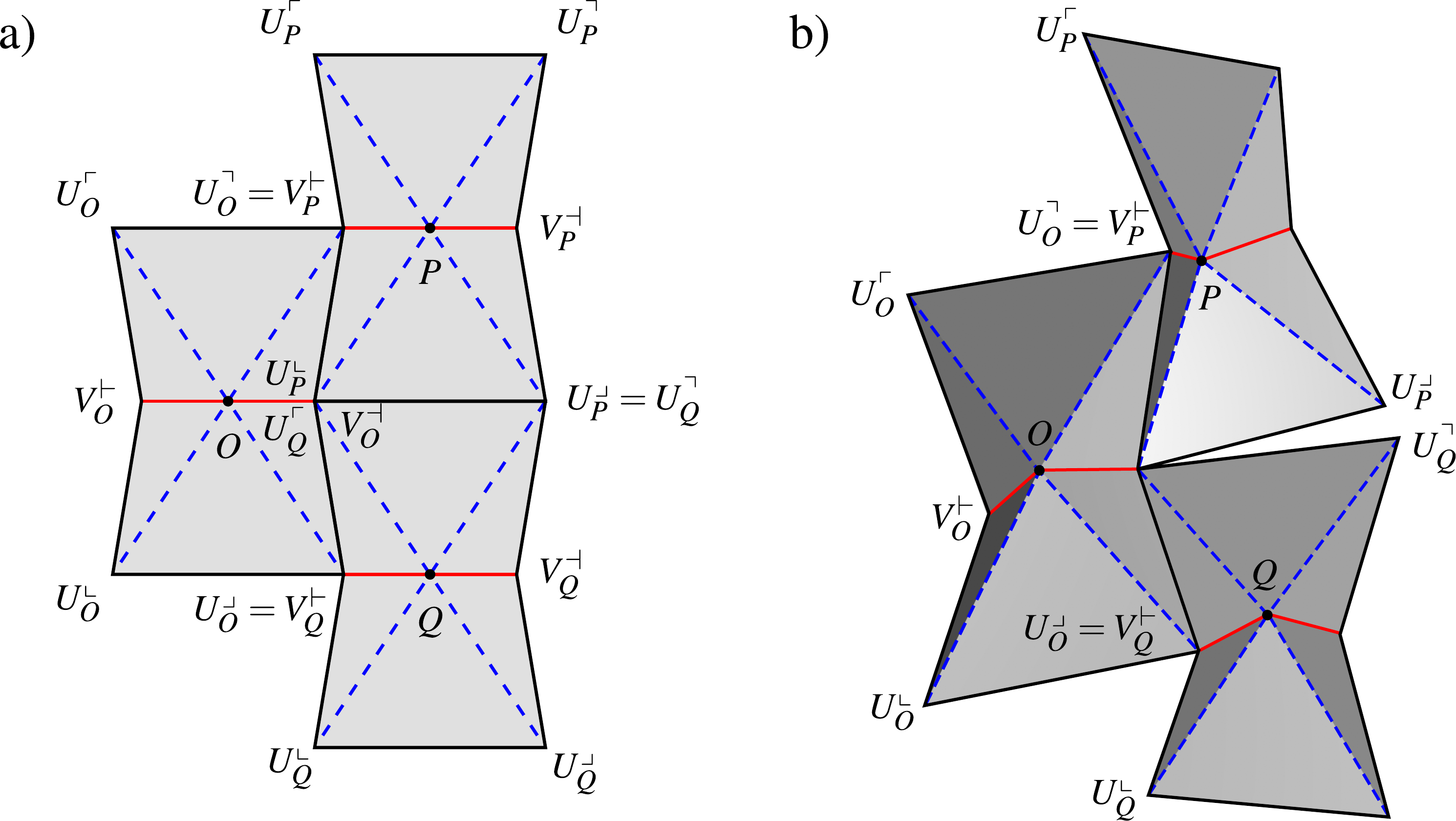}
\caption{Attaching two (congruent) neighboring cells: a) Flat configuration; b) Folded configuration; rotations of cells about common edges still required to match $\Uru_P$ and $\Uro_Q$}
\label{fig:attachtwo}
\end{figure}

 This subsection follows the assembly construction of \cite[Section 4]{robertz_iass2022} and recapitulates it for completeness in the present notation.
The essential step in constructing the semi-symmetric waterbomb tessellation
is to assemble three copies of the base cell as shown in Figure~\ref{fig:attachtwo},
where the edges $\Vr_O \Uro_O = \Ulu_P \Vl_P$
and $\Uru_O \Vr_O = \Vl_Q \Ulo_Q$ joining a pair of triangles can be thought of as hinges.

The aim is to determine the values of the two dihedral angles, namely $\sigma$ for the ``hinge'' $\Vr_O \Uro_O = \Ulu_P \Vl_P$
and $\rho$ for the ``hinge'' $\Uru_O \Vr_O = \Vl_Q \Ulo_Q$, as functions of $\lena$, $\lenb$, $\lenc$, $\foldanglebeta$, $\foldanglegamma$, such that the pair of triangles $P\Ulu_P\Uru_P$ and $Q\Uro_Q\Ulo_Q$
join along a common edge $\Ulu_P\Uru_P = \Ulo_Q\Uro_Q$ as well.

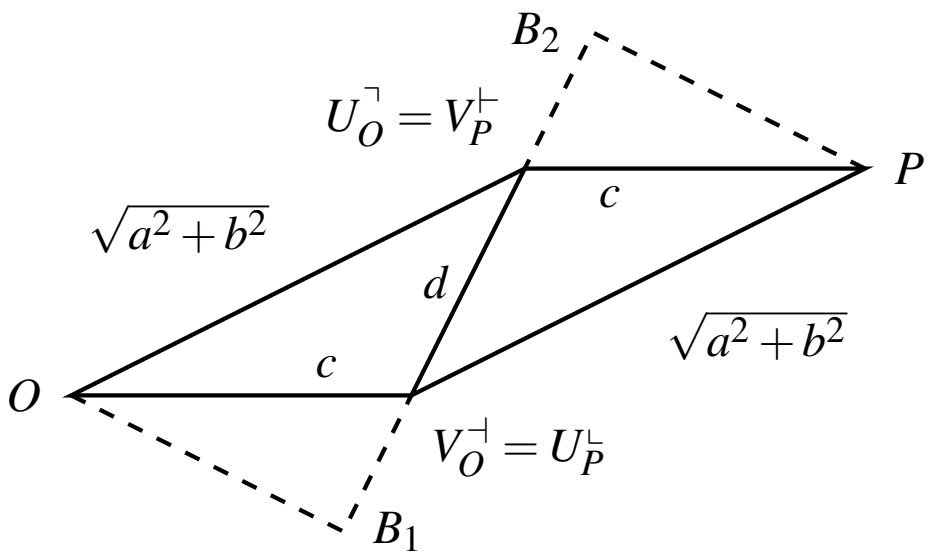
\begin{figure}
\centering 
\begin{tikzpicture}[thick,scale=0.75]
\coordinate (PP) at (7,0);
\coordinate (Vr) at (3,-2);
\coordinate (Vl) at (4,0);
\coordinate (OO) at (0,-2);
\coordinate (B2) at (4.6,1.2);
\coordinate (B1) at (2.4,-3.2);

\draw[solid] (PP) -- node [near end,below] {$\lenc$} (Vl) -- node [midway,left] {$\lend$} (Vr) -- node [midway,below right=2pt and 2pt] {$\sqrt{\lena^2+\lenb^2}$} cycle;
\draw[solid] (Vr) -- node [near start,above] {$\lenc$} (OO) -- node [midway,above left=2pt and 2pt] {$\sqrt{\lena^2+\lenb^2}$} (Vl);
\draw[dashed] (PP) -- (B2) -- (Vl);
\draw[dashed] (OO) -- (B1) -- (Vr);

\node (RO) [left=0.25em of OO] {$O$};
\node (LP) [right=0.25em of PP] {$P$};
\node (RVl) [above left=1pt of Vl] {$\Uro_O = \Vl_P$};
\node (LUro) [below right=1pt of Vr] {$\Vr_O = \Ulu_P$};
\node (AB2) [left=0.25em of B2] {$B_2$};
\node (BB1) [right=0.25em of B1] {$B_1$};
\end{tikzpicture}%
\caption{Orthogonal projections $B_1$ and $B_2$ of $O$ and $P$ on the (diagonal) axis of rotation,  reproduced from \cite{robertz_iass2022}}
\label{fig:orthproj}
\end{figure}

Assuming that the waterbomb cell shown on the left in Figures~\ref{fig:attachtwo}~(a) and (b) is in a fixed position with central vertex $O$ at $[0, 0, 0]$, we first determine the coordinates of the central vertices $P$ and $Q$
of the two neighboring cells.
Let $B_1$ and $B_2$ be the orthogonal projections of $O$ and $P$, respectively, on the line spanned by the common edge of the triangles $O\Vr_O\Uro_O$ and $P\Ulu_P\Vl_P$
as indicated in Figure~\ref{fig:orthproj}.

We choose normal vectors to the common edge as $OB_1$ and $O\Uro_O \times \Vr_O\Uro_O$, which results after normalization into
\[
n_1 \, = \, \frac{1}{\lend} \, \big[ \,
(\lena-\lenc) \cos(\foldanglegamma) \cos(\dihedral_1) + \lenb \sin(\foldanglegamma),
-(\lena-\lenc) \sin(\dihedral_1), -(\lena-\lenc) \sin(\foldanglegamma) \cos(\dihedral_1) + \lenb \cos(\foldanglegamma)
\, \big]
\]
and
\[
n_2 \, = \, \big[ \, \cos(\foldanglegamma) \sin(\dihedral_1), \cos(\dihedral_1), -\sin(\foldanglegamma) \sin(\dihedral_1) \, \big]\,.
\]
We use a rational parametrization in $t$ to describe $P$ as
\[
P \, = \, B_2 + |n_1| \, \left( -\frac{1-t^2}{1+t^2} \, n_1 + \frac{2 \, t}{1+t^2} \, n_2 \right), \qquad t \, = \, \tan\left( \frac{\sigma}{2} \right),
\]
so that $O\Vr_O P\Uro_O$ is a flat parallelogram if and only if $t = 0$,
and we obtain explicitly
\[
\begin{array}{c}
\displaystyle
P \, = \, \frac{1}{\lend^2 \, (1+t^2)} \, \Big[
2 \, \lenb \, \lenc \, \lend \cos(\foldanglegamma) \sin(\dihedral_1) \, t + \left( (\lena+\lenc) \sin(\foldanglegamma) - \lenb \cos(\foldanglegamma) \cos(\dihedral_1) \right) \lend^2 \, t^2 \, +\\[0.2em]
\qquad \qquad (\lena^2+\lenb^2-\lenc^2) \left( (\lena-\lenc) \sin(\foldanglegamma) - \lenb \cos(\foldanglegamma) \cos(\dihedral_1) \right),\\[0.2em]
\, 2 \, \lenb \, \lenc \, \lend \cos(\dihedral_1) \, t + (\lend^2 \, t^2 + \lena^2 + \lenb^2 - \lenc^2) \, \lenb \sin(\dihedral_1),\\[0.2em]
\qquad \qquad \qquad \quad -2 \, \lenb \, \lenc \, \lend \sin(\foldanglegamma) \sin(\dihedral_1) \, t + \left( (\lena+\lenc) \cos(\foldanglegamma) + \lenb \sin(\foldanglegamma) \cos(\dihedral_1) \right) \lend^2 \, t^2 \, +\\[0.2em]
\qquad \qquad (\lena^2+\lenb^2-\lenc^2) \left( (\lena-\lenc) \cos(\foldanglegamma) + \lenb \sin(\foldanglegamma) \cos(\dihedral_1) \right)
\Big]\,.
\end{array}
\]
The analogous construction with $u = \tan(\rho/2)$ yields the coordinates of $Q$ as follows:
\[
\begin{array}{c}
\displaystyle
Q \, = \, \frac{1}{\lend^2 \, (1+u^2)} \, \Big[
2 \, \lenb \, \lenc \, \lend \cos(\foldanglegamma) \sin(\dihedral_3) \, u + \left( (\lena+\lenc) \sin(\foldanglegamma) - \lenb \cos(\foldanglegamma) \cos(\dihedral_3) \right) \lend^2 \, u^2 \, +\\[0.2em]
\qquad \qquad (\lena^2+\lenb^2-\lenc^2) \left( (\lena-\lenc) \sin(\foldanglegamma) - \lenb \cos(\foldanglegamma) \cos(\dihedral_3) \right),\\[0.2em]
\, -2 \, \lenb \, \lenc \, \lend \cos(\dihedral_3) \, u - (\lend^2 \, u^2 + \lena^2 + \lenb^2 - \lenc^2) \, \lenb \sin(\dihedral_3),\\[0.2em]
\qquad \qquad \qquad \quad -2 \, \lenb \, \lenc \, \lend \sin(\foldanglegamma) \sin(\dihedral_3) \, u + \left( (\lena+\lenc) \cos(\foldanglegamma) + \lenb \sin(\foldanglegamma) \cos(\dihedral_3) \right) \lend^2 \, u^2 \, +\\[0.2em]
\qquad \qquad (\lena^2+\lenb^2-\lenc^2) \left( (\lena-\lenc) \cos(\foldanglegamma) + \lenb \sin(\foldanglegamma) \cos(\dihedral_3) \right)
\Big]\,.
\end{array}
\]
Next, given the parametrization of $\Uru_P$ as vertex of the attached cell with central vertex $P$ and of $\Uro_Q$ of the attached cell with central vertex $Q$,
the distance between $\Uru_P$ and the point $\Uro_O$ of the cell with central vertex $O$ is
\[
|\Uru_P - \Uro_O| \, \, = \, \,
|\Uru_P - \Vl_P| \, \, = \, \,
\sqrt{\lend^2 + 4 \, \lenc \, \sin(\foldanglegamma) \, (\lena \, \sin(\foldanglegamma) - \lenb \cos(\foldanglegamma) \cos(\dihedral_4))}\,,
\]
whereas the distance between $\Uro_Q$ and the point $\Uru_O$ of the cell with central vertex $O$ is
\[
|\Uro_Q - \Uru_O| \, \, = \, \,
|\Uro_Q - \Vl_Q| \, \, = \, \,
\sqrt{\lend^2 + 4 \, \lenc \, \sin(\foldanglegamma) \, (\lena \, \sin(\foldanglegamma) - \lenb \cos(\foldanglegamma) \cos(\dihedral_2))}\,.
\]
Now, the two triangles $\Vr_O P \Uru_P$ and $\Vr_O Q \Uro_Q$ join along a common edge if and only if
their vertices $\Uru_P$ and $\Uro_Q$ are in the same position. We refer to that common location as $R$.
It is a point of intersection of the three spheres
\begin{itemize}
\item $S^{(1)}$ with center $\Uro_O$ and radius $|\Uru_P - \Uro_O|$ determined above,
\item $S^{(2)}$ with center $\Uru_O$ and radius $|\Uro_Q - \Uru_O|$ determined above,
\item $S^{(3)}$ with center $V^{\dashv}_O$ and radius $2 \lena$.
\end{itemize}
By solving the corresponding system of equations symbolically, we find two points of intersection $R^{(1)}$ and $R^{(2)}$ depending on $\lena$, $\lenb$, $\lenc$, $\foldanglegamma$ and $\dihedral_1$, \ldots, $\dihedral_4$.
Finally, in order to determine $t = \tan(\sigma/2)$, $u = \tan(\rho/2)$,
we solve the equations
\[
|P - R^{(i)}|^2 \, \, = \, \, \lena^2 + \lenb^2, \qquad
|Q - R^{(i)}|^2 \, \, = \, \, \lena^2 + \lenb^2, \qquad i = 1, 2,
\]
resulting in four candidate solutions overall, one of which is given by
\[
\begin{array}{rcl}
\displaystyle
\tan\left( \frac{\rho}{2} \right) \!\! & \!\! = \!\! & \!\!
\Big[ \, -\lenb \, (\lena-\lenc) \, (\cos(\dihedral_1+\dihedral_3) - 1) \, \sqrt{W}
- \lenb \, \Big( \,
\lenc \, (\cos(\dihedral_2) \, \sqrlene - \cos(\dihedral_4) \, \sqrlend) \sin(\dihedral_1+\dihedral_3) \sin(2 \, \gamma)\\[1em]
\multicolumn{3}{r}{+ \, \lena \, \lenb \sin(\dihedral_1+\dihedral_3) \, (\cos(\dihedral_1+\dihedral_3) - 1) \, (2 \, \lena + \lenc \, (\cos(2 \, \gamma) - 1)) \, \Big) - (\cos(\dihedral_1+\dihedral_3) - 1) \, (\sqrlend + \sqrlene) \,
\sqrt{S_{\rho}} \, \Big] \, \Big/}\\[1em]
\multicolumn{3}{r}{\Big[ \, \lend \, (-\lenb \sin(\dihedral_1+\dihedral_3) \, \sqrt{W} - T_{\rho}) \, \Big]}
\end{array}
\]
and
\[
\begin{array}{rcl}
\displaystyle
\tan\left( \frac{\sigma}{2} \right) \!\! & \!\! = \!\! & \!\!
\Big[ \, -\lenb \, (\lena-\lenc) \, (\cos(\dihedral_1+\dihedral_3) - 1) \, \sqrt{W}
- \lenb \, \Big( \,
\lenc \, (\cos(\dihedral_4) \, \sqrlene - \cos(\dihedral_2) \, \sqrlend) \sin(\dihedral_1+\dihedral_3) \sin(2 \, \foldanglegamma)\\[1em]
\multicolumn{3}{r}{+ \, \lena \, \lenb \sin(\dihedral_1+\dihedral_3) \, (\cos(\dihedral_1+\dihedral_3) - 1) \, (2 \, \lena + \lenc \, (\cos(2 \, \foldanglegamma) - 1)) \, \Big) - (\cos(\dihedral_1+\dihedral_3) - 1) \, (\sqrlend + \sqrlene) \,
\sqrt{S_{\sigma}} \, \Big] \, \Big/}\\[1em]
\multicolumn{3}{r}{ \, \Big[ \, \lend \, (-\lenb \sin(\dihedral_1+\dihedral_3) \, \sqrt{W} - T_{\sigma}) \, \Big]}
\end{array}
\]
where we use the abbreviations
\[
\sqrlend \, := \, (\lena-\lenc)^2 + \lenb^2\,, \qquad
\lend \, := \, \sqrt{\sqrlend}\,, \qquad
\sqrlene \, := \, (\lena-\lenc)^2 + \lenb^2 \cos(\dihedral_1+\dihedral_3)\,,
\]
as well as
\[
\begin{array}{rcl}
W \! & \! := \! & \! 2 \, ( 1 -\cos(\dihedral_1+\dihedral_3) ) \, \Big[ \,
(\lena^2 - \cos(\dihedral_2) \cos(\dihedral_4) \, \lenb^2) \, \lenc^2 \sin^2(2 \, \foldanglegamma) \, -\\[1em]
& & \!\!\!\!\!\!\!\!\!\!
\lena \, \lenb \, \lenc \, (\cos(\dihedral_2) + \cos(\dihedral_4)) \, (2 \, \lena + \lenc \, (\cos(2 \, \foldanglegamma) - 1)) \sin(2 \, \foldanglegamma) \, -\\[1em]
& & \!\!\!\!\!\!\!\!\!\!
2 \, \lena^2 \, \lenc \, (2 \, \lena - \lenc) \, (\cos(2 \, \foldanglegamma) + 1) + 2 \, \lena^2 \, \lenb^2 \, (\cos(\dihedral_1+\dihedral_3) + 1) \, \Big]
- \sqrlend \, (\cos(\dihedral_2) - \cos(\dihedral_4))^2 \, \lenc^2 \sin^2(2 \, \foldanglegamma)
\end{array}
\]
and
\[
\begin{array}{rcl}
S_{\rho} \! & \! := \! & \!
4 \, \lena \, \lenb^2 \, (\lenb \cos(\dihedral_2) \sin(2 \, \foldanglegamma) + \lena \, (\cos(2 \, \foldanglegamma) + 1))
- (\lena^2 + \lenb^2) \, (\lenb \cos(\dihedral_2) \sin(2 \, \foldanglegamma) + \lena \, (\cos(2 \, \foldanglegamma) + 1))^2
\end{array}
\]
and
\[
\begin{array}{rcl}
T_{\rho} \! & \! := \! & \!
(\cos(\dihedral_1+\dihedral_3) - 1) \, \Big[ \, \lena \left( \lenb^2
 \left( \cos(\dihedral_1+\dihedral_3) + 1 \right) + 2 \, \lena \, (\lena - \lenc) \right) (\cos(2 \, \foldanglegamma) + 1)\\[1em]
& & + \, \lenb \left( (\sqrlend + \sqrlene) \cos(\dihedral_2) + (\lena - \lenc) \, \lenc \, (\cos(\dihedral_2) + \cos(\dihedral_4)) \right) \sin(2 \, \foldanglegamma)
- 2 \, \lena \, \lenb^2 \, (\cos(\dihedral_1+\dihedral_3)+1) \, \Big]
\end{array}
\]
as well as
\[
\begin{array}{rcl}
S_{\sigma} \! & \! := \! & \!
4 \, \lena \, \lenb^2 \, (\lenb \cos(\dihedral_4) \sin(2 \, \foldanglegamma) + \lena \, (\cos(2 \, \foldanglegamma) + 1))
- (\lena^2 + \lenb^2) \, (\lenb \cos(\dihedral_4) \sin(2 \, \foldanglegamma) + \lena \, (\cos(2 \, \foldanglegamma) + 1))^2
\end{array}
\]
and
\[
\begin{array}{rcl}
T_{\sigma} \! & \! := \! & \!
(\cos(\dihedral_1+\dihedral_3) - 1) \, \Big[ \, \lena \left( \lenb^2
 \left( \cos(\dihedral_1+\dihedral_3) + 1 \right) + 2 \, \lena \, (\lena - \lenc) \right) (\cos(2 \, \foldanglegamma) + 1)\\[1em]
& & + \, \lenb \left( (\sqrlend + \sqrlene) \, \cos(\dihedral_4) + (\lena - \lenc) \, \lenc \, (\cos(\dihedral_2) + \cos(\dihedral_4)) \right) \sin(2 \, \foldanglegamma)
- 2 \, \lena \, \lenb^2 \, (\cos(\dihedral_1+\dihedral_3)+1) \, \Big].
\end{array}
\]

\section{Cylindrical global geometries from waterbomb cell}\label{sec:folded_geometries}
 Uniform tessellations of (semi-)symmetric waterbomb cells form cylindrical global geometries; Theorem~\ref{thm:global_curvature} makes this statement precise. In this section we investigate these geometries and provide parameter studies for symmetric and semi-symmetric foldings, building on \cite{chudoba2021, robertz_iass2022}. We also give a new explicit algebraic description of the cylinders and derive the conditions for planar configurations.

All statements about axes and inclinations use the reference frame fixed in Section~\ref{sec:parametrization} and Figure~\ref{fig:pattern_semisym}: \(O\) is the origin, \(\Vl,\Vr\) lie in the \(xz\)-plane, and the \(z\)-axis is their symmetry axis.

With the compatibility of the waterbomb cells derived in the previous sections, we can define $\Tro,\Tru \in \mathrm{SE}(3)$ as the two neighbor-cell isometries defined by the attachment conditions \begin{equation*}
    \Tro(\Vl)=\Uro,
    \qquad
    \Tro(\Ulu)=\Vr,
\end{equation*}
and
\begin{equation*}
    \Tru(\Ulo)=\Vr,
    \qquad
    \Tru(\Vl)=\Uru.
\end{equation*}
Thus, \(\Tro\) attaches the vertices \(\Vl\) and \(\Ulu\) of a new cell to the vertices \(\Uro\) and \(\Vr\), respectively, of an existing cell, whereas \(\Tru\) attaches the vertices \(\Ulo\) and \(\Vl\) of a new cell to the vertices \(\Vr\) and \(\Uru\), respectively, of an existing cell. By the compatibility established above, the two isometries commute,
\begin{equation}
    \Tro\Tru=\Tru\Tro.
    \label{eq:commuting-tessellation-generators}
\end{equation}
Write
\begin{equation*}
    \Tro(X)
    =
    \Qro X+\tro,
    \qquad
    \Qro\in\mathrm{SO}(3),
    \quad
    \tro\in\mathbb{R}^3.
\end{equation*}
Let
\[
    \Qro
    =
    \bigl(q_{ij}\bigr)_{i,j=1,\ldots,3}.
\]
We first state the curvature result in a form that isolates its main
geometric conclusion. The subsequent technical theorem gives the axis
and explicit representations of the corresponding cylinder.

\begin{figure}[h]
\centering
    \includegraphics[width=0.9\textwidth]{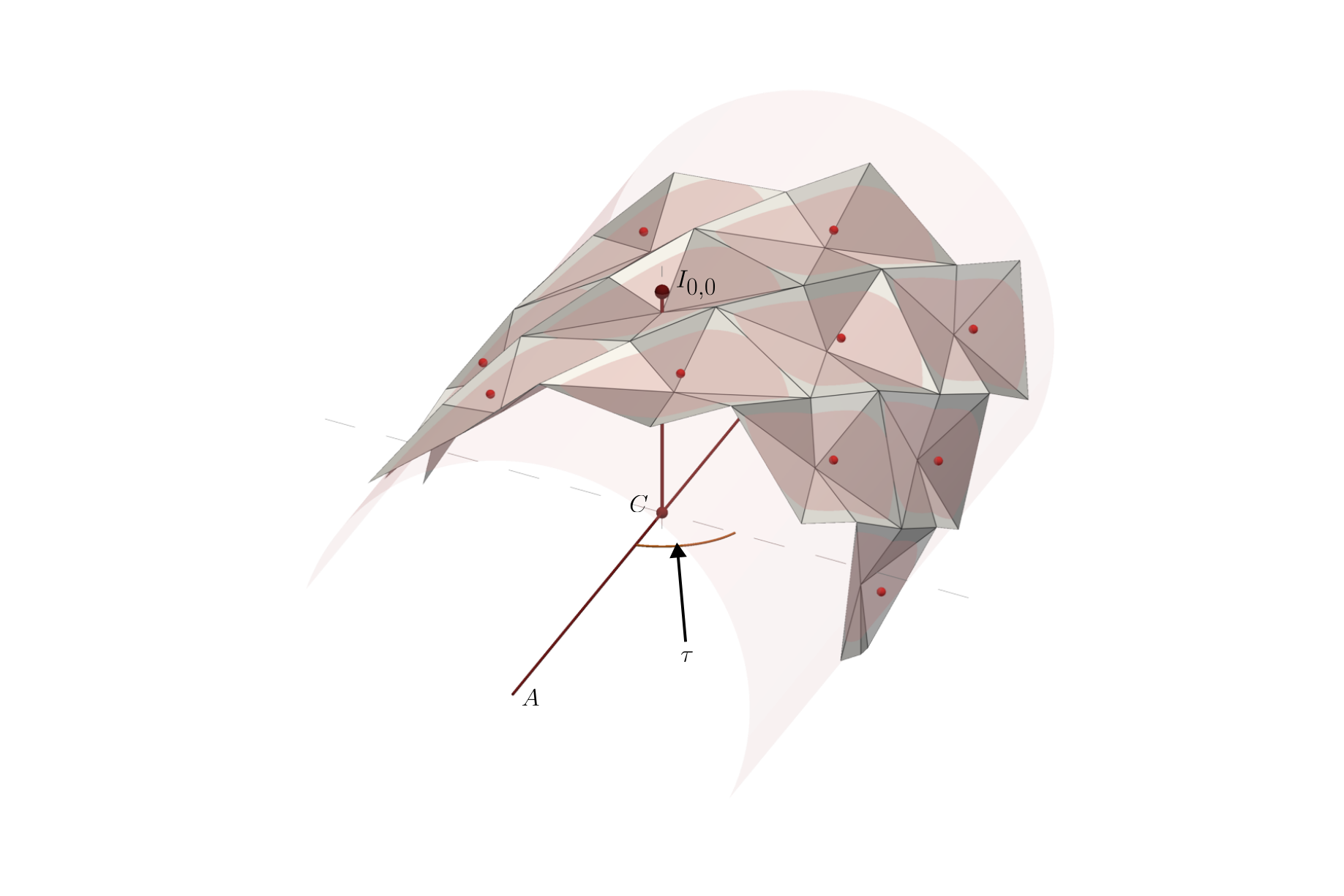}
\caption{Curvature cylinder of a uniform semi-symmetric waterbomb tessellation. The gray cells form the tessellated shell, and the highlighted points \(I_{m,n}\) lie on the light-red cylinder \(C_I\). The cylinder has horizontal axis \(A\) through \(C\) and radius \(r\). The angle \(\tau\) measures the rotation of \(A\) about the \(z\)-axis from the \(x\)-directed cylinder axis of the symmetric case. The configuration shown has \(a=1\), \(b=2\), \(c=1\), \(\gamma=1.25\), and \(\Delta\beta=-0.08\).}
\label{fig:curvature_derivation}
\end{figure}

\begin{samepage}
\begin{theorem}[Curvature of a uniform semi-symmetric waterbomb tessellation]
\label{thm:global_curvature}
Consider the uniform tessellation generated by the commuting
neighbor-cell isometries \(\Tro\) and \(\Tru\) above, and assume the
non-degenerate case
\[
    q_{32}^2+q_{13}^2>0.
\]
In the reference waterbomb cell \(\mathcal{B}_{0,0}\), let \(I\) be the
intersection of \(\overline{\Vl\Vr}\) with the \(z\)-axis. In the
parametrization above,
    \begin{equation*}
        I
        =
        (0,0,c\cos\gamma)^{\mathsf T}.
    \end{equation*}
The corresponding point in each cell of the tessellation is
    \begin{equation*}
        I_{m,n}
        :=
        \Tro^{\,m}\Tru^{\,n}(I),
        \qquad
        (m,n)\in\mathbb{Z}^2.
    \end{equation*}
There is a horizontal line \(A\), intersecting the
\(z\)-axis at \(C=(0,0,z_C)^{\mathsf T}\), such that all points
\(I_{m,n}\) have the same distance
    \begin{equation*}
        r
        =
        \operatorname{dist}(I,A)
        =
        \left|
            c\cos\gamma-z_C
        \right|
    \end{equation*}
from \(A\). Provided \(r>0\), the points \(I_{m,n}\) therefore
lie on a common circular cylinder \(C_I\). One of its principal, as illustrated in
Figure~\ref{fig:curvature_derivation}. One of its principal curvatures vanishes, while the magnitude of the other is
    \begin{equation*}
        \kappa
        =
        \frac{1}{r}
        =
        \frac{1}{
            \left|
                c\cos\gamma-z_C
            \right|
        }.
    \end{equation*}
\end{theorem}
\end{samepage}

\begin{theorem}[Technical description of the curvature cylinder]
\label{thm:global-cylinder-I}
Under the hypotheses and with the notation of
Theorem~\ref{thm:global_curvature}, define
\begin{equation}
    \tau
    :=
    \operatorname{atan2}
    \bigl(
        q_{13},
        q_{32}
    \bigr)
    \pmod{\pi},
    \label{eq:def-tau-global-cylinder}
\end{equation}
and
\begin{equation*}
    a_\tau
    :=
    \begin{pmatrix}
        \cos\tau\\
        \sin\tau\\
        0
    \end{pmatrix},
    \qquad
    b_\tau
    :=
    \begin{pmatrix}
        -\sin\tau\\
        \cos\tau\\
        0
    \end{pmatrix},
    \qquad
    e_z
    :=
    \begin{pmatrix}
        0\\
        0\\
        1
    \end{pmatrix}.
\end{equation*}
Furthermore, set
\begin{equation*}
    h_{\urcorner}
    :=
    a_\tau^{\mathsf T}\tro
\end{equation*}
and
\begin{equation}
    z_C
    :=
    \frac{
        \bigl(
            \tro-h_{\urcorner}a_\tau
        \bigr)^{\mathsf T}
        \bigl(
            e_z-\Qro e_z
        \bigr)
    }{
        \left\|
            e_z-\Qro e_z
        \right\|^2
    }.
    \label{eq:def-zC-global-cylinder}
\end{equation}

In the non-collapsed case \(r>0\), the cylinder \(C_I\) from
Theorem~\ref{thm:global_curvature} has the following explicit
description (see Figure~\ref{fig:curvature_derivation}):

\begin{enumerate}
    \item
    The center of the distinguished circular cross-section containing
    \(I\) is
    \begin{equation*}
        C
        =
        \begin{pmatrix}
            0\\
            0\\
            z_C
        \end{pmatrix},
    \end{equation*}
    and the cylinder axis is
    \begin{equation*}
        A
        =
        \left\{
            C+s\,a_\tau
            :
            s\in\mathbb{R}
        \right\}.
    \end{equation*}

    \item
    An explicit parametrization of \(C_I\) is
    \begin{equation*}
        X(s,\varphi)
        =
        C
        +
        s\,a_\tau
        +
        r
        \left(
            \cos\varphi\,e_z
            +
            \sin\varphi\,b_\tau
        \right),
        \qquad
        s\in\mathbb{R},
        \quad
        \varphi\in[0,2\pi).
    \end{equation*}
    Equivalently,
    \begin{equation}
        \begin{aligned}
            x(s,\varphi)
            &=
            s\cos\tau
            -
            r\sin\varphi\sin\tau,
            \\
            y(s,\varphi)
            &=
            s\sin\tau
            +
            r\sin\varphi\cos\tau,
            \\
            z(s,\varphi)
            &=
            z_C+r\cos\varphi.
        \end{aligned}
        \label{eq:explicit-global-cylinder-components}
    \end{equation}

    \item
    The corresponding implicit representation is
    \begin{equation}
        \left\|
            \left(
                \mathrm{Id}
                -
                a_\tau a_\tau^{\mathsf T}
            \right)
            (X-C)
        \right\|^2
        =
        r^2,
        \label{eq:implicit-global-cylinder-vector}
    \end{equation}
    or, in Cartesian coordinates,
    \begin{equation}
        \left(
            -x\sin\tau+y\cos\tau
        \right)^2
        +
        \left(
            z-z_C
        \right)^2
        =
        r^2.
        \label{eq:implicit-global-cylinder}
    \end{equation}
\end{enumerate}
\end{theorem}

Before proving the theorem in Section \ref{sec:proof_curvature} we present parameter studies showing the impact of the design parameters on the global geometry for the symmetric and semi-symmetric foldings.

\subsection{Patterns realizable from symmetric base cell}\label{sec:global_geometry_sym}

This subsection builds on the symmetric curvature study of \cite{chudoba2021} and relates it to Theorem~\ref{thm:global_curvature}. For symmetric foldings, the cylinder axis is parallel to the \(x\)-axis and its circular cross-section lies in the \(yz\)-plane. Table~\ref{tab:param_study_sym} illustrates how the folding state \(\gamma\) and the dimensionless cell parameters \(\eta=b/a\) and \(\zeta=c/a\) affect the resulting global shell geometry and its curvature magnitude \(\kappa\). In particular, \(\eta\) controls the vertical slenderness of the base cell, while \(\zeta\) controls the relative length of its central fold. The zero-curvature entry shows that the crease pattern can also form a planar panel with a two-dimensional rib pattern.

When a signed curvature is used, we orient the circular cross-section by the upward normal at the master cell and call the curvature positive when the center of curvature lies below the shell and negative when it lies above; reversing the orientation reverses this sign. By contrast, \(\kappa\) in Theorem~\ref{thm:global_curvature} and the values in Table~\ref{tab:param_study_sym} denote non-negative curvature magnitudes. Because every uniform pattern considered here is cylindrical, its principal curvatures are \(0\) and \(\pm\kappa\), and its Gaussian curvature is zero. The two principal curvatures therefore cannot be controlled independently within this uniform-cell family.

To construct structural members with a sufficient local bending moment capacity, the height of the cross section, that is $z^\dashv = c \cos(\gamma)$, is a relevant design parameter. An appealing feature of the waterbomb tessellation is the possibility to adjust the effective cross-sectional height in a certain range independently of the global shell curvature by appropriately changing the base parameters $a,b,c$.

For a symmetric folding, the cylinder axis \(A\) from
Theorem~\ref{thm:global_curvature} is parallel to the \(x\)-axis, and
the distinguished circular cross-section lies in the \(yz\)-plane.
Its center is therefore \(C=(0,0,z_C)^{\mathsf T}\). As sketched in
Figure~\ref{fig:generation}(b), \(C\) has the same distance from
\[
    I
    =
    \frac{1}{2}
    \left(
        V_\gamma^\dashv+V_\gamma^\vdash
    \right)
    =
    (0,0,c\cos\gamma)^{\mathsf T}
\]
and from the midpoint
\[
    M_U
    :=
    \frac{1}{2}
    \left(
        U_\gamma^\urcorner+U_\gamma^\ulcorner
    \right)
    =
    (0,\yro,\zro)^{\mathsf T}.
\]
Consequently,
\[
    \lVert C-I\rVert^2
    =
    \lVert C-M_U\rVert^2,
\]
or, equivalently,
\[
    \left(z_C-c\cos\gamma\right)^2
    =
    (\yro)^2+\left(z_C-\zro\right)^2.
\]
Using \((\yro)^2+(\zro)^2=b^2\) and solving for \(z_C\) gives
\[
    z_C
    =
    \frac{b^2-c^2\cos^2\gamma}
    {2\left(\zro-c\cos\gamma\right)}
    =
    \frac{\cos\gamma}{2}
    \frac{b^2-c^2\cos^2\gamma}
    {a(1-\sin\gamma)-c\cos^2\gamma}.
\]
In the notation of Theorem~\ref{thm:global_curvature}, the cylinder
radius in the symmetric case is therefore
\[
    r_{\mathrm{sym}}
    =
    r
    =
    \left|c\cos\gamma-z_C\right|
    =
    \frac{\cos\gamma}{2}
    \left|
        \frac{
            b^2+c^2\cos^2\gamma-2ac(1-\sin\gamma)
        }{
            a(1-\sin\gamma)-c\cos^2\gamma
        }
    \right|.
\]
Hence,
\[
    \kappa_{\mathrm{sym}}
    =
    \frac{1}{r_{\mathrm{sym}}}.
\]

\subsection{Patterns realizable from semi-symmetric base cell}
As in the symmetric case, tessellations generated from a semi-symmetric base cell form cylindrical global geometries. Semi-symmetry additionally rotates the cylinder axis relative to the reference frame and thereby changes the rib orientation. To isolate the influence of the semi-symmetric cell parameters, the study below varies one parameter at a time about a fixed reference configuration. It extends the corresponding parametric study of \cite{robertz_iass2022} by reporting the curvature magnitudes.

For an easier adjustment of the deviation from the symmetric case, the parameter $\beta$ is decomposed into two quantities, namely $\beta_\mathrm{sym}$ and $\Delta\beta$ according to
\begin{equation*}
    \beta = \beta_\mathrm{sym} + \Delta\beta,
\end{equation*}
where $\beta_\mathrm{sym}$ represents the angle $\beta$ in the case of symmetric folding of the waterbomb cell. For $\beta = \beta_\mathrm{sym}$ the coordinates of the vertices $\Ulo$ and $\Vr$ are obtained from \cite{chudoba2021} (after correcting the sign of $\sin{\gamma}$ for the $y$- and $z$-coordinates of $\Ulo$ in Eq.\ (2) and Eq.\ (3))
\begin{equation*}
    \Ulo = \Bigg[-a,\quad \sqrt{b^2 -a^2 \bigg(\frac{1 - \sin{\foldanglegamma}}{\cos{\foldanglegamma}}\bigg)^2},\quad a \bigg(\frac{1 - \sin{\foldanglegamma}}{\cos{\foldanglegamma}}\bigg)  \Bigg],\qquad \Vr = \Bigg[c\sin{\foldanglegamma},\quad 0,\quad c\cos{\foldanglegamma}\Bigg]. \\
\end{equation*}
Using
\begin{equation*}
    \cos{\beta} = \frac{\Ulo \cdot \Vr}{|\Ulo| \,  |\Vr|},
\end{equation*}
we obtain
\begin{equation*}
    \beta_\mathrm{sym} = \arccos{\bigg[\frac{a  (1 - 2 \sin(\foldanglegamma))}{\sqrt{a^2 + b^2}}\bigg]}.
\end{equation*}
The parameter $\Delta\beta$ controls the deviation from the symmetric case, e.g., $\Delta\beta = 0$ corresponds to a symmetric cell folding.

We use the dimensionless parameters $\eta=b/a$ and $\zeta=c/a$ and prescribe a reference configuration $(\foldanglegamma,\eta,\zeta,\Delta\beta)$. Varying one parameter at a time produces the shell geometries summarized in Table~\ref{tab:param_study_semi_sym}. The study demonstrates a range of cross-sectional heights and curvature magnitudes, while $\Delta\beta$ controls the direction and magnitude of the shell twist. The admissible algebraic range excludes two parameter values used in the symmetric study; the limiting rows in Table~\ref{tab:param_study_semi_sym} therefore use $\foldanglegamma=0.06$ and $\foldanglegamma=1.25$ instead of $\foldanglegamma=0$ and $\foldanglegamma=1.3$, respectively.

\begin{table*}[t]
\caption{Parametric study showing the effect of different symmetric cell parameters on the global shell geometries and their curvature magnitude $\kappa = 1/r$,  extended from \cite{robertz_iass2022} with additional information on curvature.}
\label{tab:param_study_sym}
\centering
\begin{tblr}{
  colspec = {llcccc},
  stretch = 2.0,
  column{3-6} = {c},
  cell{1}{1-6} = {font=\small},
  rows = {m},
}
\toprule
Variation & \renewcommand{\arraystretch}{1.0} \begin{tabular}[c]{@{}l@{}} Parameters \\ \footnotesize $(\foldanglegamma, \eta, \zeta)$\end{tabular} & Curvature & Front View & Top View & Perspective \\
\midrule
Reference & \renewcommand{\arraystretch}{1.0} \begin{tabular}[c]{@{}l@{}} $\foldanglegamma=1.050$ \\ $\eta=2.00$ \\ $\zeta=1.00$ \end{tabular} & 0.1161 & 
\includegraphics[height=0.9\paramheight, valign=m]{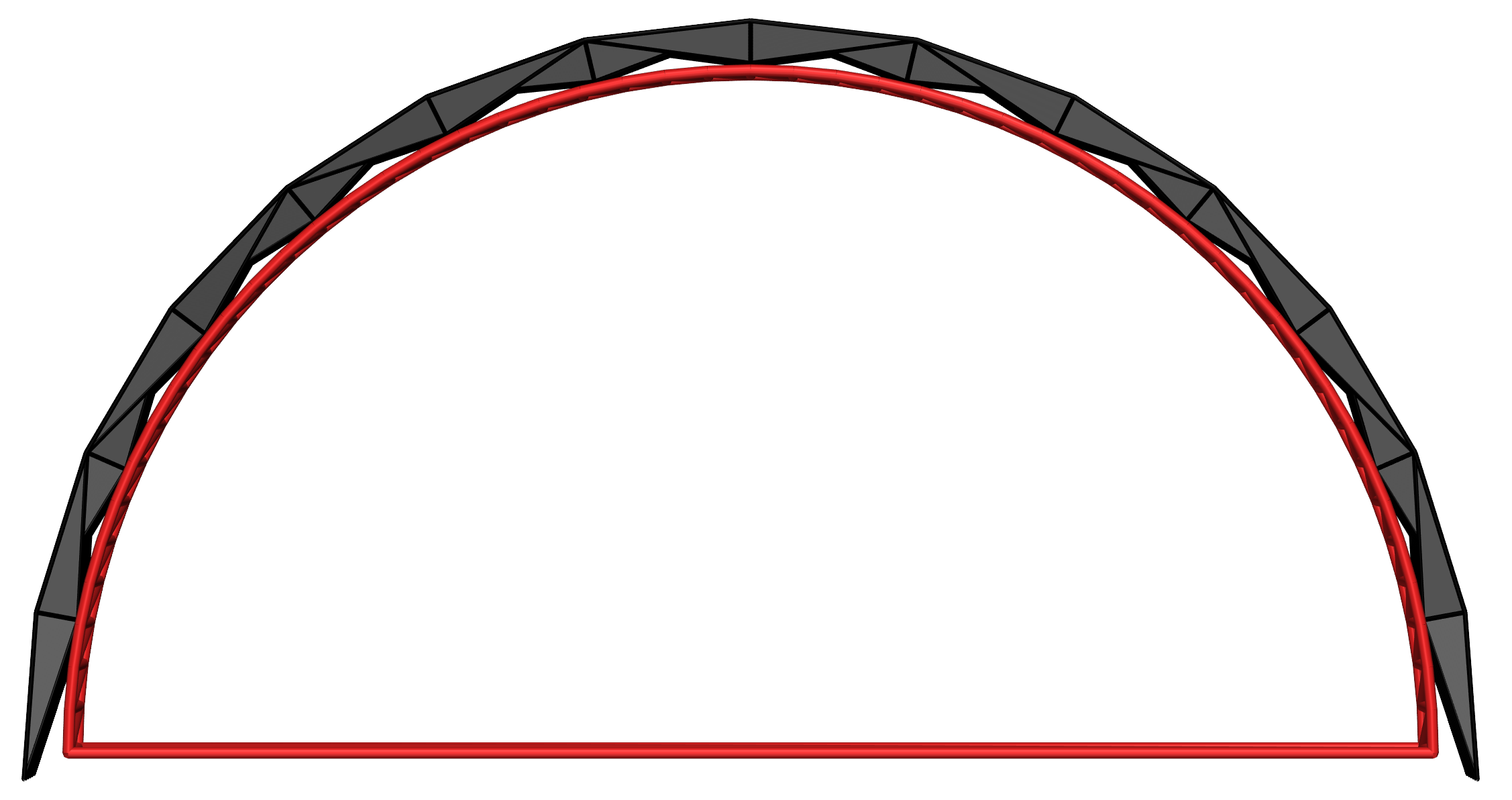} & 
\includegraphics[height=\paramheight, valign=m]{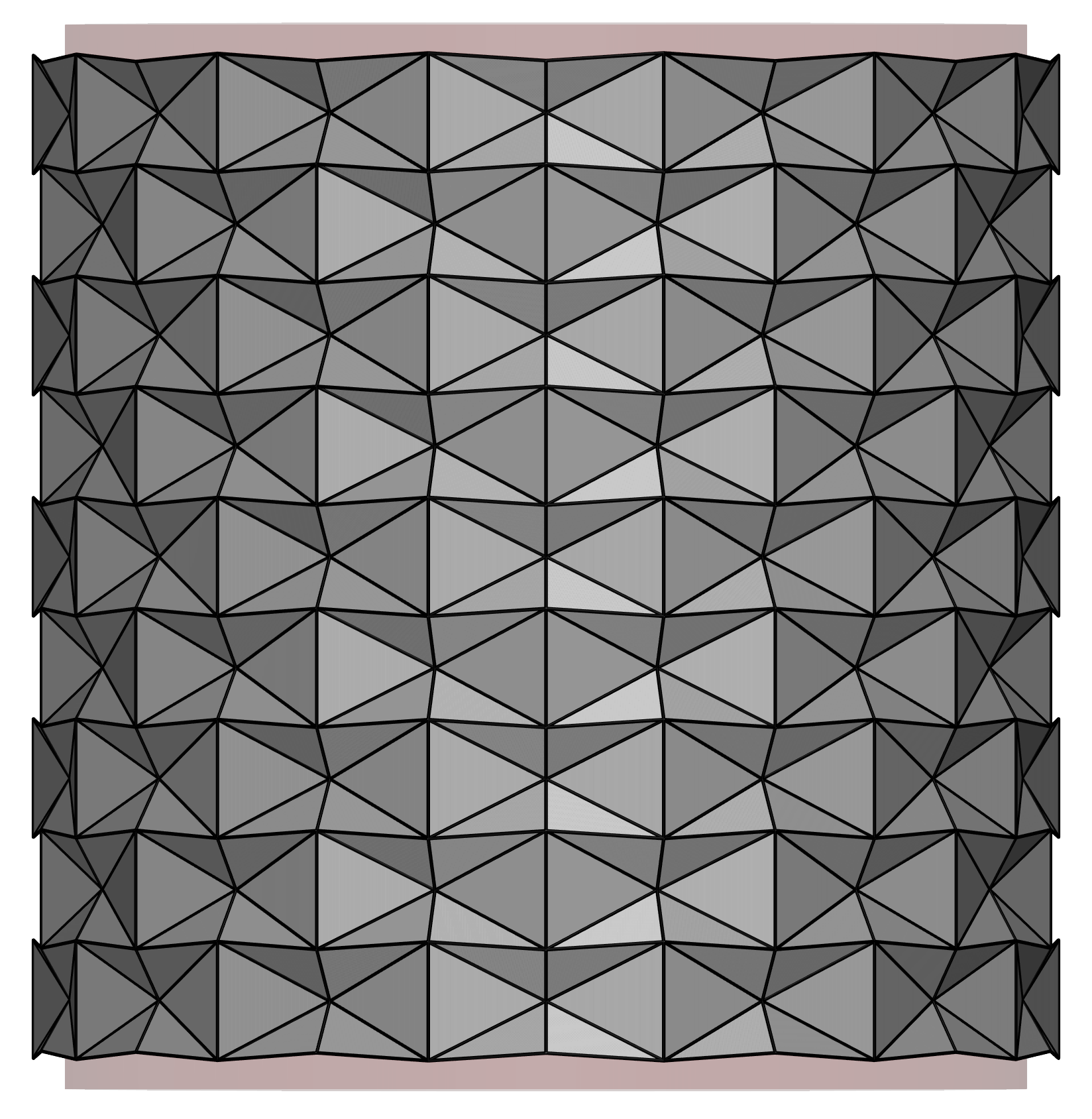} & 
\includegraphics[height=\paramheight, valign=m]{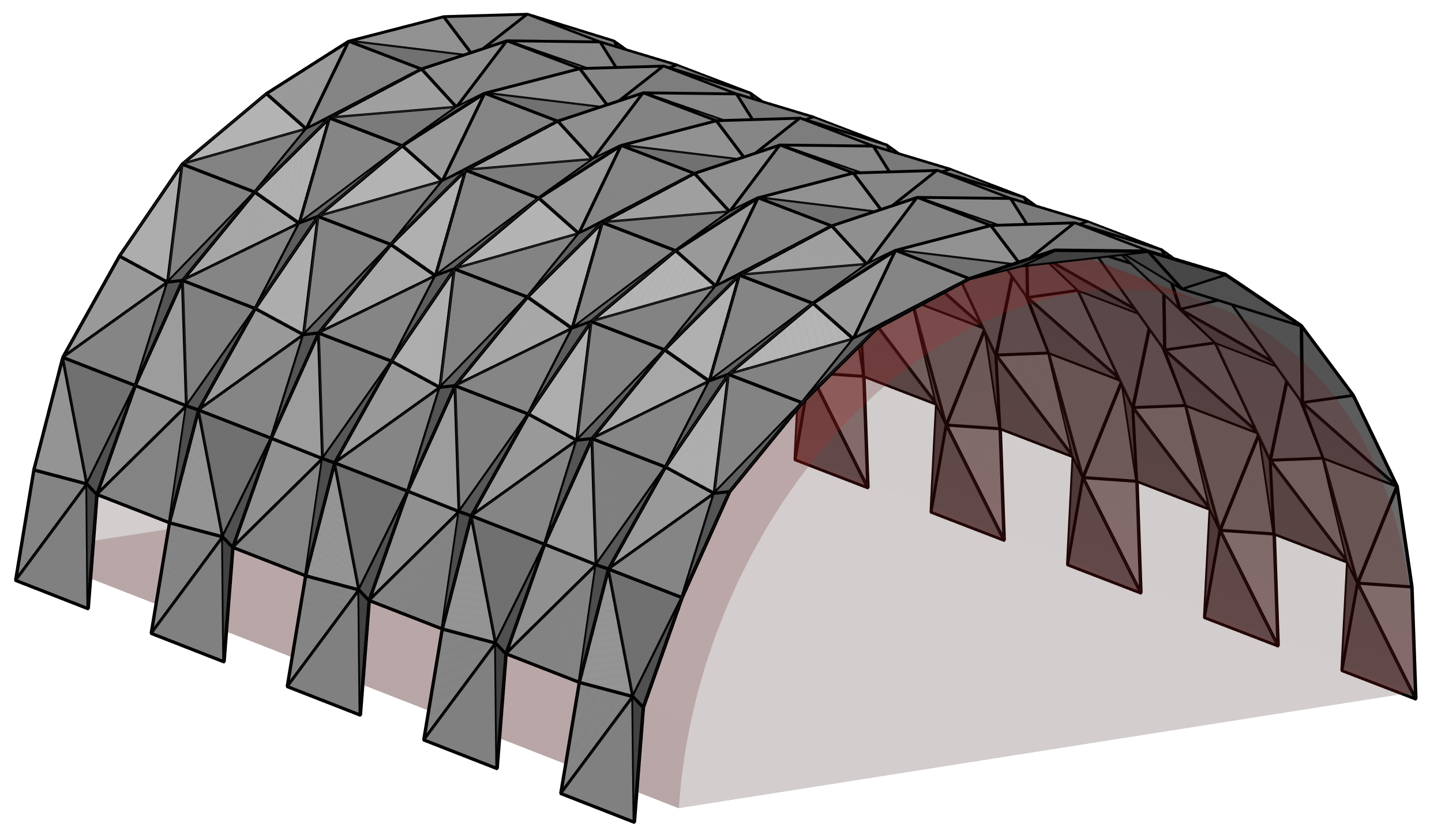} \\
\midrule
\SetCell[r=3]{m} $\gamma$ effect & \renewcommand{\arraystretch}{1.0} \begin{tabular}[c]{@{}l@{}} $\foldanglegamma=0.000$ \\ $\eta=2.00$ \\ $\zeta=1.00$ \end{tabular} & 0.0000 & 
\includegraphics[height=0.9\paramheight, valign=m]{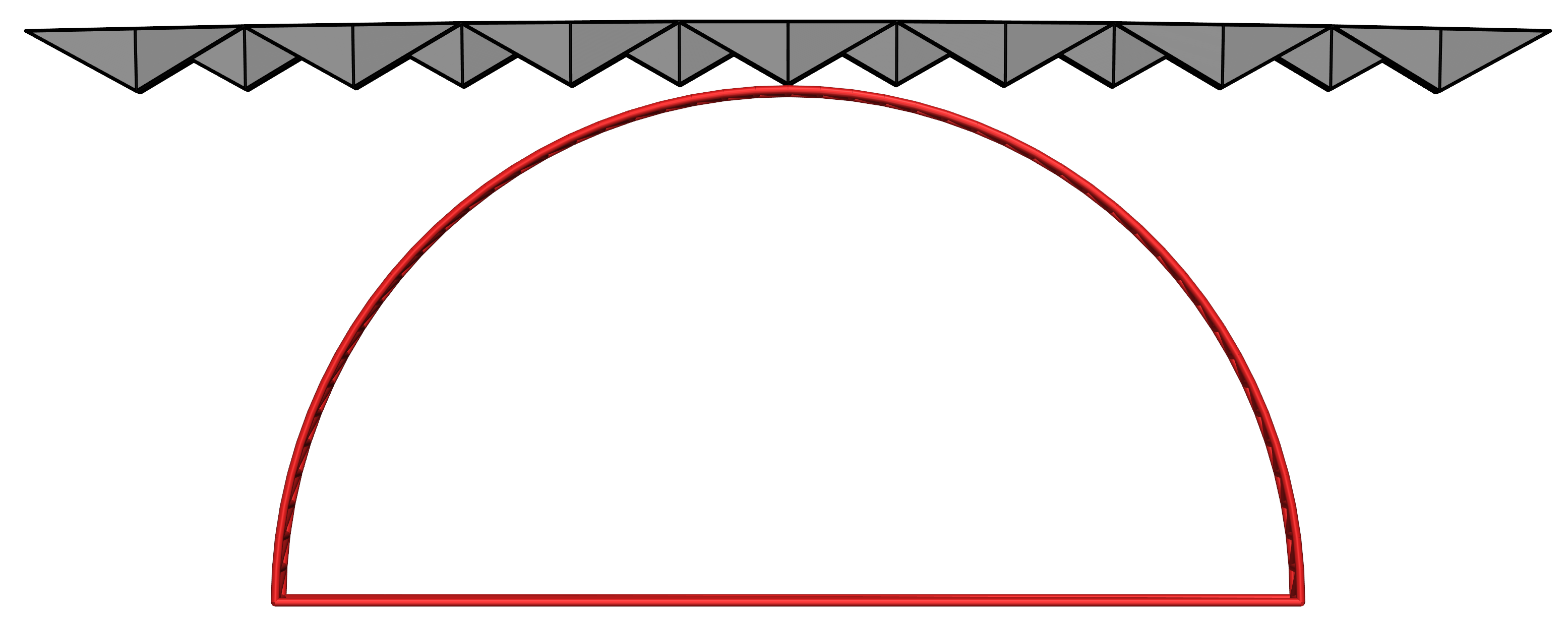} & 
\includegraphics[height=\paramheight, valign=m]{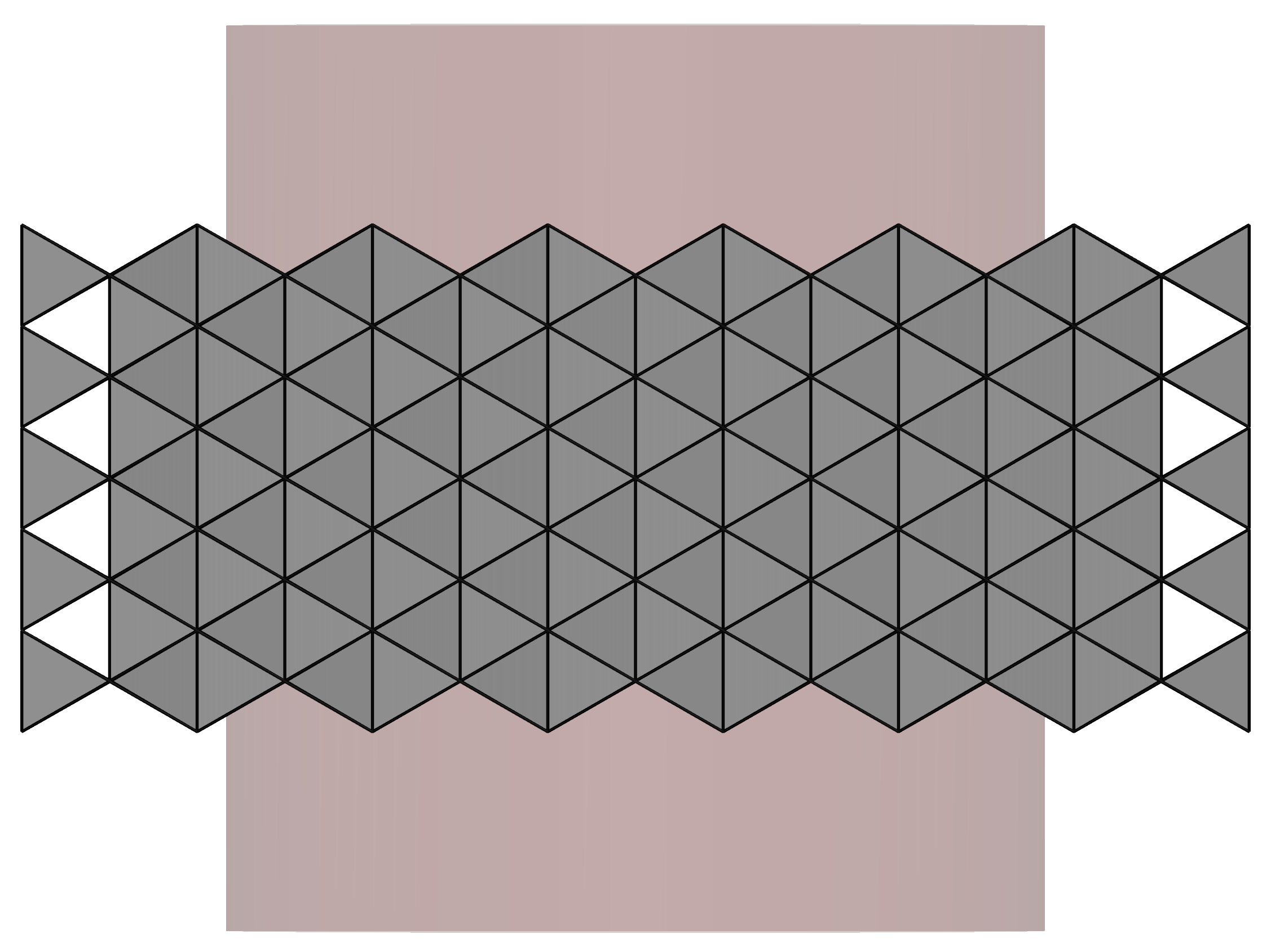} & 
\includegraphics[height=\paramheight, valign=m]{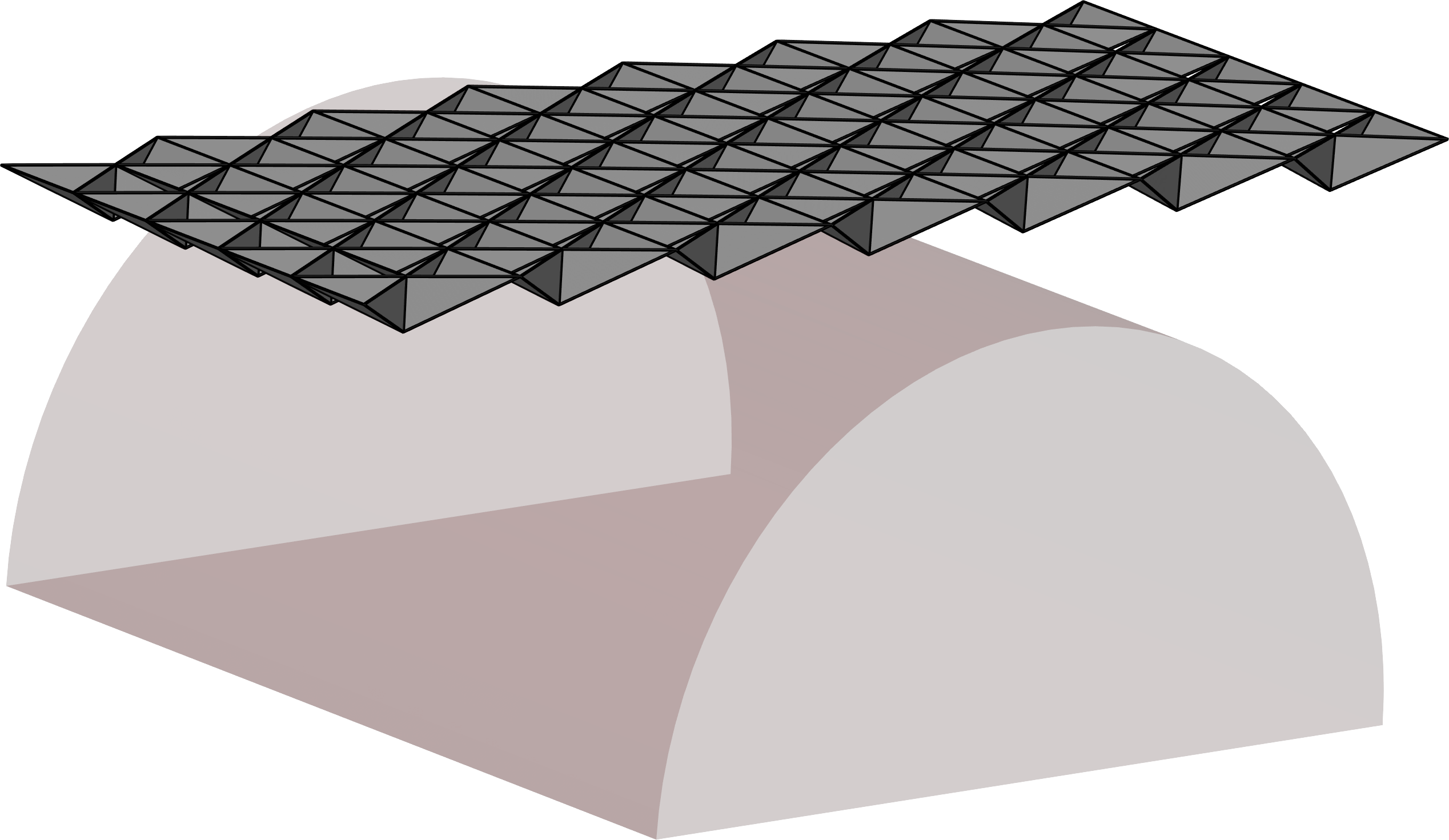} \\
 & \renewcommand{\arraystretch}{1.0} \begin{tabular}[c]{@{}l@{}} $\foldanglegamma=0.750$ \\ $\eta=2.00$ \\ $\zeta=1.00$ \end{tabular} & 0.1522 & 
\includegraphics[height=0.9\paramheight, valign=m]{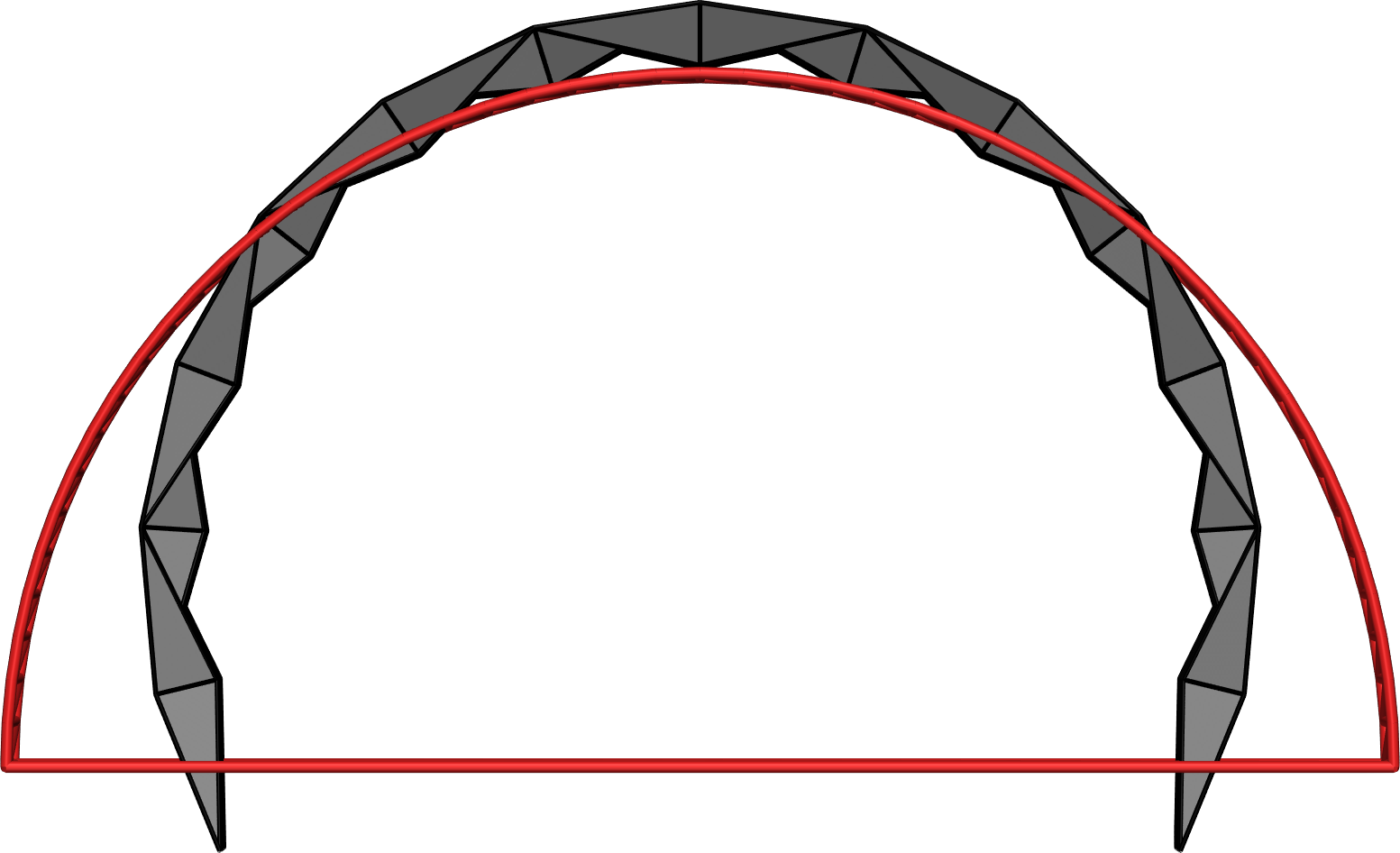} & 
\includegraphics[height=\paramheight, valign=m]{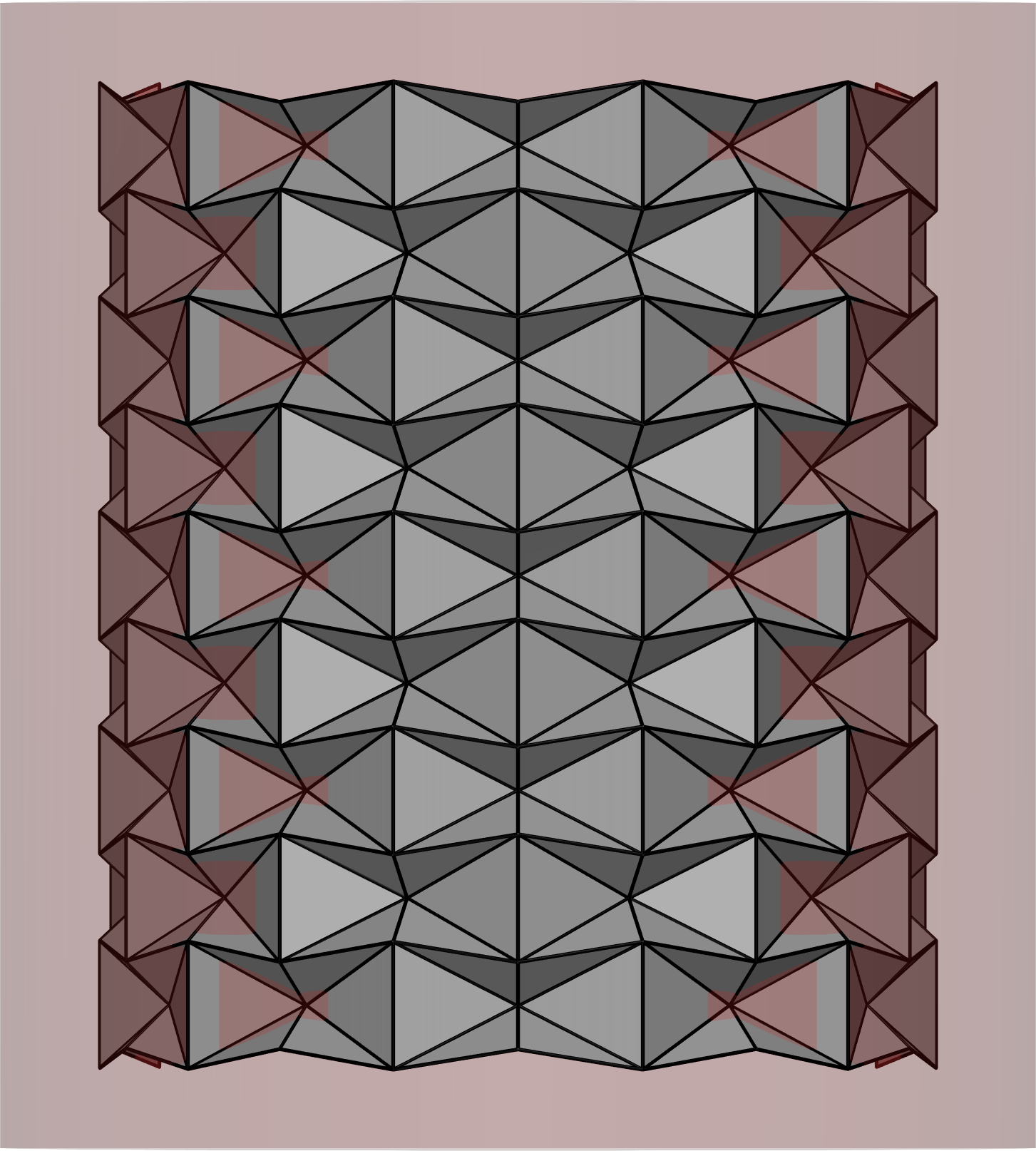} & 
\includegraphics[height=\paramheight, valign=m]{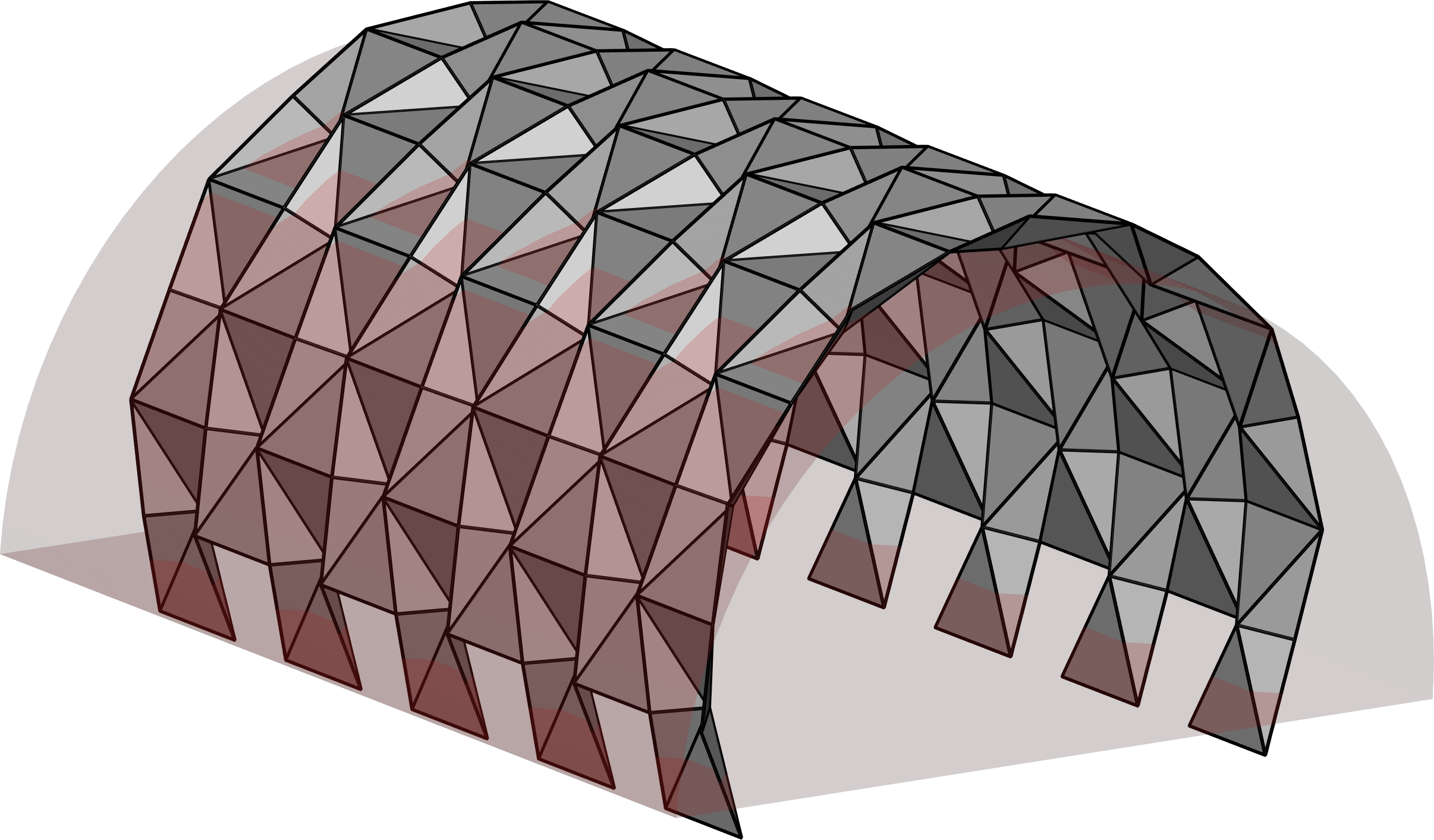} \\
 & \renewcommand{\arraystretch}{1.0} \begin{tabular}[c]{@{}l@{}} $\foldanglegamma=1.250$ \\ $\eta=2.00$ \\ $\zeta=1.00$ \end{tabular} & 0.0768 & 
\includegraphics[height=0.9\paramheight, valign=m]{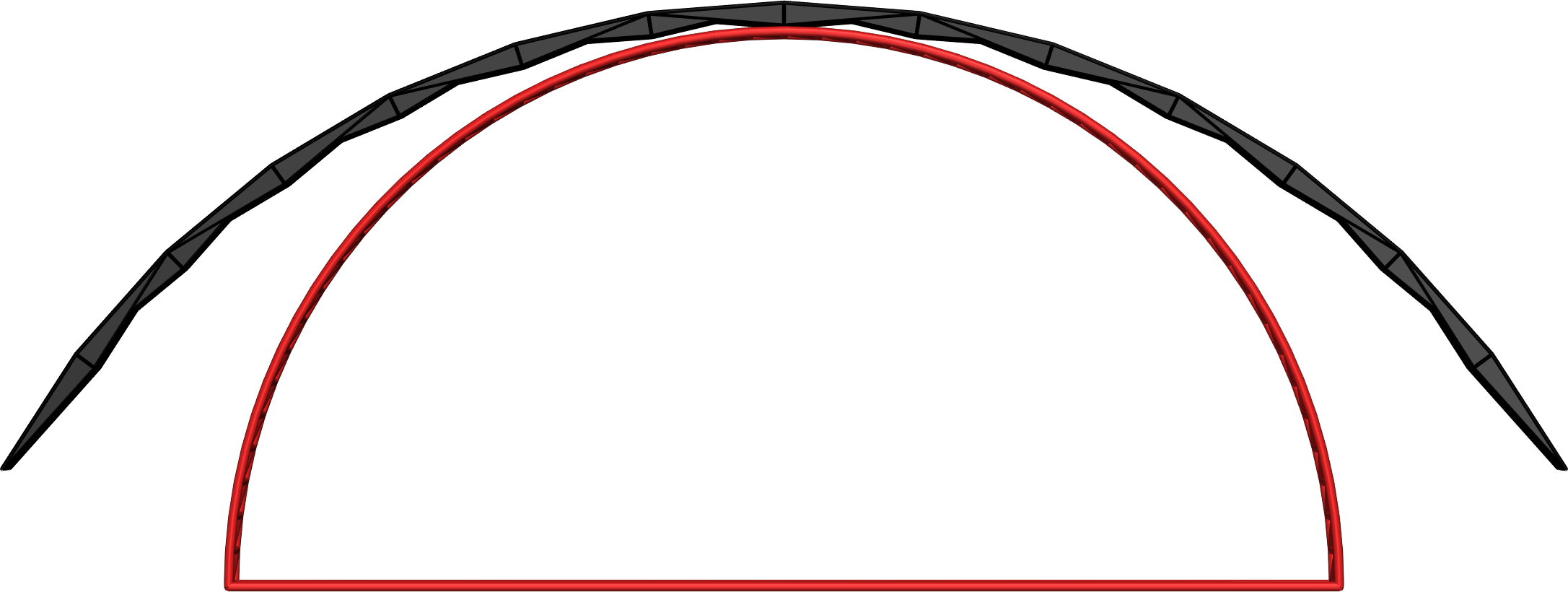} & 
\includegraphics[height=\paramheight, valign=m]{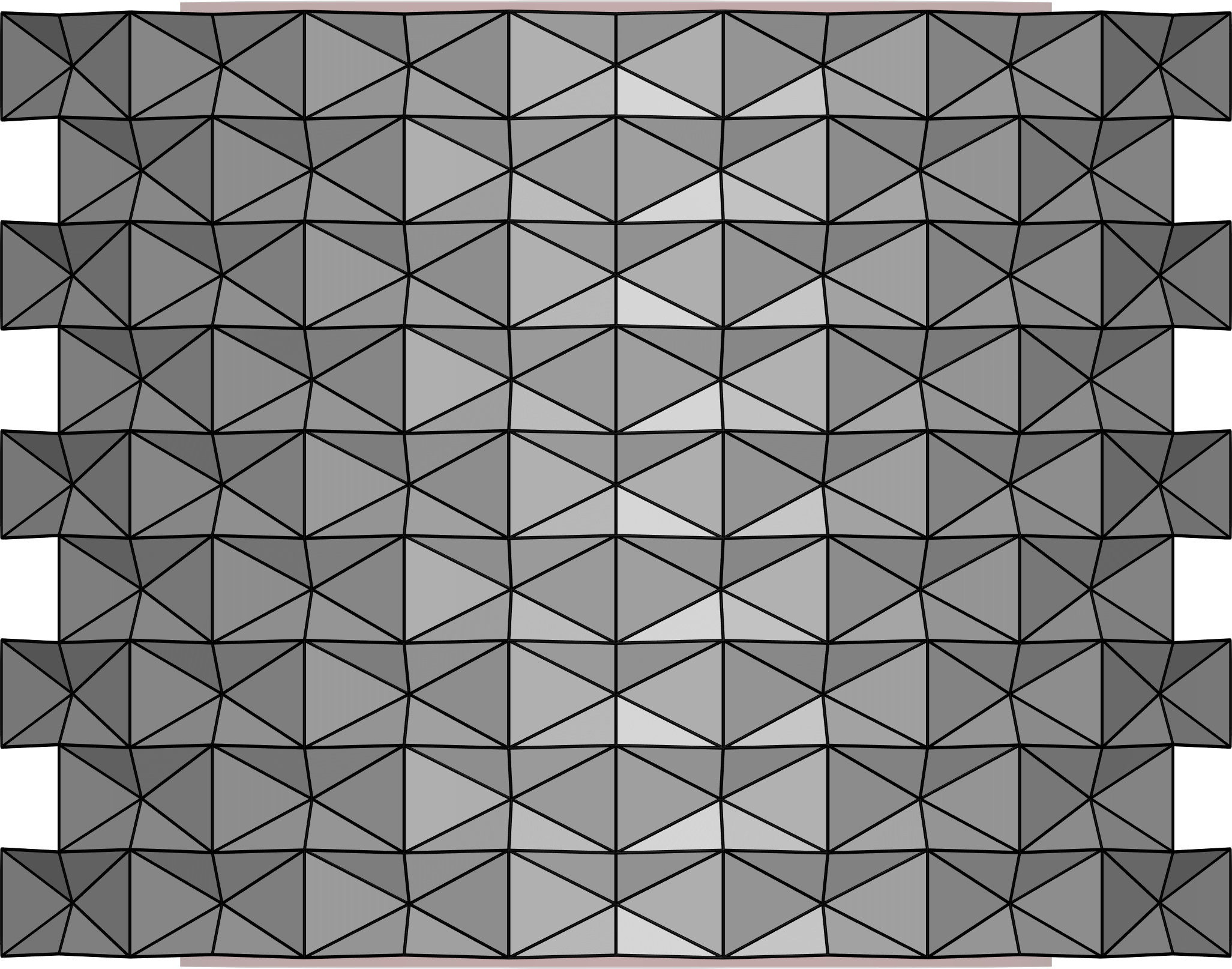} & 
\includegraphics[height=\paramheight, valign=m]{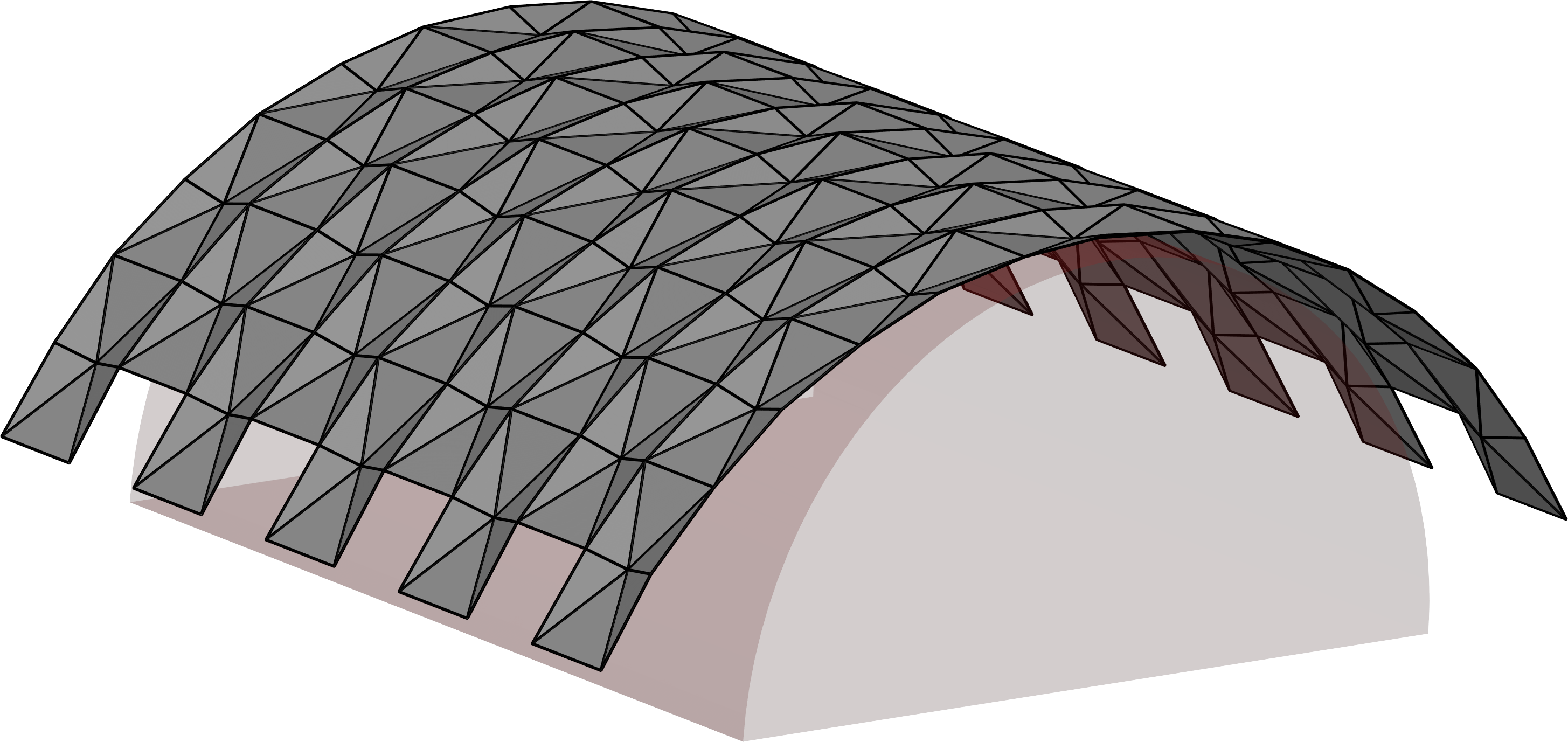} \\
\midrule
\SetCell[r=2]{m} $\eta$ effect & \renewcommand{\arraystretch}{1.0} \begin{tabular}[c]{@{}l@{}} $\foldanglegamma=1.050$ \\ $\eta=1.50$ \\ $\zeta=1.00$ \end{tabular} & 0.2071 & 
\includegraphics[height=0.9\paramheight, valign=m]{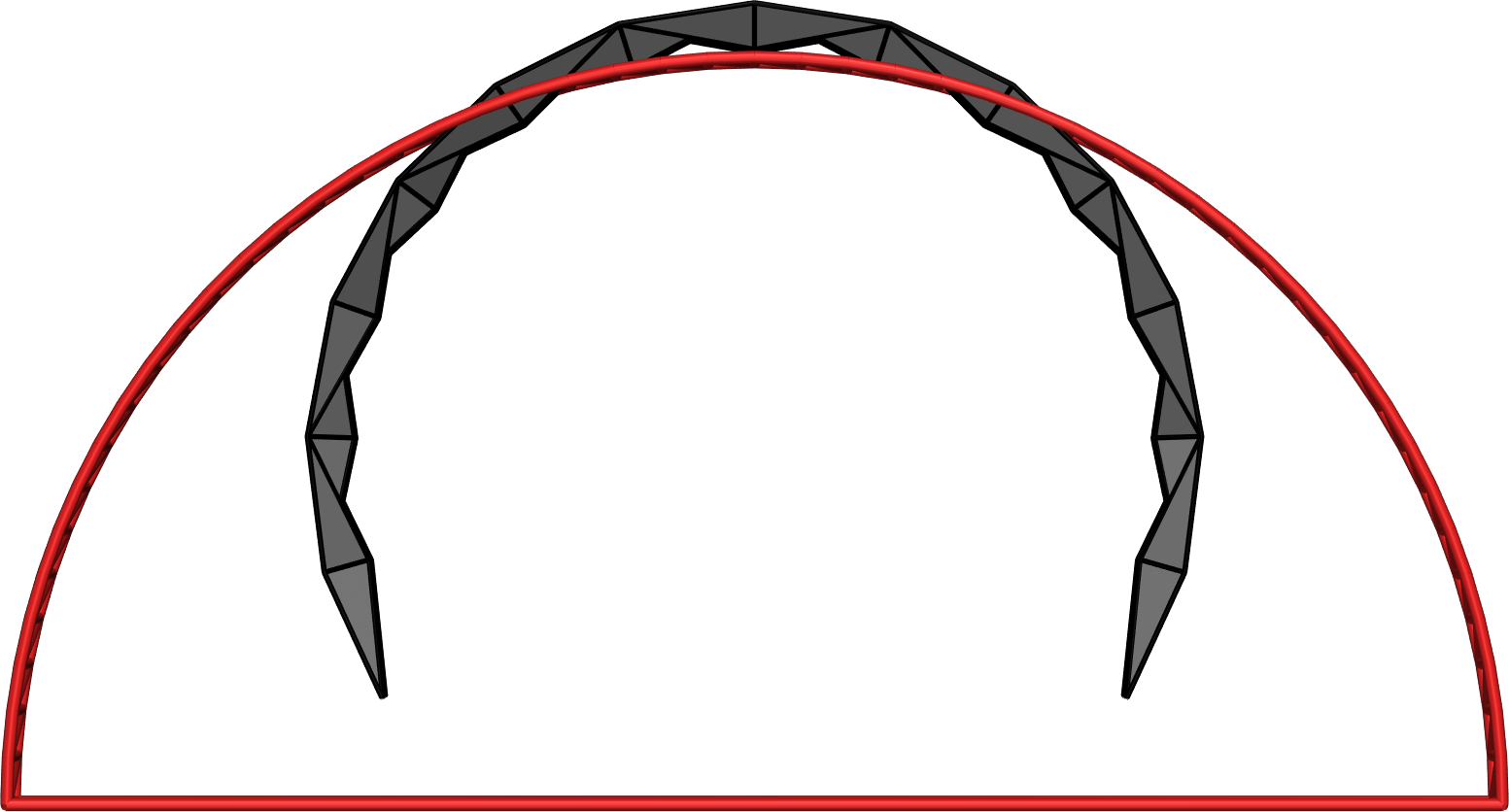} & 
\includegraphics[height=\paramheight, valign=m]{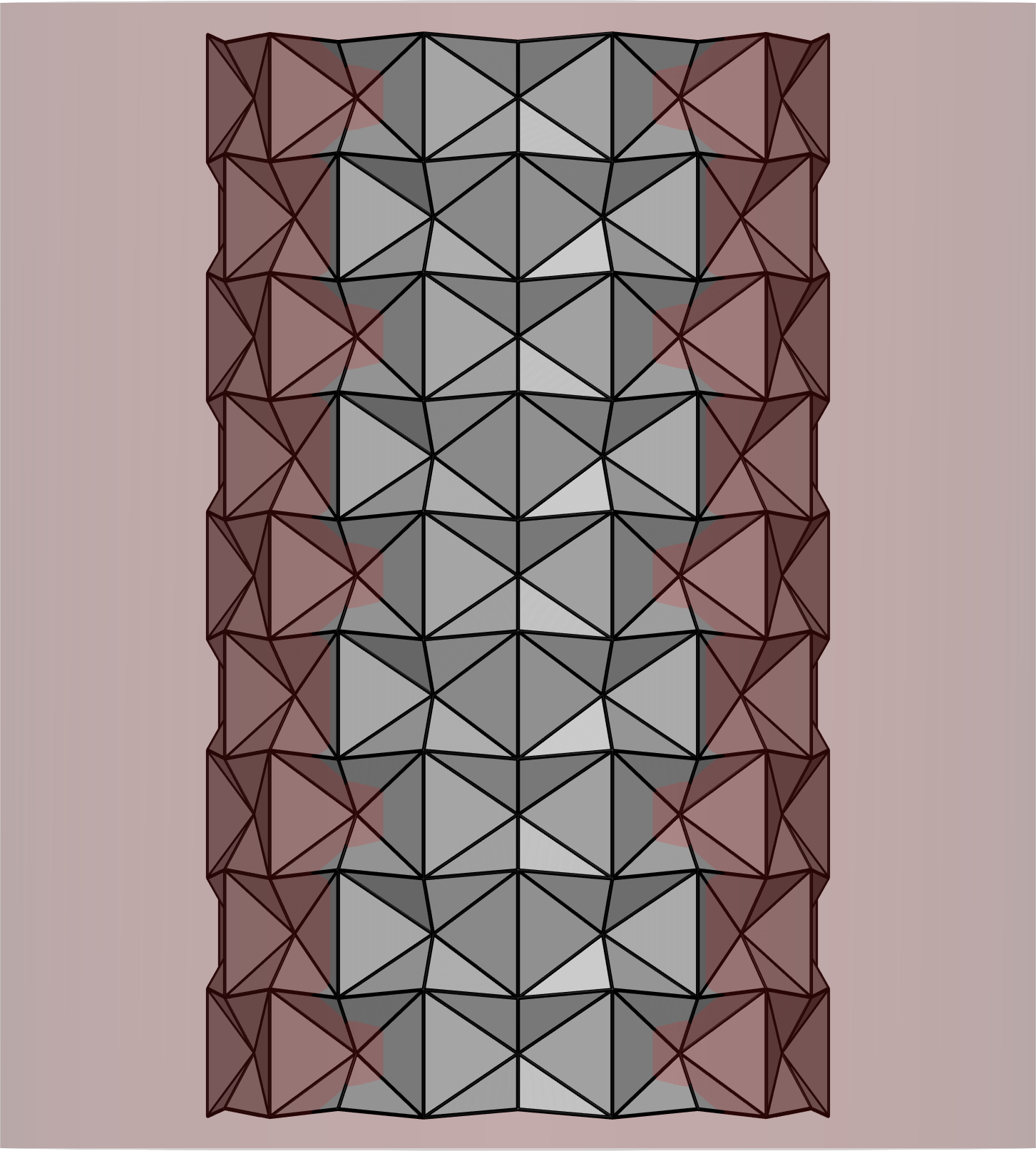} & 
\includegraphics[height=\paramheight, valign=m]{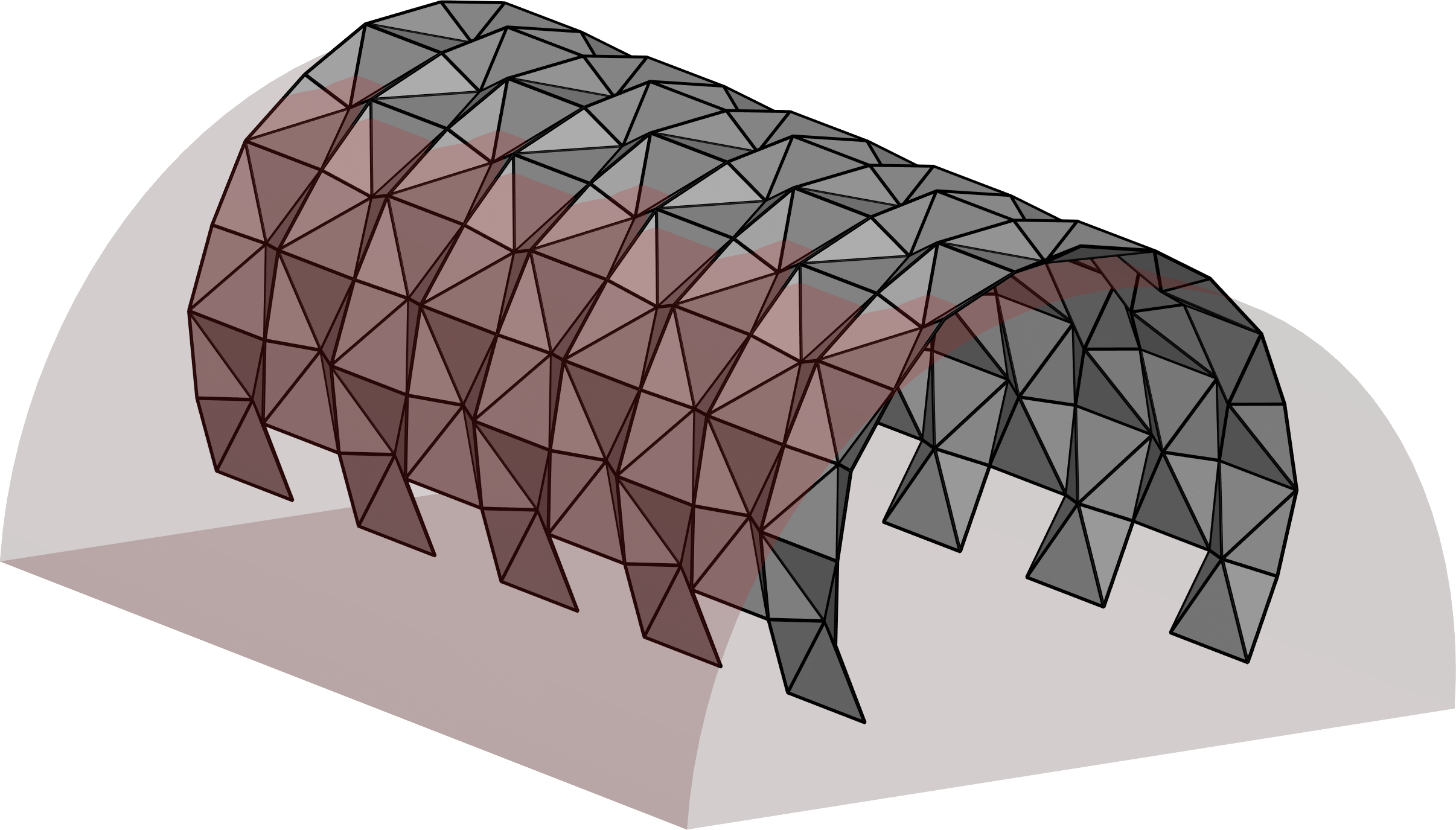} \\
 & \renewcommand{\arraystretch}{1.0} \begin{tabular}[c]{@{}l@{}} $\foldanglegamma=1.050$ \\ $\eta=2.50$ \\ $\zeta=1.00$ \end{tabular} & 0.0742 & 
\includegraphics[height=0.9\paramheight, valign=m]{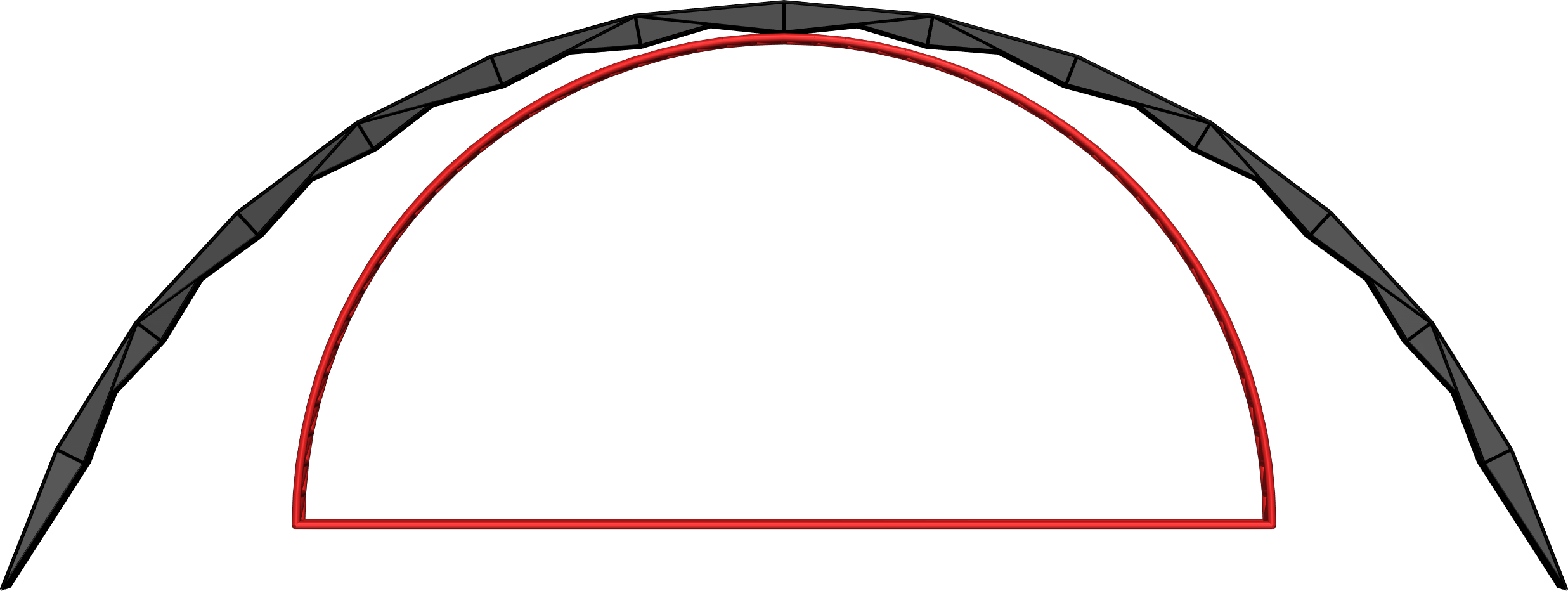} & 
\includegraphics[height=\paramheight, valign=m]{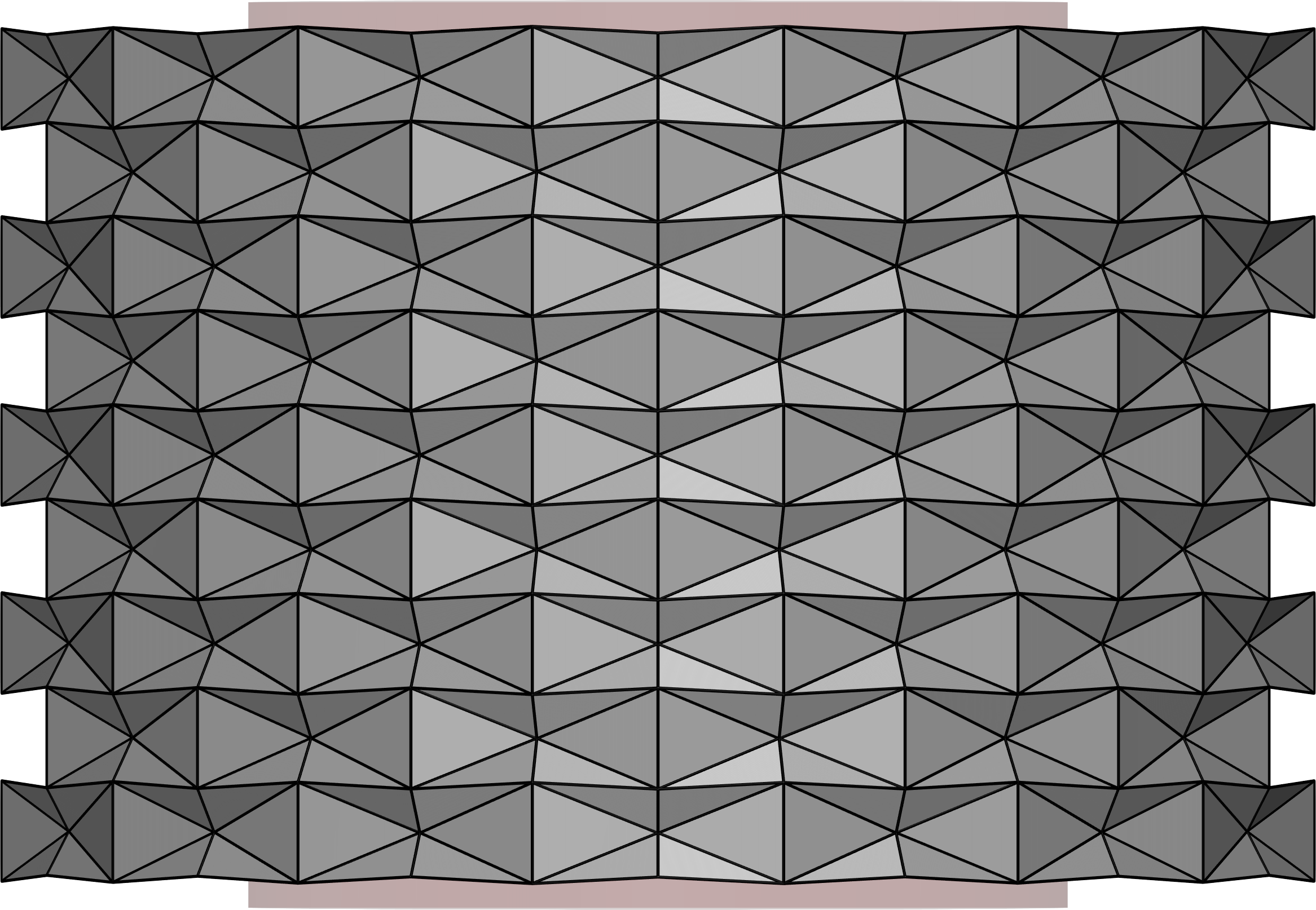} & 
\includegraphics[height=\paramheight, valign=m]{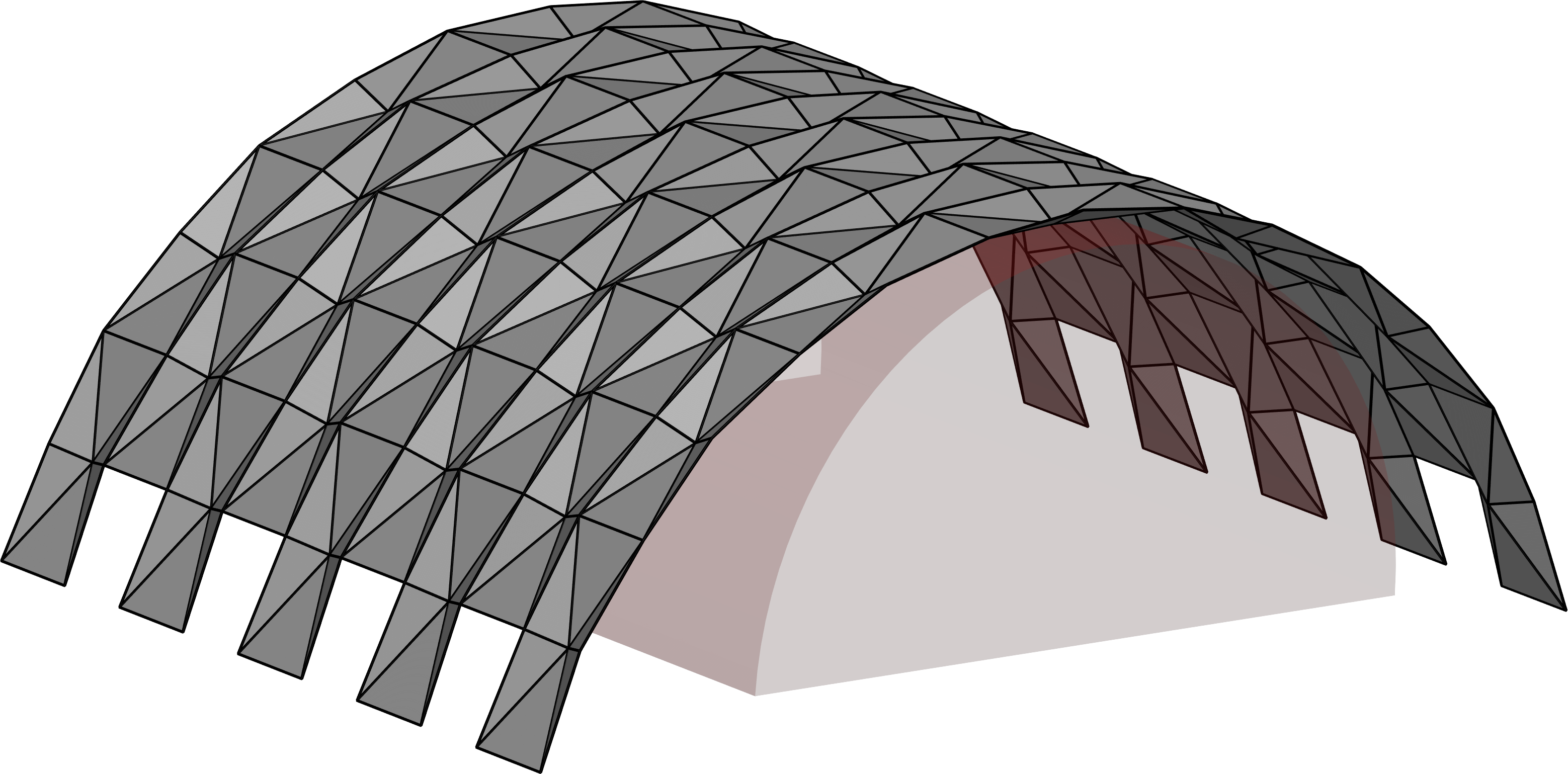} \\
\midrule
\SetCell[r=2]{m} $\zeta$ effect & \renewcommand{\arraystretch}{1.0} \begin{tabular}[c]{@{}l@{}} $\foldanglegamma=1.050$ \\ $\eta=2.00$ \\ $\zeta=0.70$ \end{tabular} & 0.0416 & 
\includegraphics[height=0.9\paramheight, valign=m]{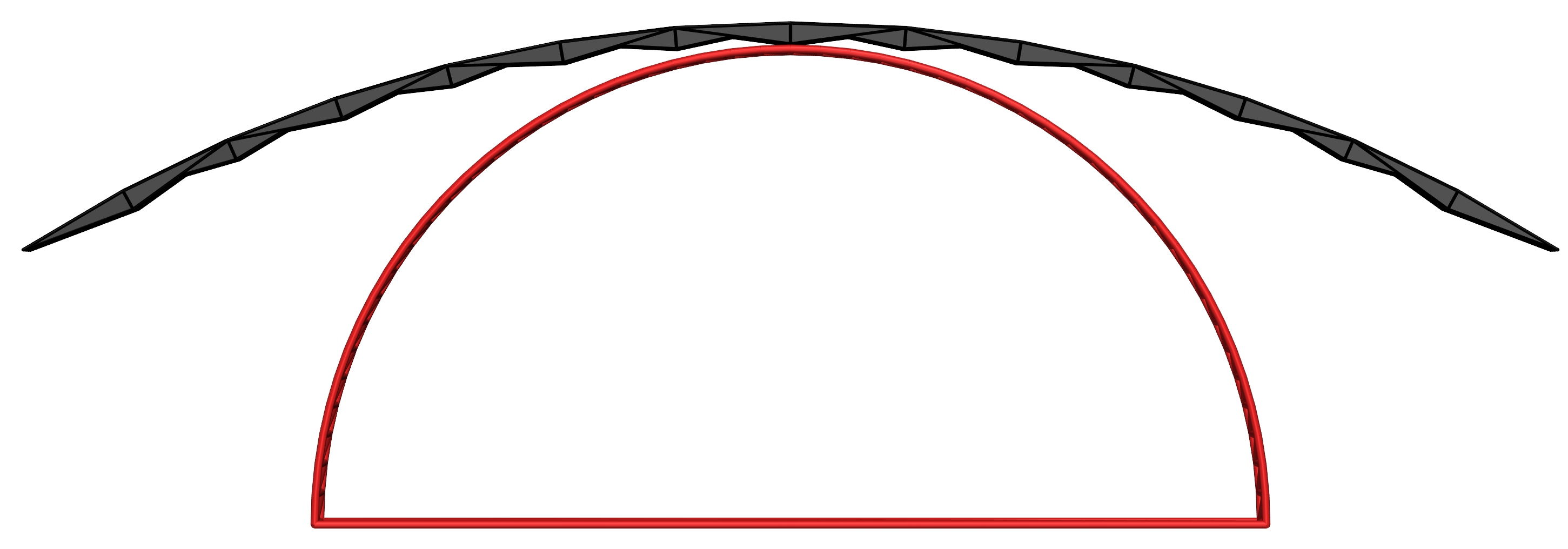} & 
\includegraphics[height=\paramheight, valign=m]{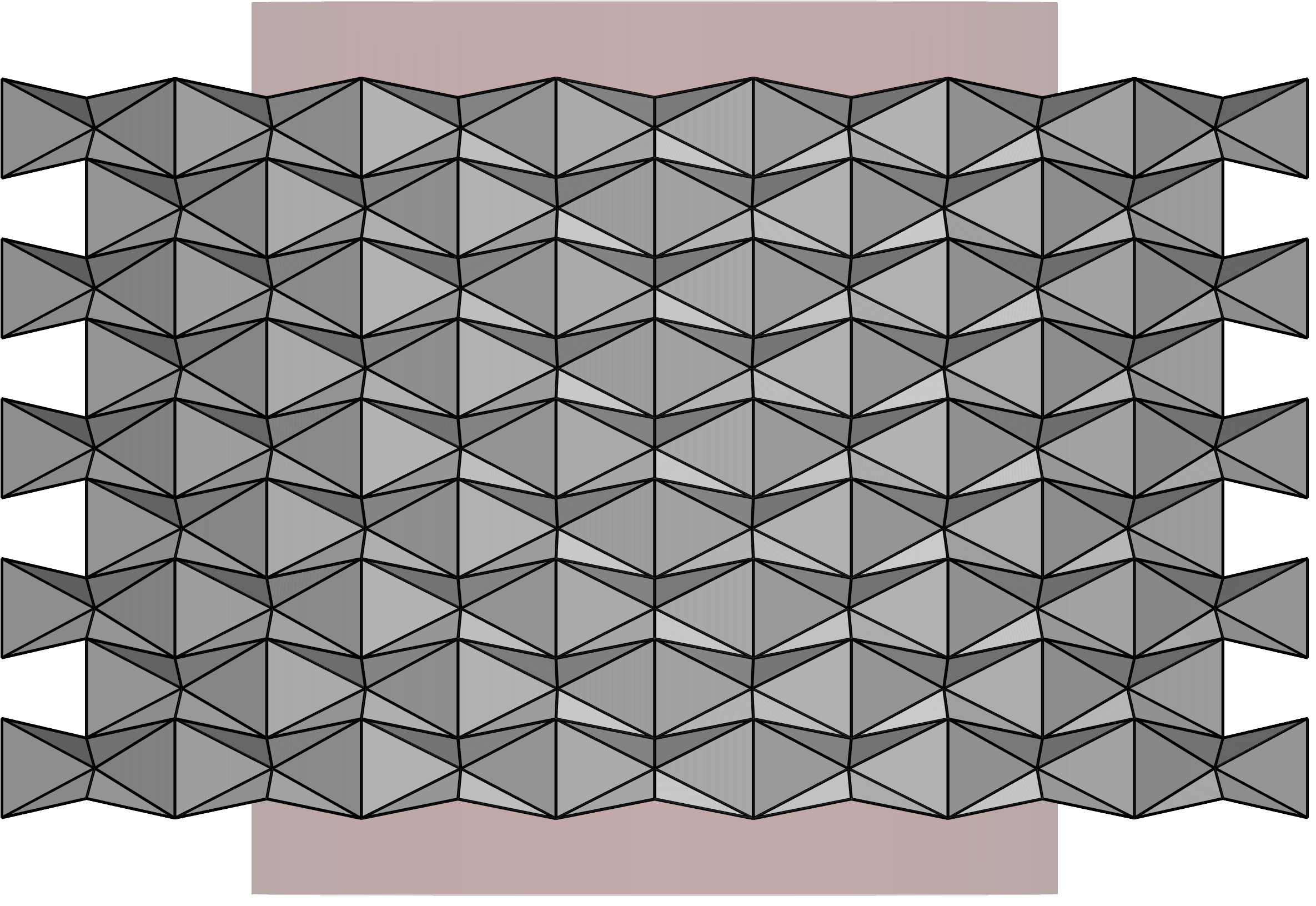} & 
\includegraphics[height=\paramheight, valign=m]{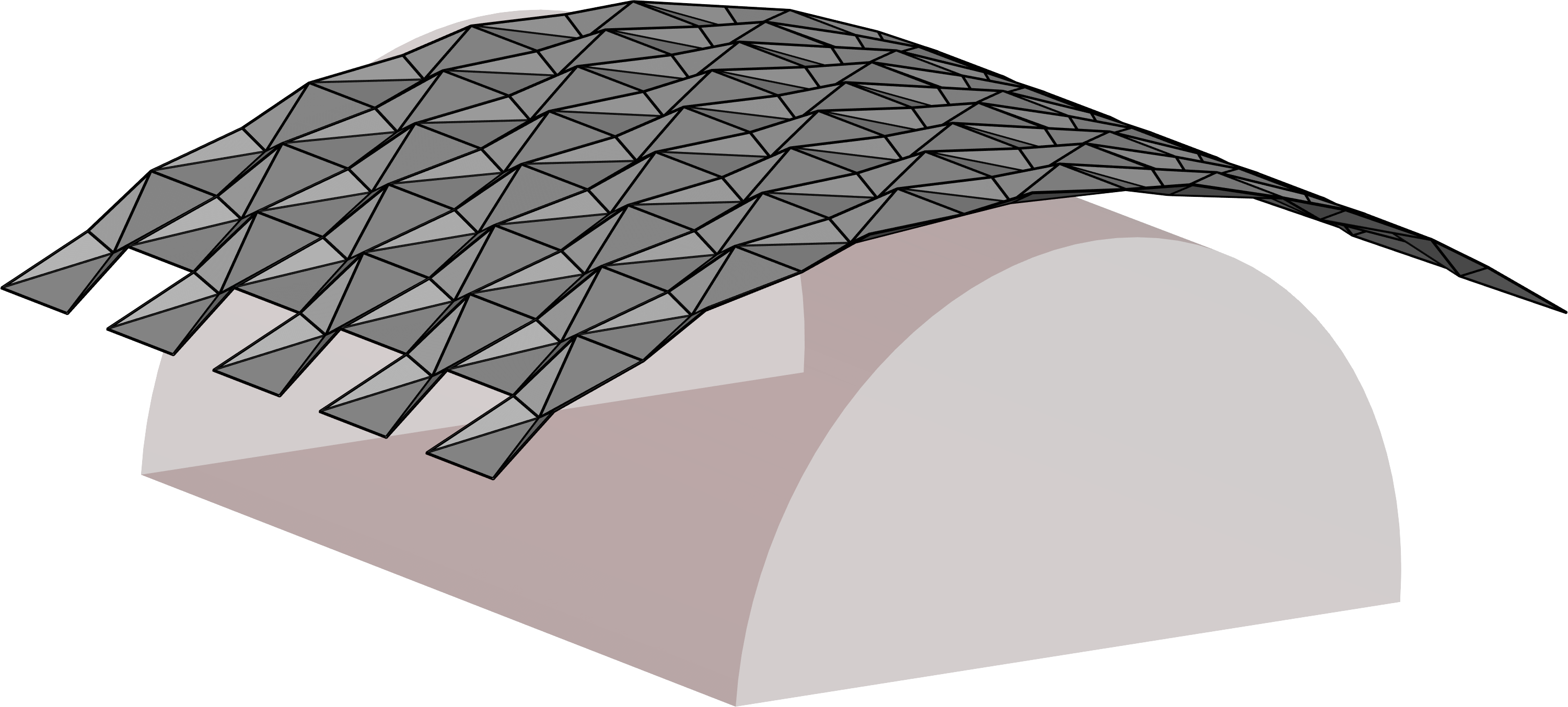} \\
 & \renewcommand{\arraystretch}{1.0} \begin{tabular}[c]{@{}l@{}} $\foldanglegamma=1.050$ \\ $\eta=2.00$ \\ $\zeta=1.30$ \end{tabular} & 0.1868 & 
\includegraphics[height=0.9\paramheight, valign=m]{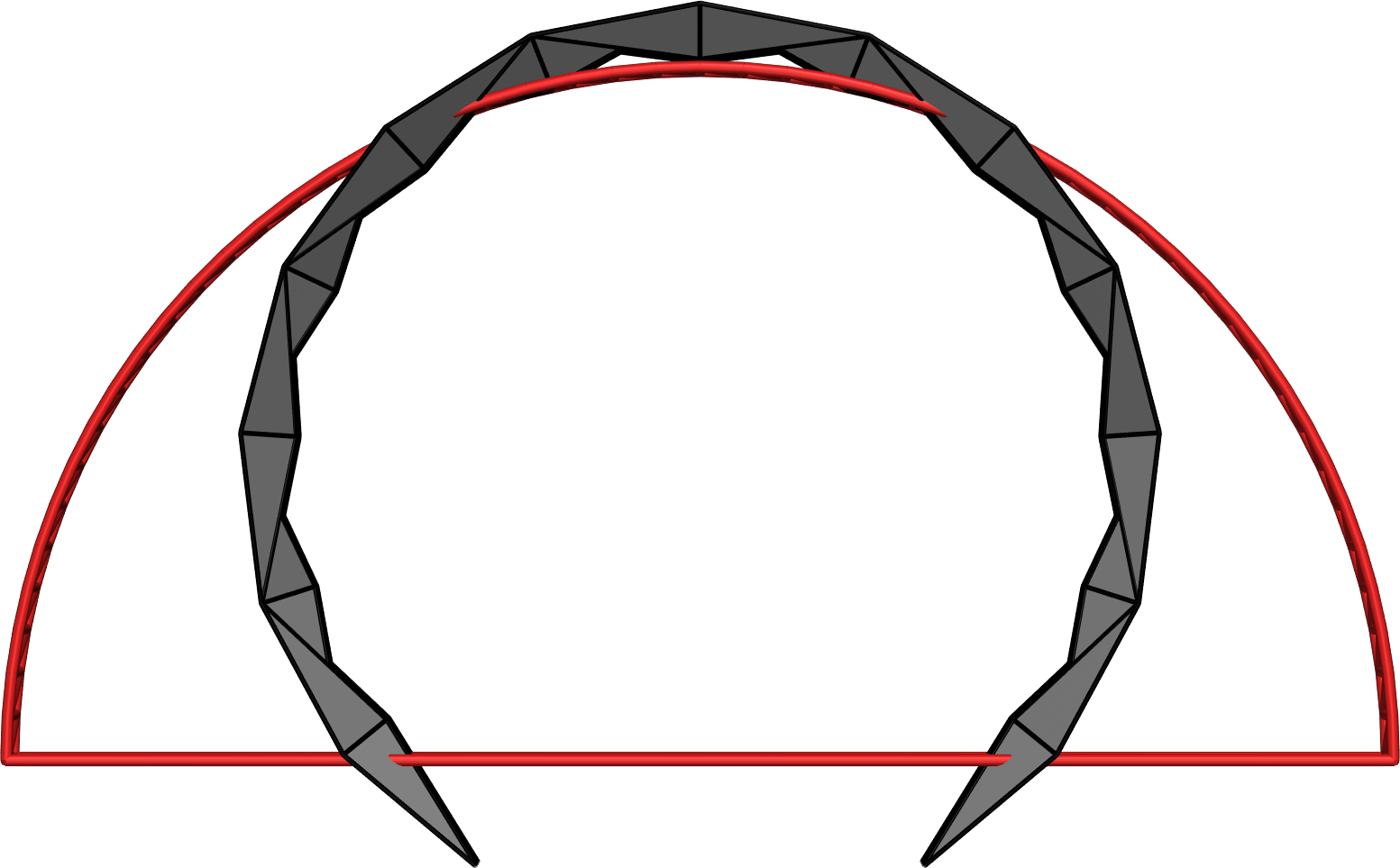} & 
\includegraphics[height=\paramheight, valign=m]{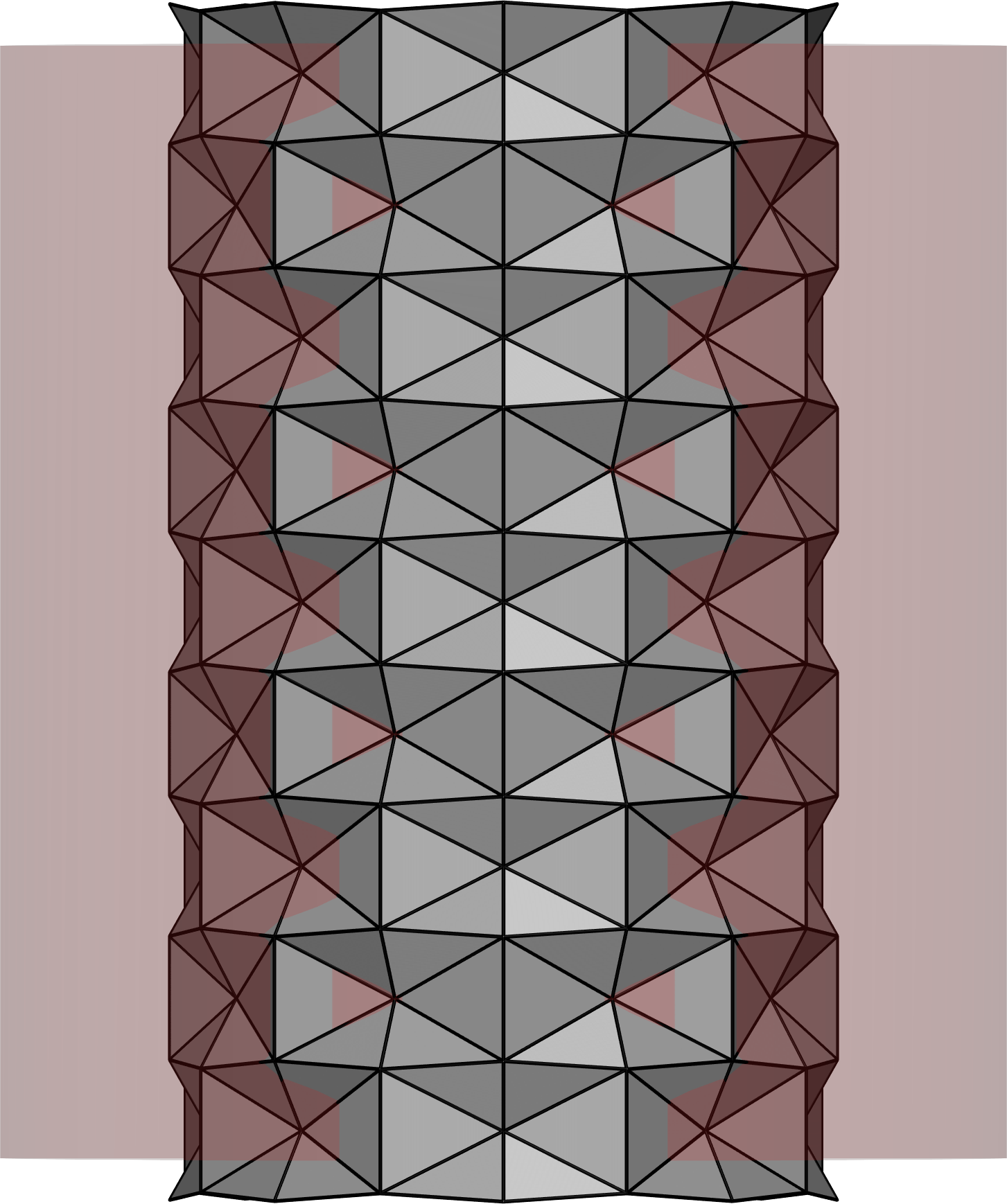} & 
\includegraphics[height=\paramheight, valign=m]{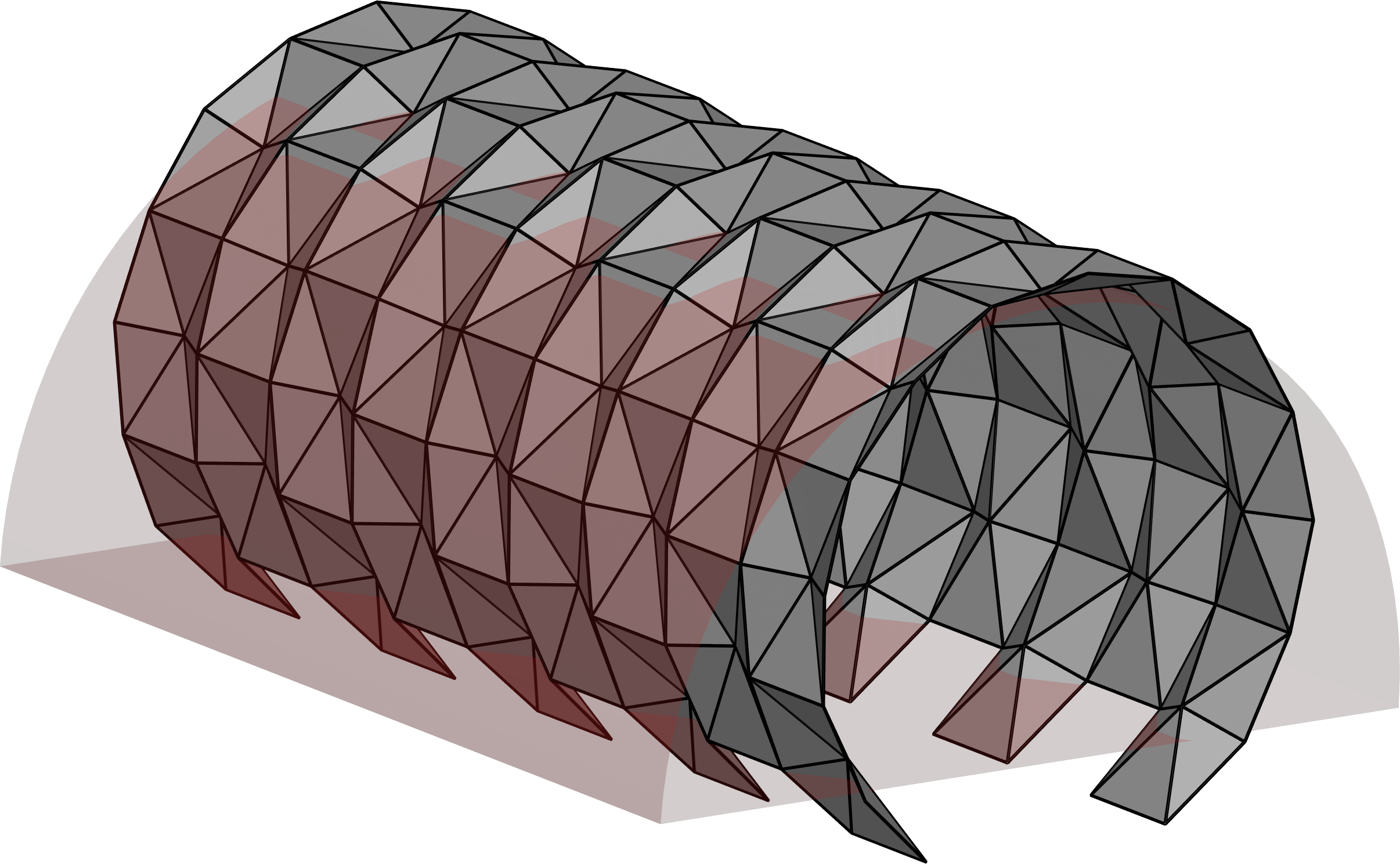} \\
\bottomrule
\end{tblr}
\end{table*}

\begin{table*}[t]
\caption{Parametric study showing the effect of different semi-symmetric cell parameters ($\Delta\beta=-0.04$) on the global shell geometries and their curvature magnitude $\kappa = 1/r$,  extended from \cite{robertz_iass2022} by additional information on curvature.}
\label{tab:param_study_semi_sym}
\centering
\small
\begin{tblr}{
  colspec = {llcccc},
  stretch = 2.0,
  column{3-6} = {c},
  cell{1}{1-6} = {font=\small},
  rows = {m},
}
\toprule
Variation & \renewcommand{\arraystretch}{1.0} \begin{tabular}[c]{@{}l@{}} Parameters \\ \footnotesize $(\foldanglegamma, \eta, \zeta)$\end{tabular} & Curvature & Front View & Top View & Perspective \\
\midrule
Reference & \renewcommand{\arraystretch}{1.0} \begin{tabular}[c]{@{}l@{}} $\foldanglegamma=1.050$ \\ $\eta=2.00$ \\ $\zeta=1.00$ \end{tabular} & 0.1252 & 
\includegraphics[height=0.9\paramheight, valign=m]{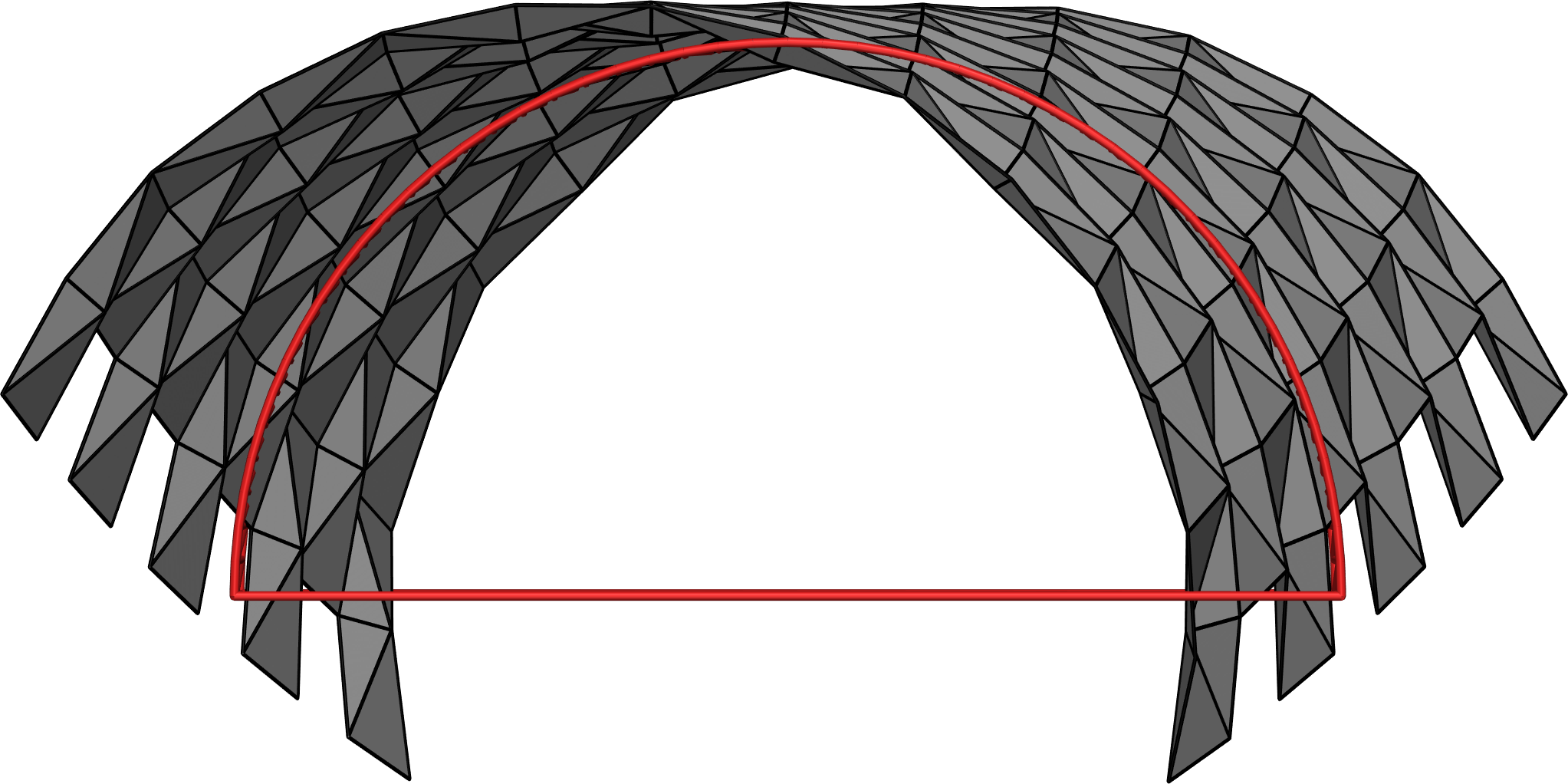} &
\includegraphics[height=\paramheight, valign=m]{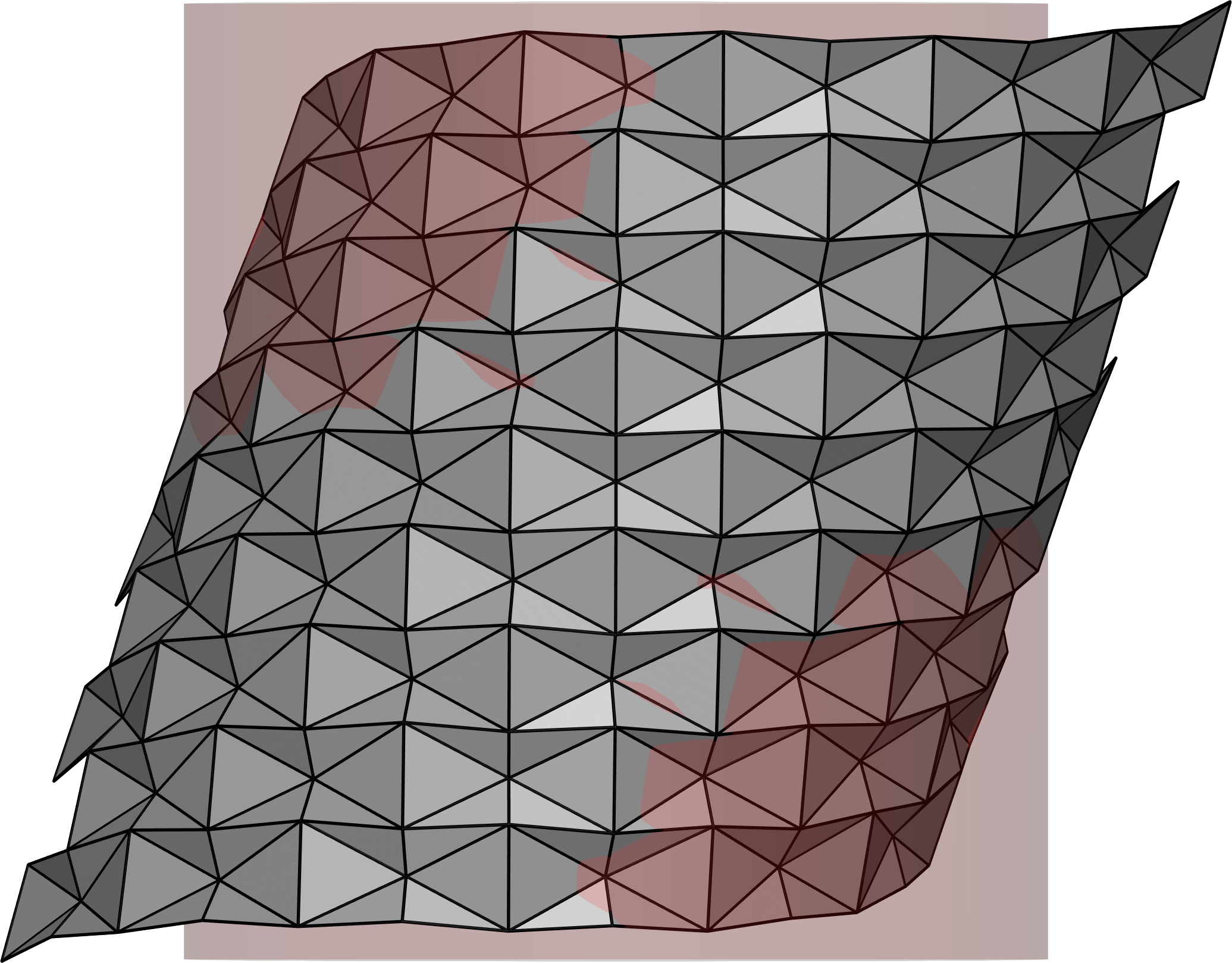} &
\includegraphics[height=\paramheight, valign=m]{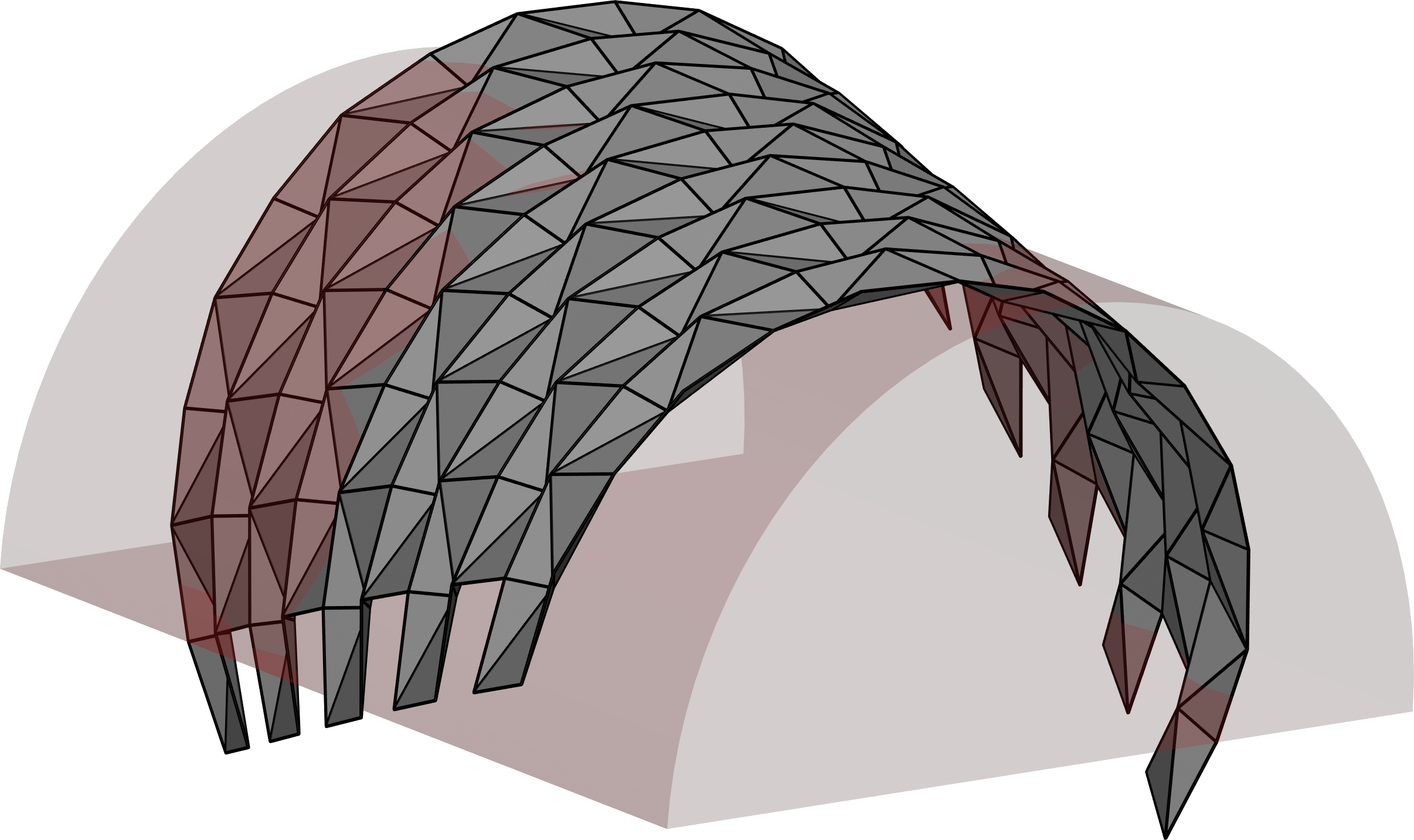} \\
\midrule
\SetCell[r=3]{m} $\gamma$ effect & \renewcommand{\arraystretch}{1.0} \begin{tabular}[c]{@{}l@{}} $\foldanglegamma=0.060$ \\ $\eta=2.00$ \\ $\zeta=1.00$ \end{tabular} & 0.0437 & 
\includegraphics[height=0.9\paramheight, valign=m]{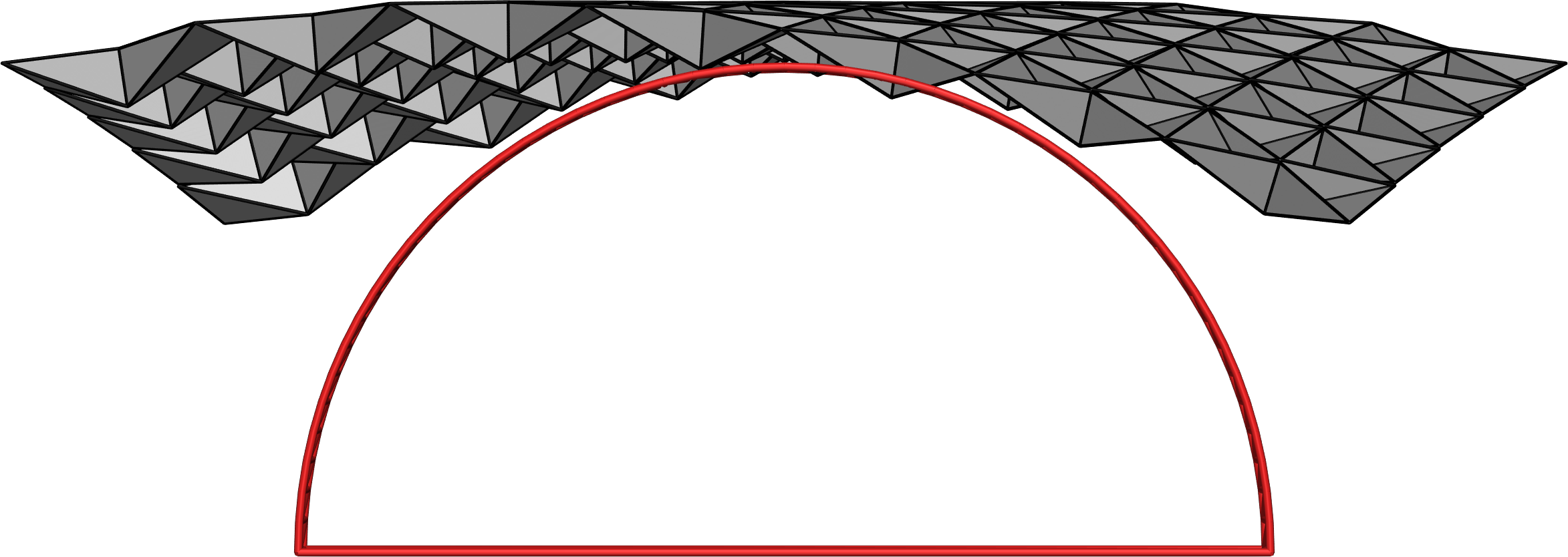} & 
\includegraphics[height=\paramheight, valign=m]{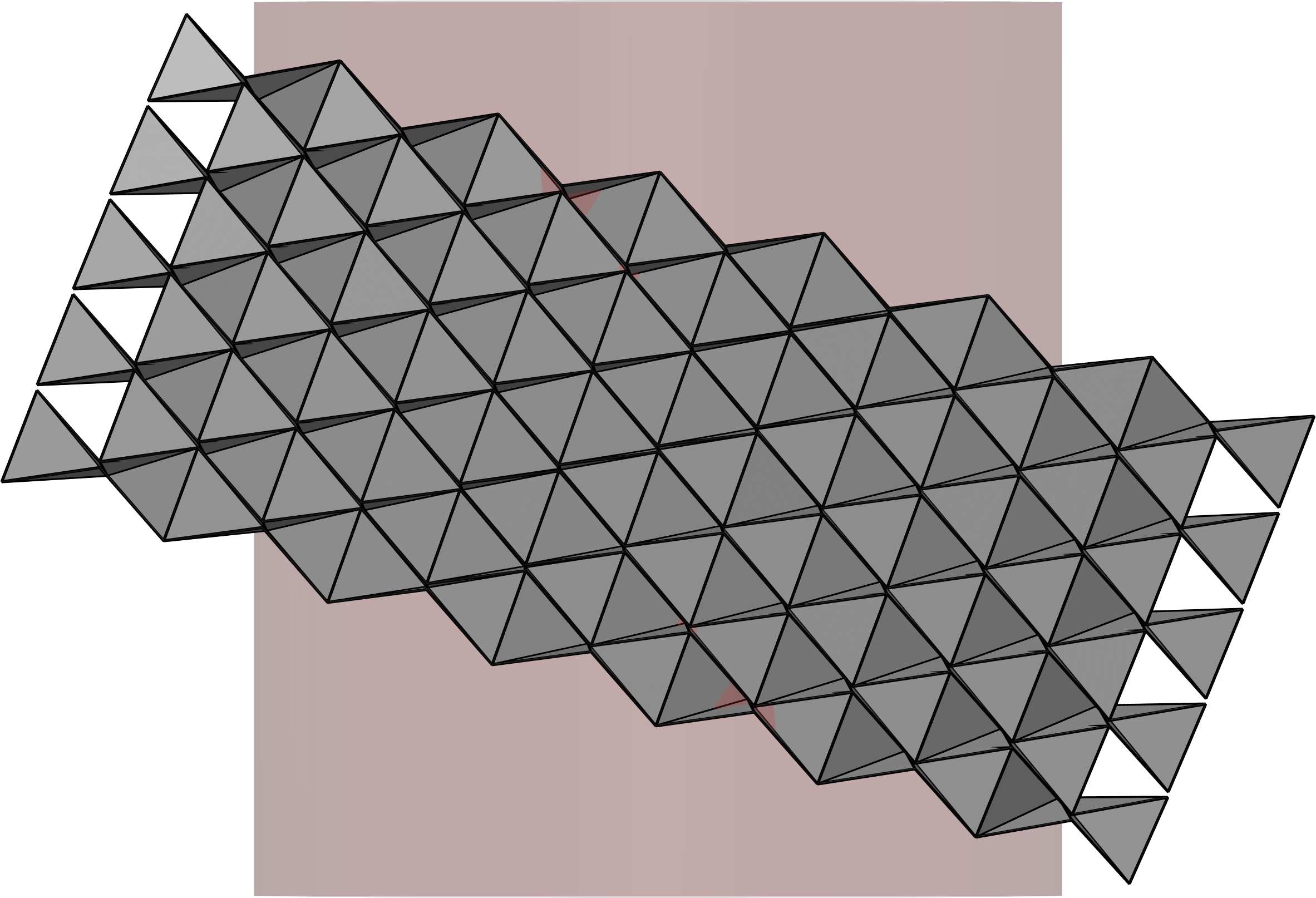} & 
\includegraphics[height=\paramheight, valign=m]{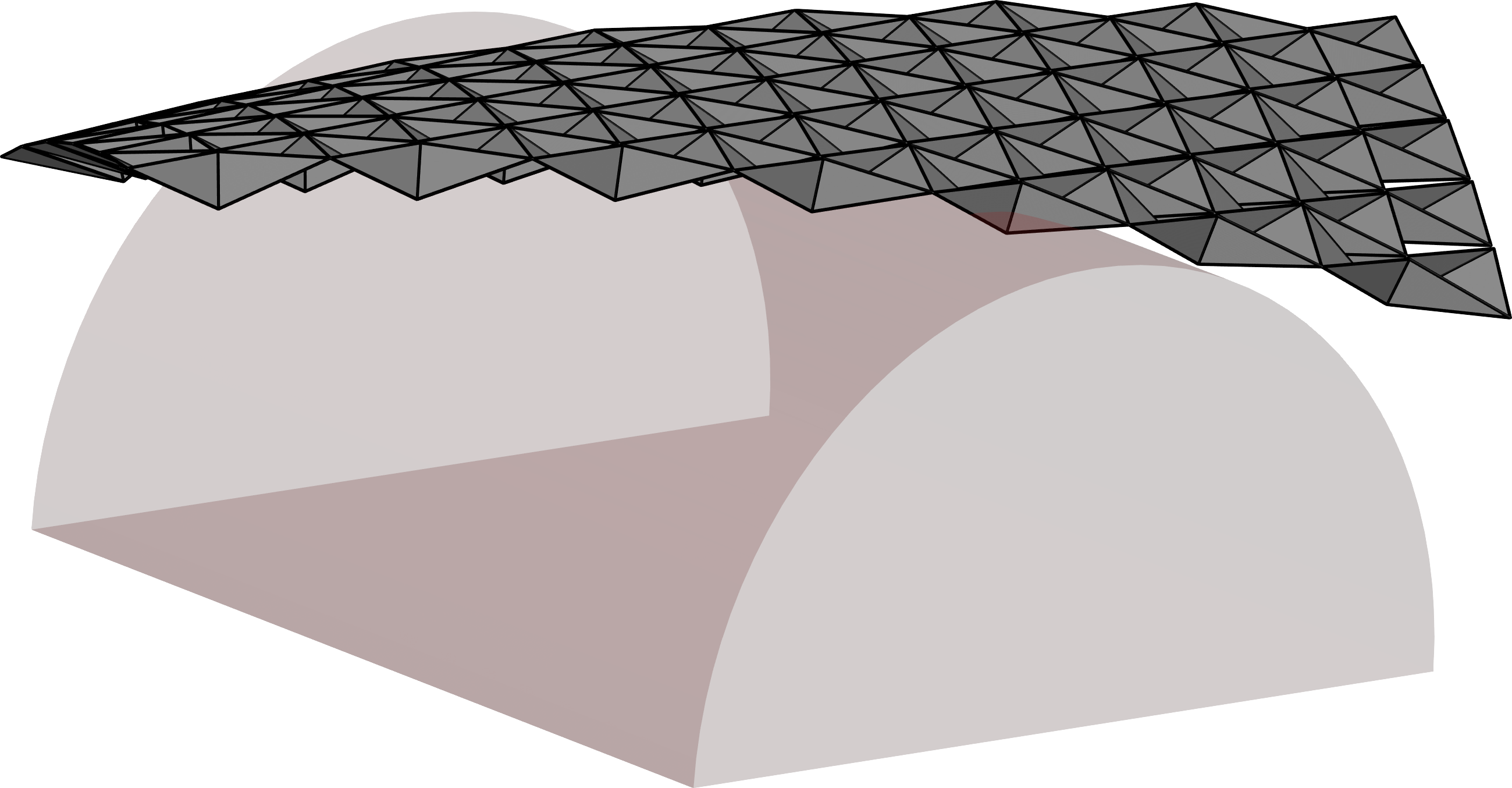} \\
 & \renewcommand{\arraystretch}{1.0} \begin{tabular}[c]{@{}l@{}} $\foldanglegamma=0.750$ \\ $\eta=2.00$ \\ $\zeta=1.00$ \end{tabular} & 0.1552 & 
\includegraphics[height=0.9\paramheight, valign=m]{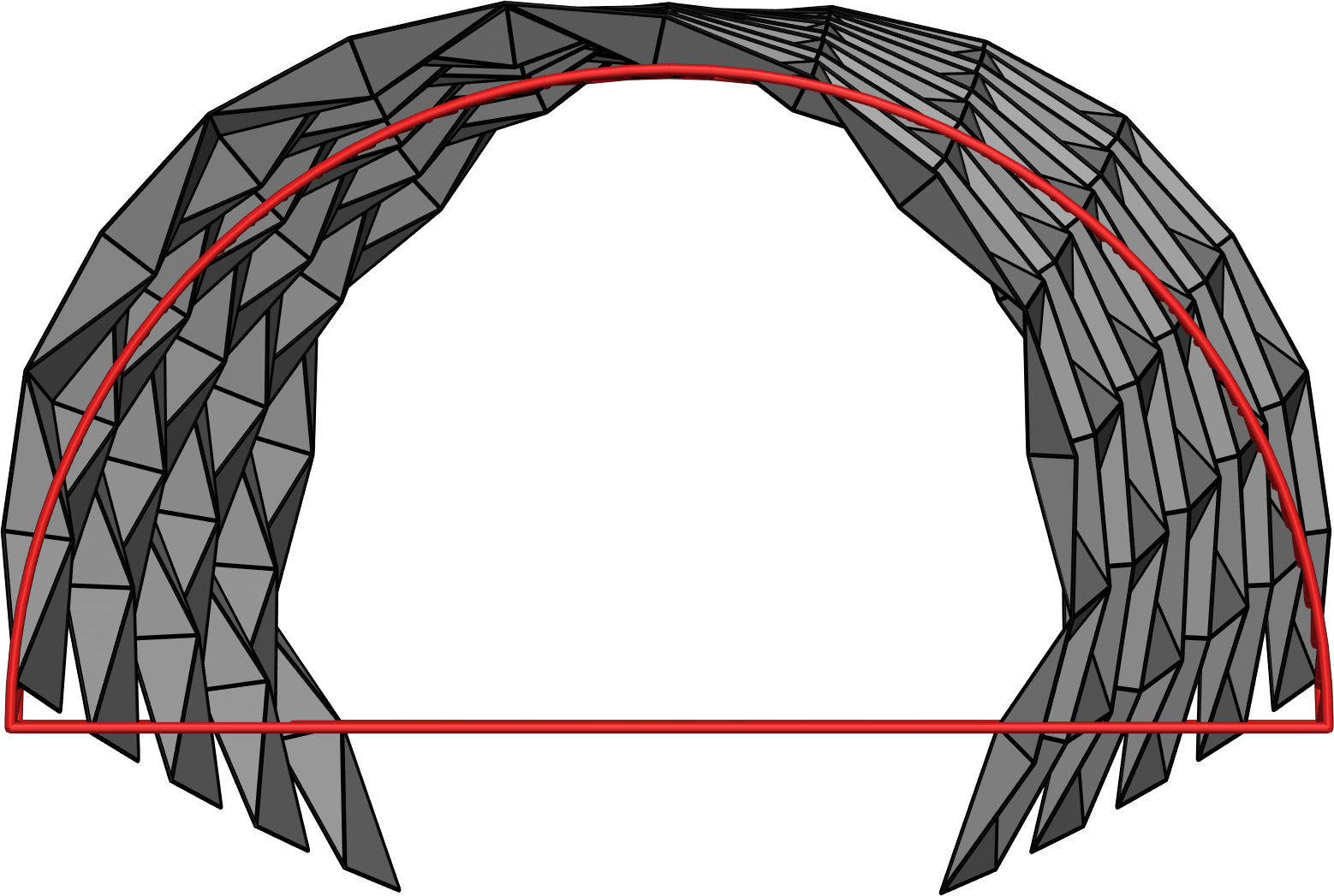} & 
\includegraphics[height=\paramheight, valign=m]{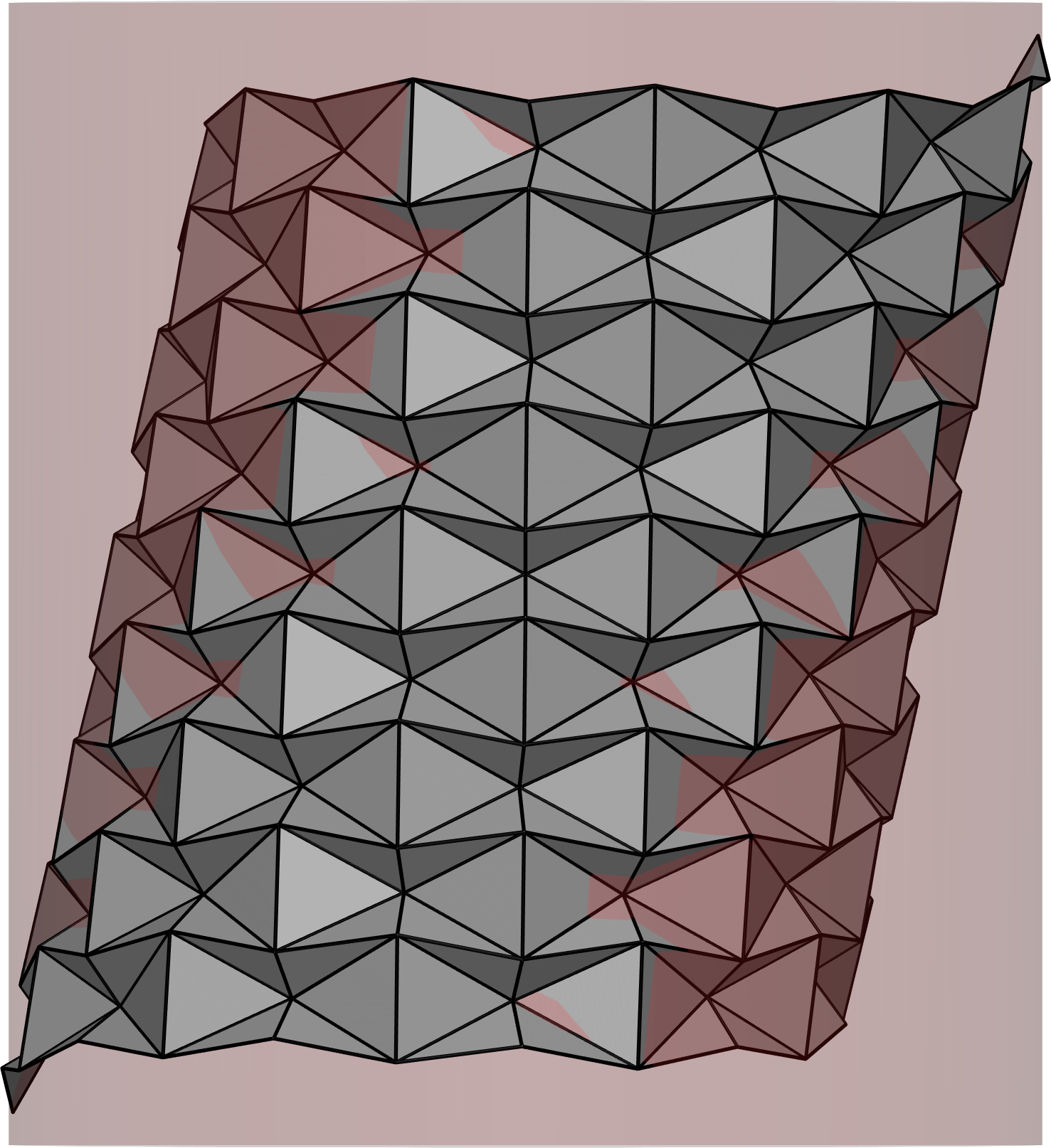} & 
\includegraphics[height=\paramheight, valign=m]{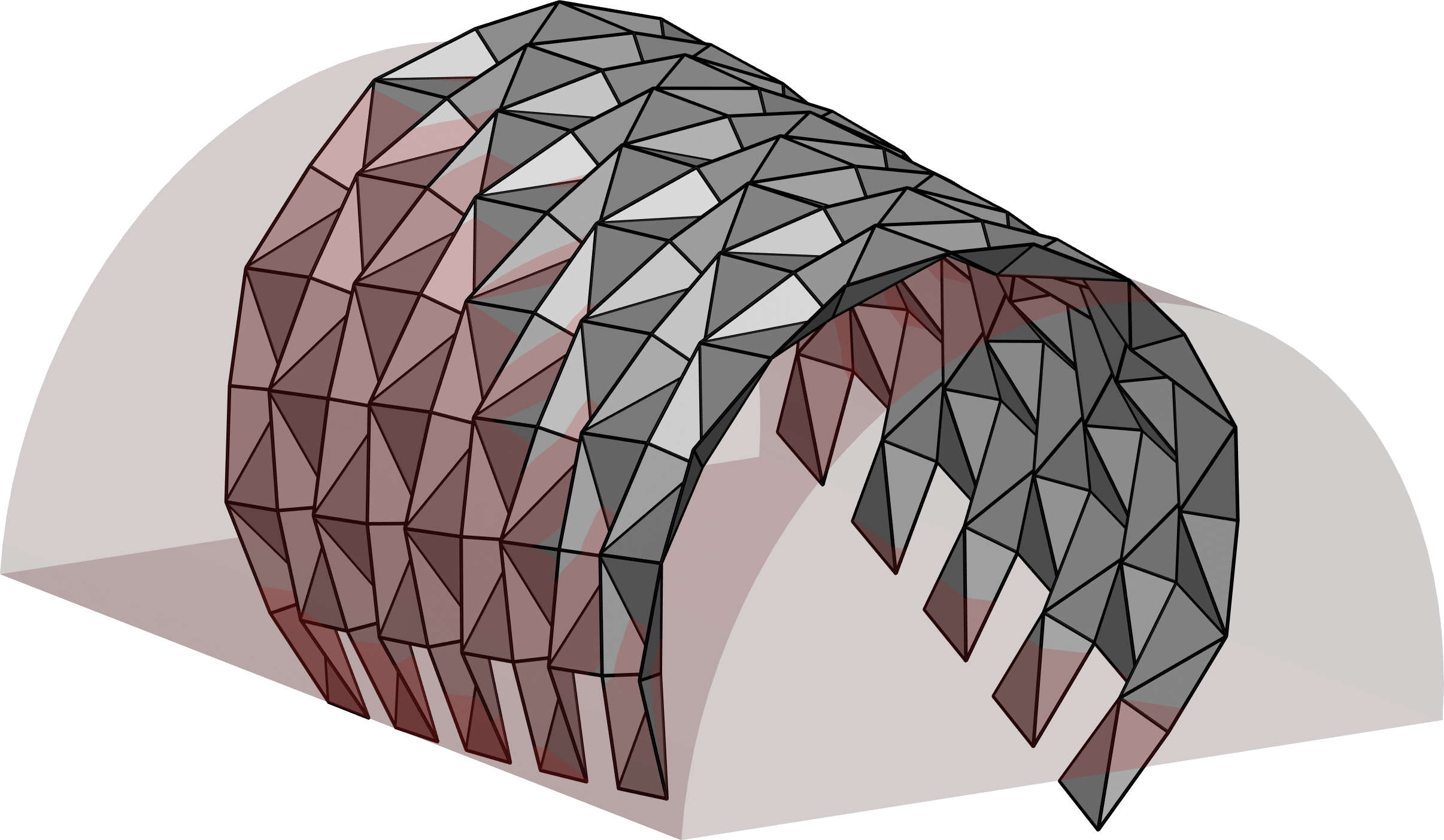} \\
 & \renewcommand{\arraystretch}{1.0} \begin{tabular}[c]{@{}l@{}} $\foldanglegamma=1.250$ \\ $\eta=2.00$ \\ $\zeta=1.00$ \end{tabular} & 0.1119 & 
\includegraphics[height=0.9\paramheight, valign=m]{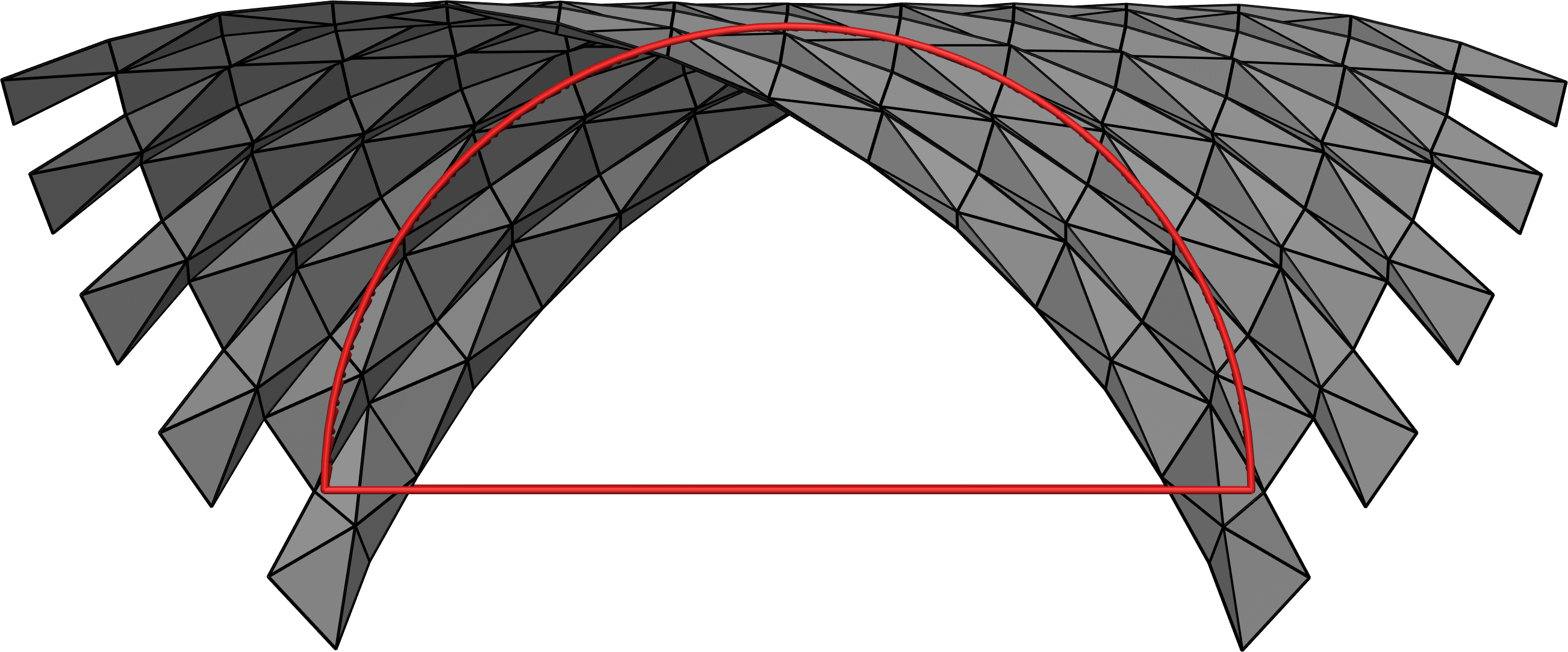} & 
\includegraphics[height=\paramheight, valign=m]{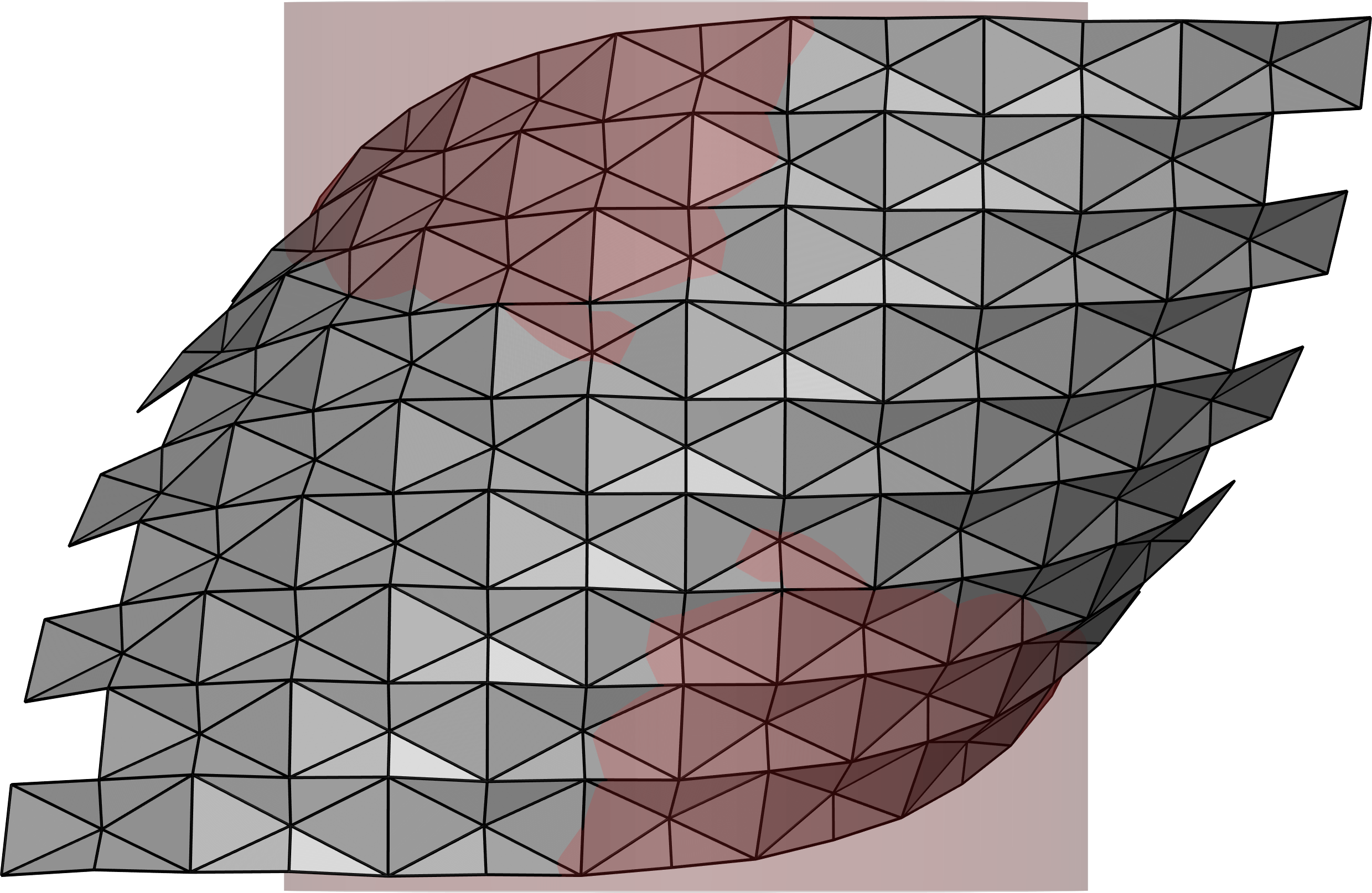} & 
\includegraphics[height=\paramheight, valign=m]{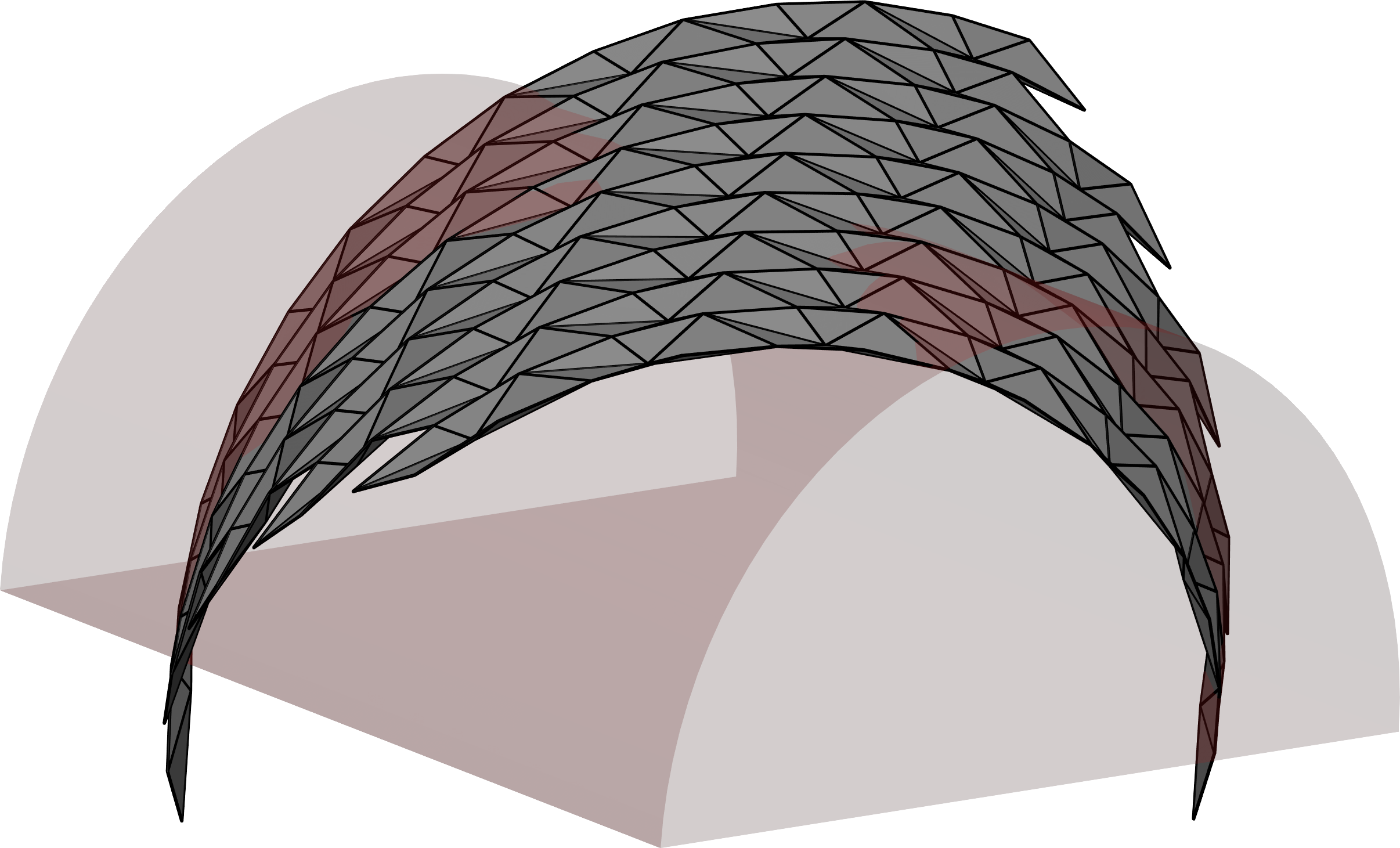} \\
\midrule
\SetCell[r=2]{m} $\eta$ effect & \renewcommand{\arraystretch}{1.0} \begin{tabular}[c]{@{}l@{}} $\foldanglegamma=1.050$ \\ $\eta=1.50$ \\ $\zeta=1.00$ \end{tabular} & 0.2104 & 
\includegraphics[height=0.9\paramheight, valign=m]{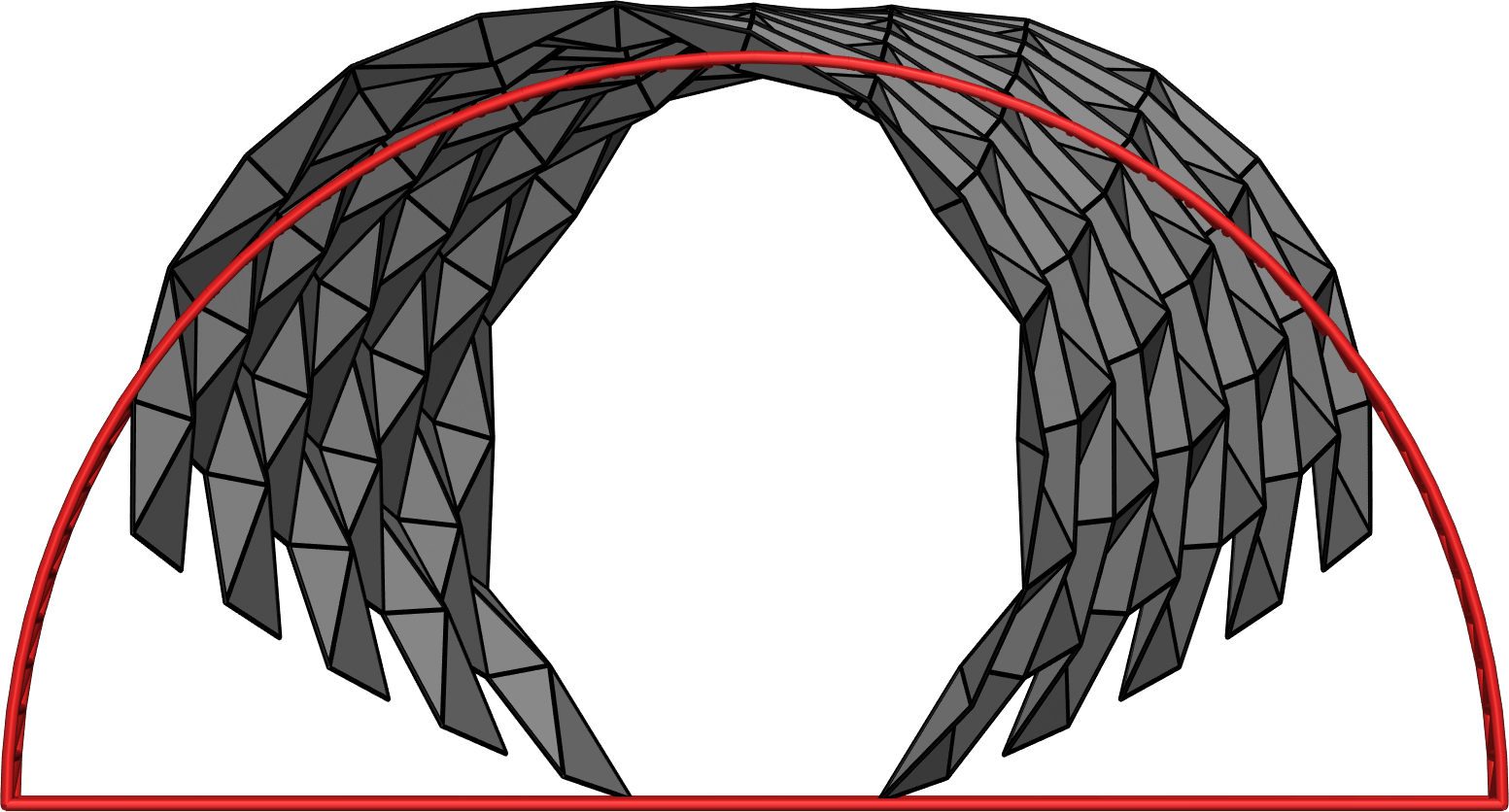} & 
\includegraphics[height=\paramheight, valign=m]{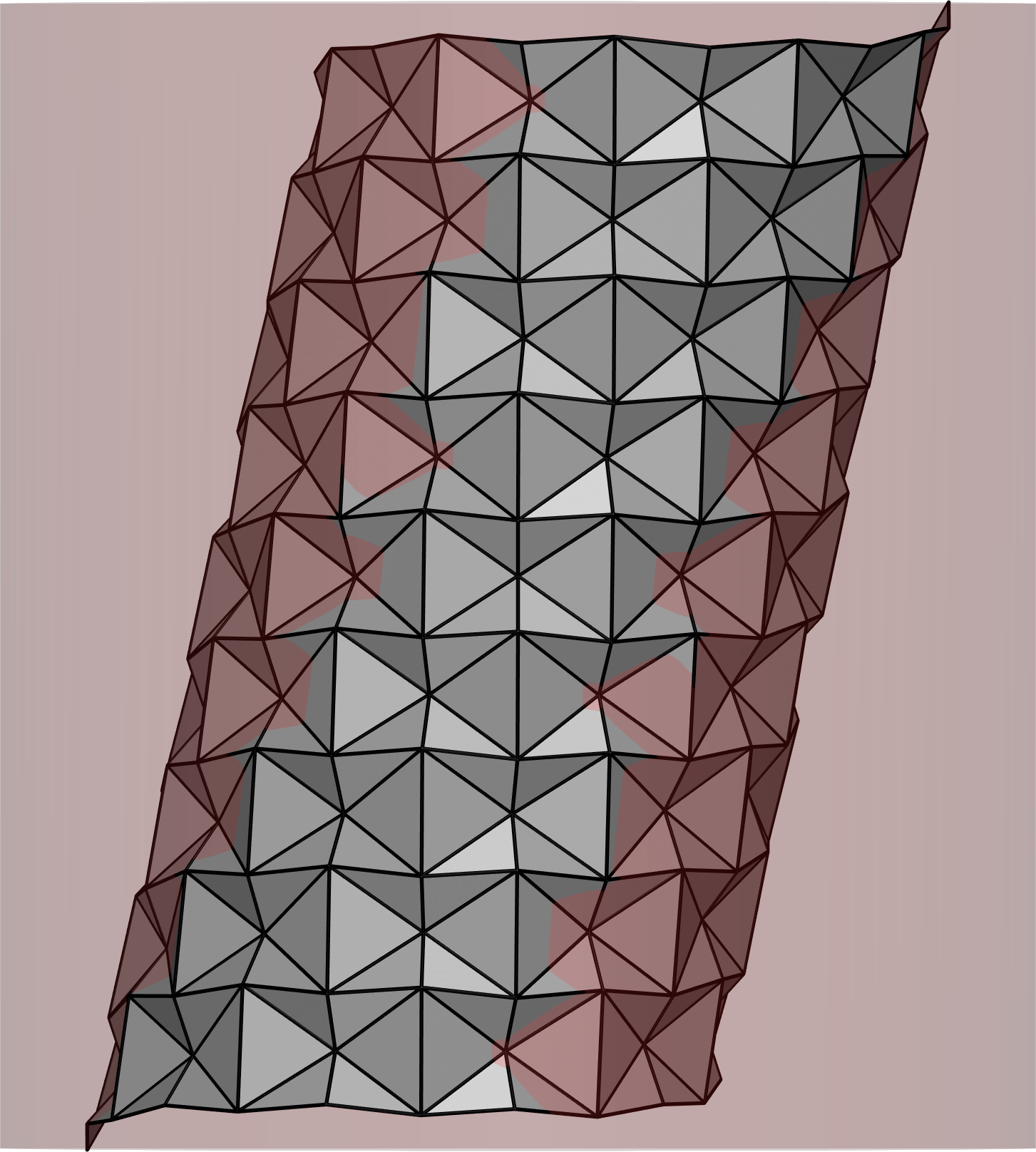} & 
\includegraphics[height=\paramheight, valign=m]{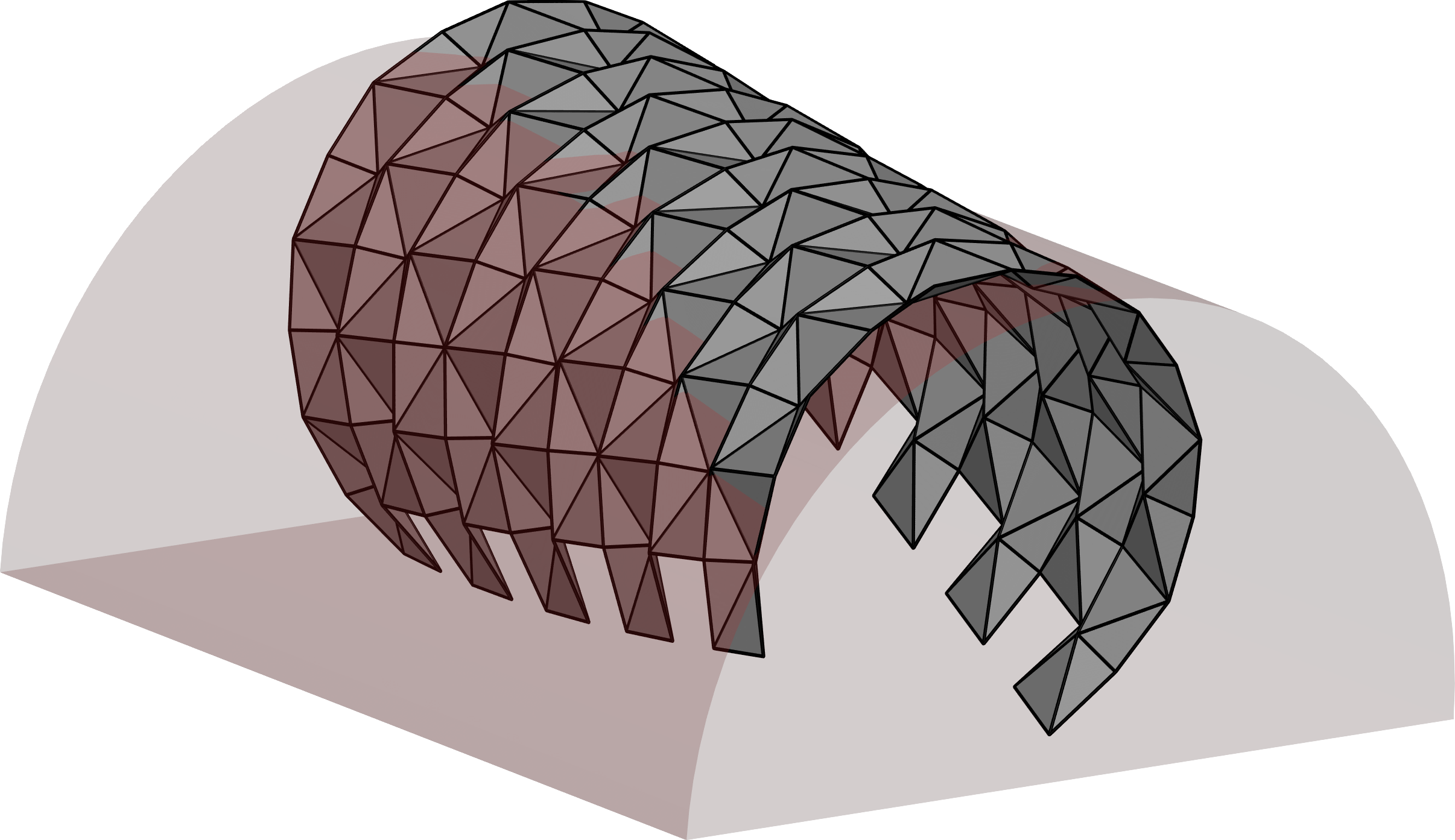} \\
 & \renewcommand{\arraystretch}{1.0} \begin{tabular}[c]{@{}l@{}} $\foldanglegamma=1.050$ \\ $\eta=2.50$ \\ $\zeta=1.00$ \end{tabular} & 0.0905 & 
\includegraphics[height=0.9\paramheight, valign=m]{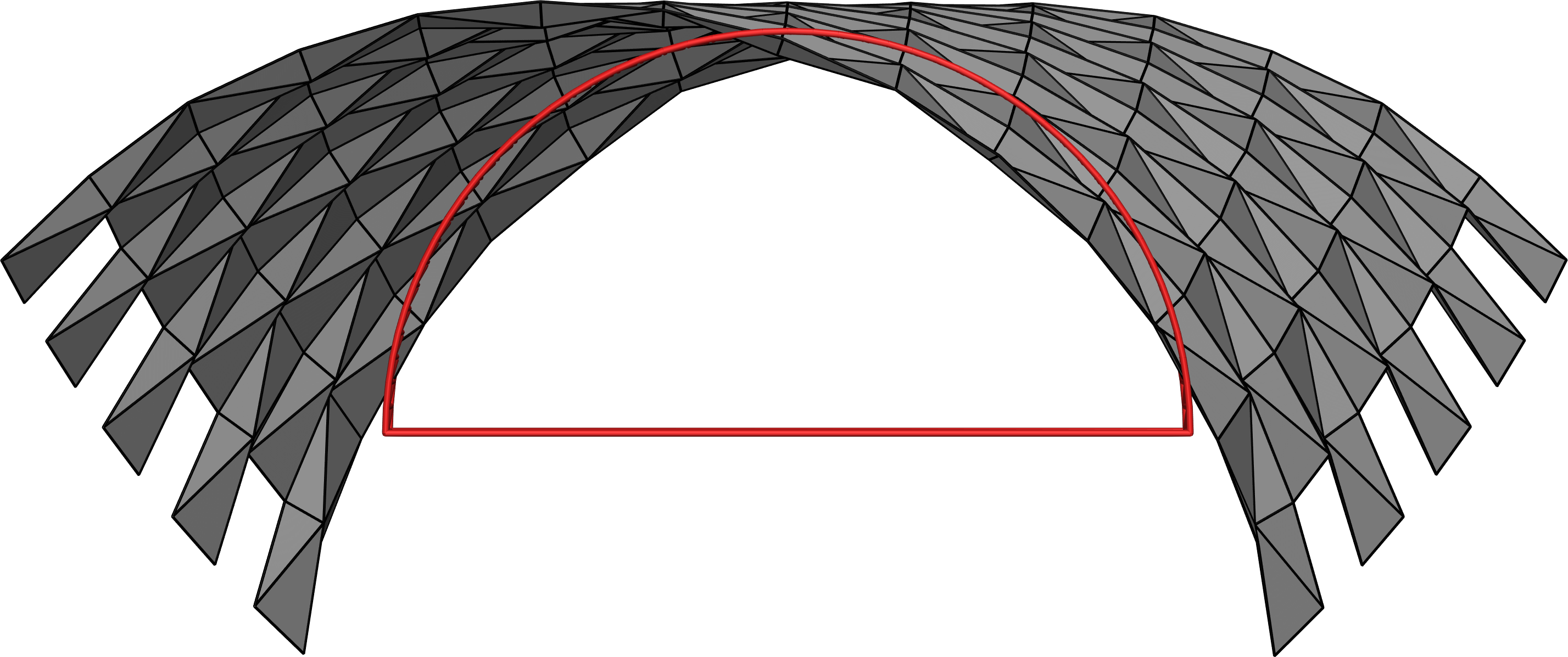} & 
\includegraphics[height=\paramheight, valign=m]{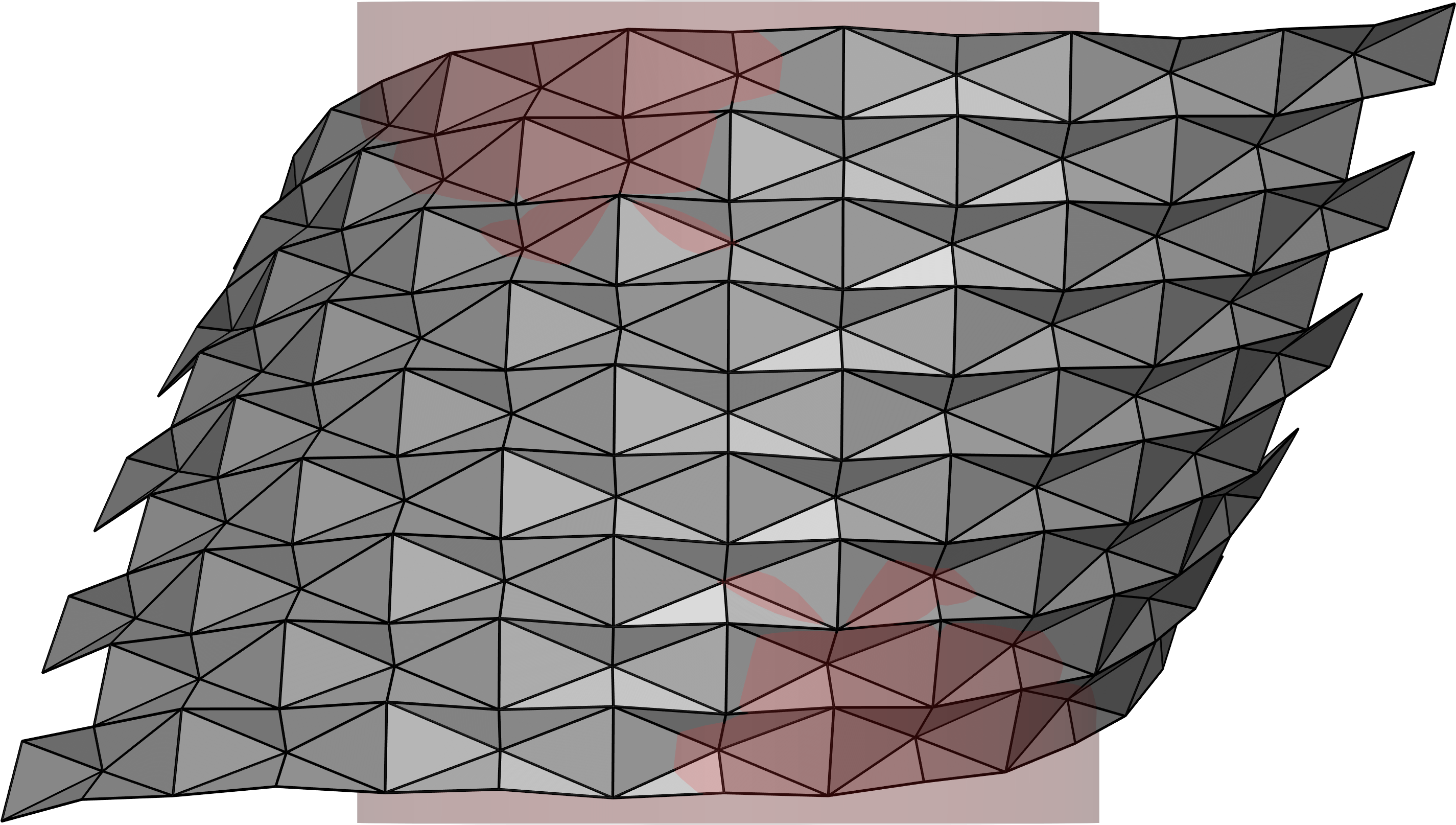} & 
\includegraphics[height=\paramheight, valign=m]{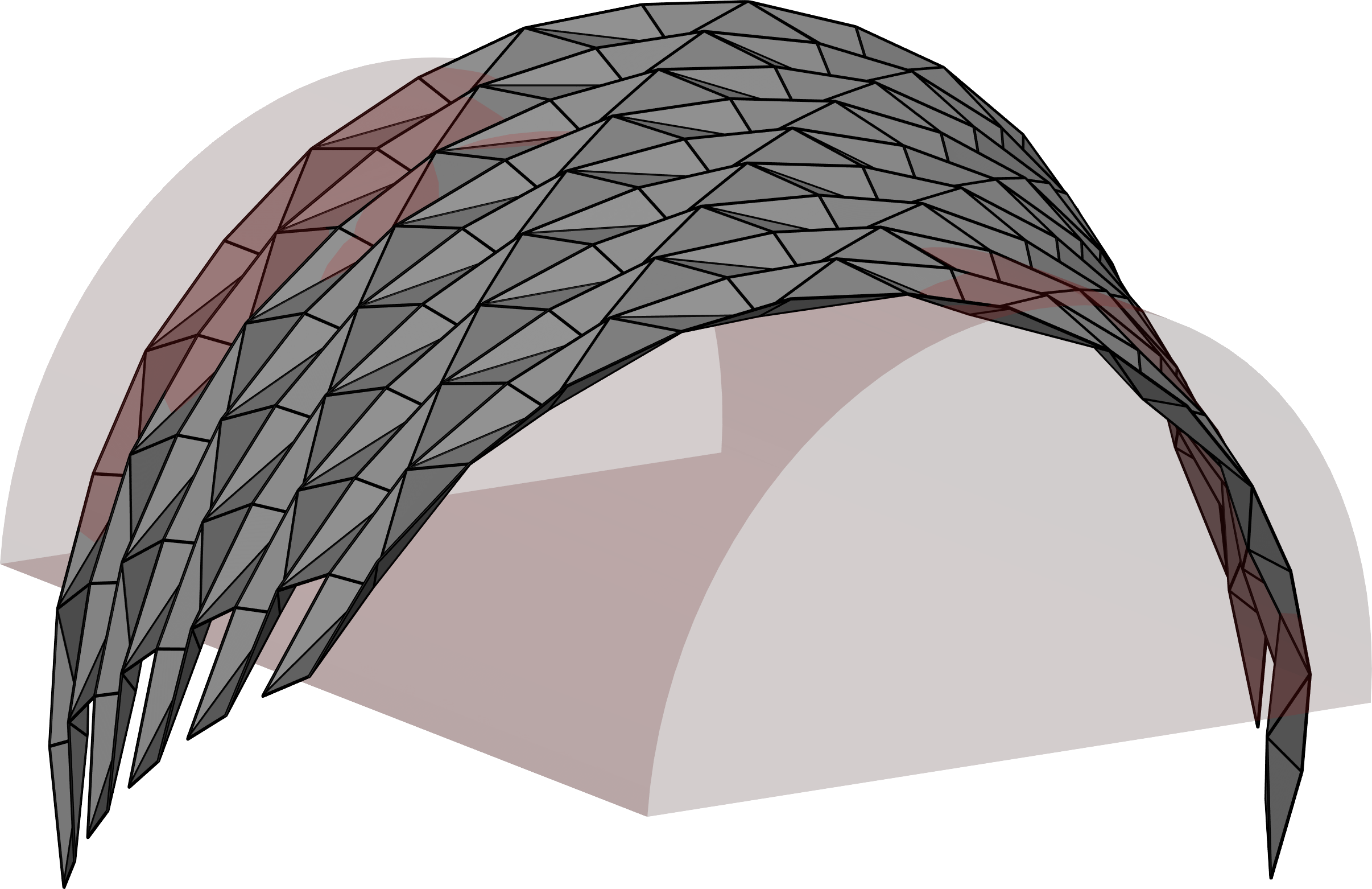} \\
\midrule
\SetCell[r=2]{m} $\zeta$ effect & \renewcommand{\arraystretch}{1.0} \begin{tabular}[c]{@{}l@{}} $\foldanglegamma=1.050$ \\ $\eta=2.00$ \\ $\zeta=0.70$ \end{tabular} & 0.0827 & 
\includegraphics[height=0.9\paramheight, valign=m]{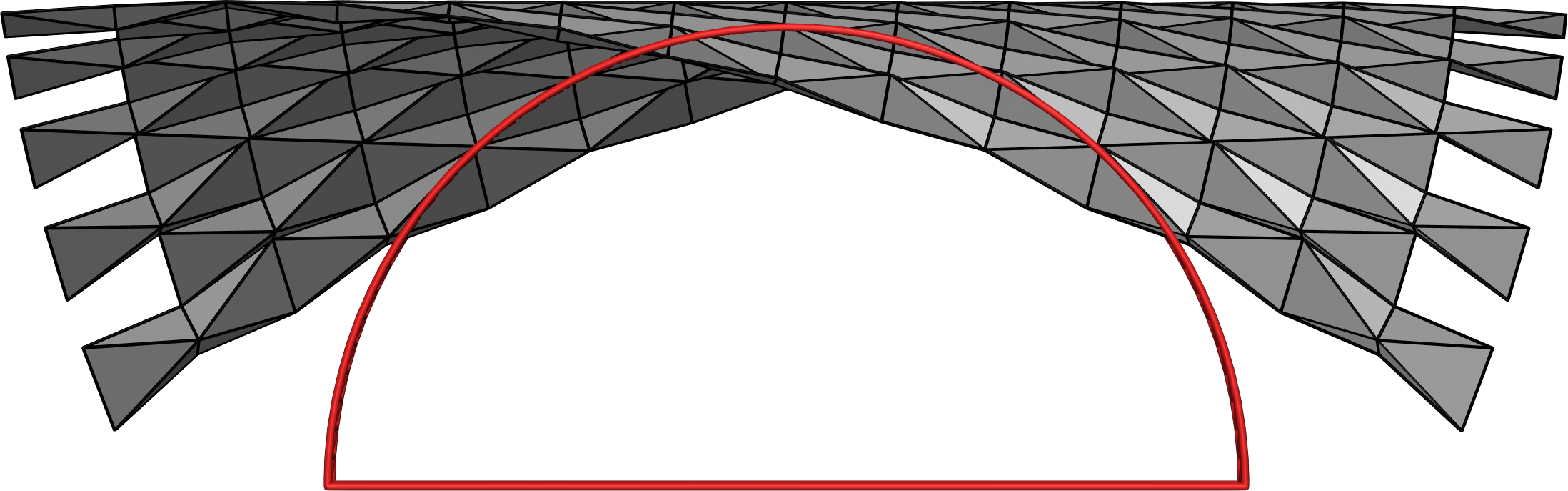} & 
\includegraphics[height=\paramheight, valign=m]{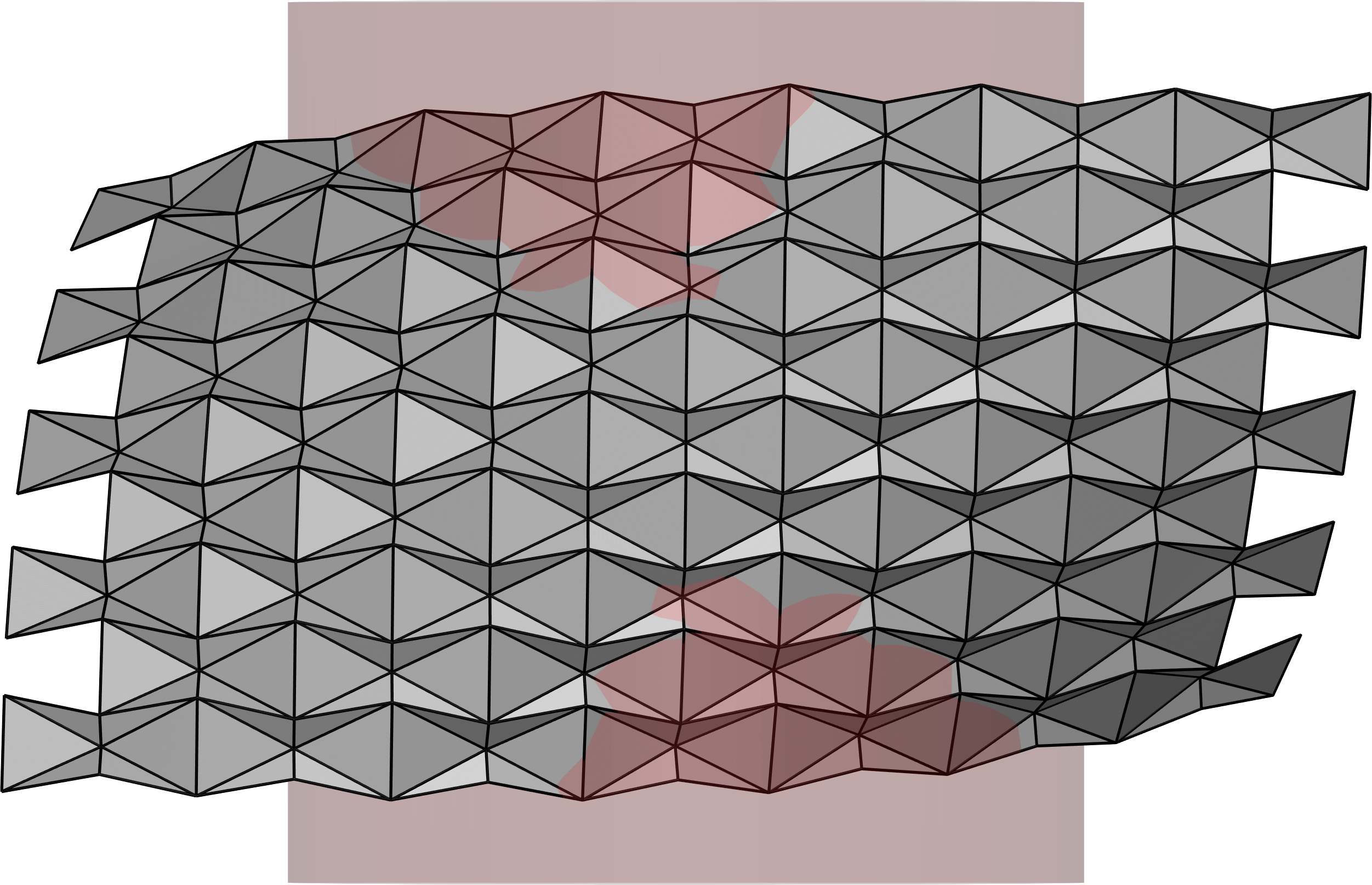} & 
\includegraphics[height=\paramheight, valign=m]{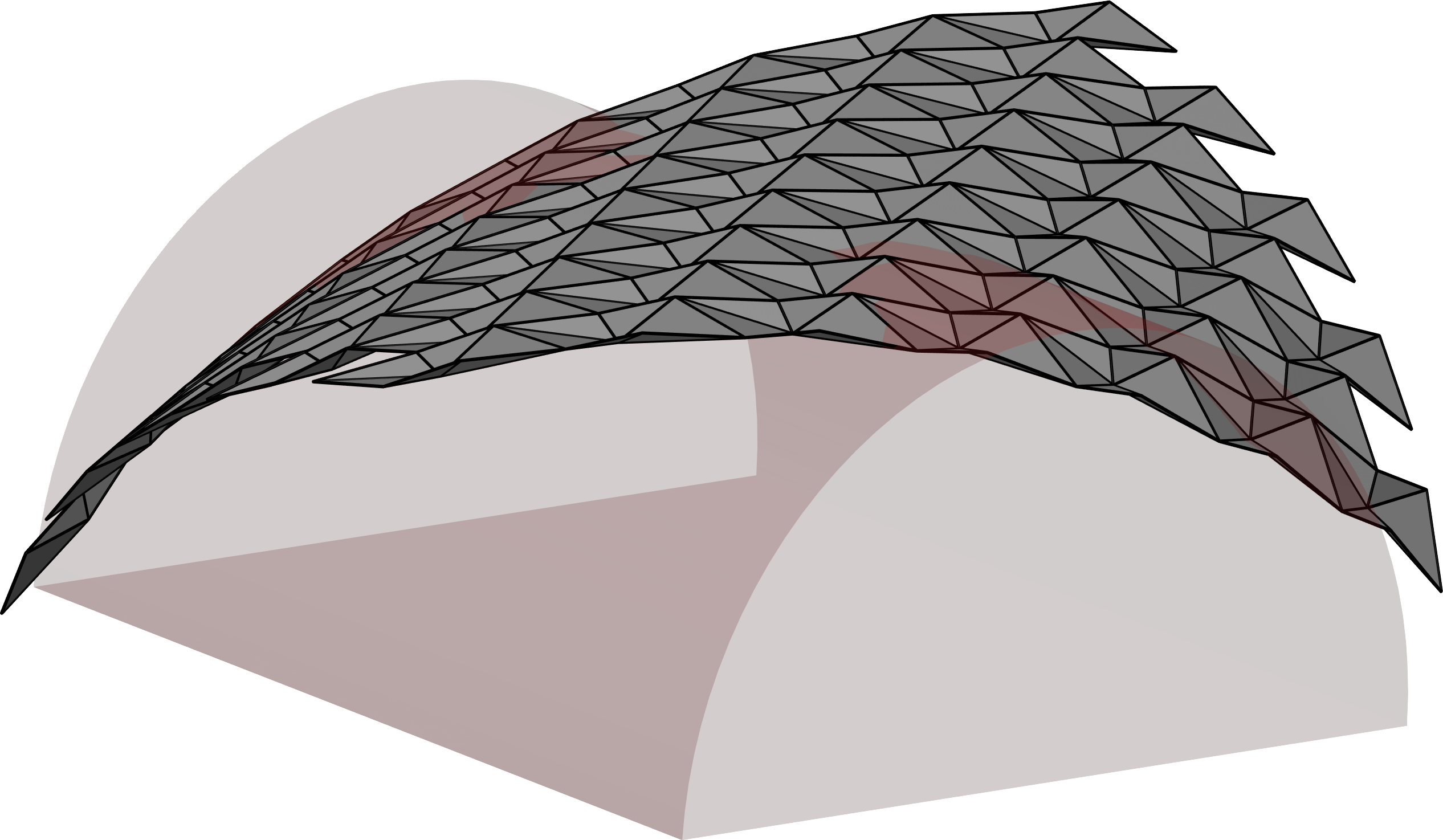} \\
 & \renewcommand{\arraystretch}{1.0} \begin{tabular}[c]{@{}l@{}} $\foldanglegamma=1.050$ \\ $\eta=2.00$ \\ $\zeta=1.30$ \end{tabular} & 0.1914 & 
\includegraphics[height=0.9\paramheight, valign=m]{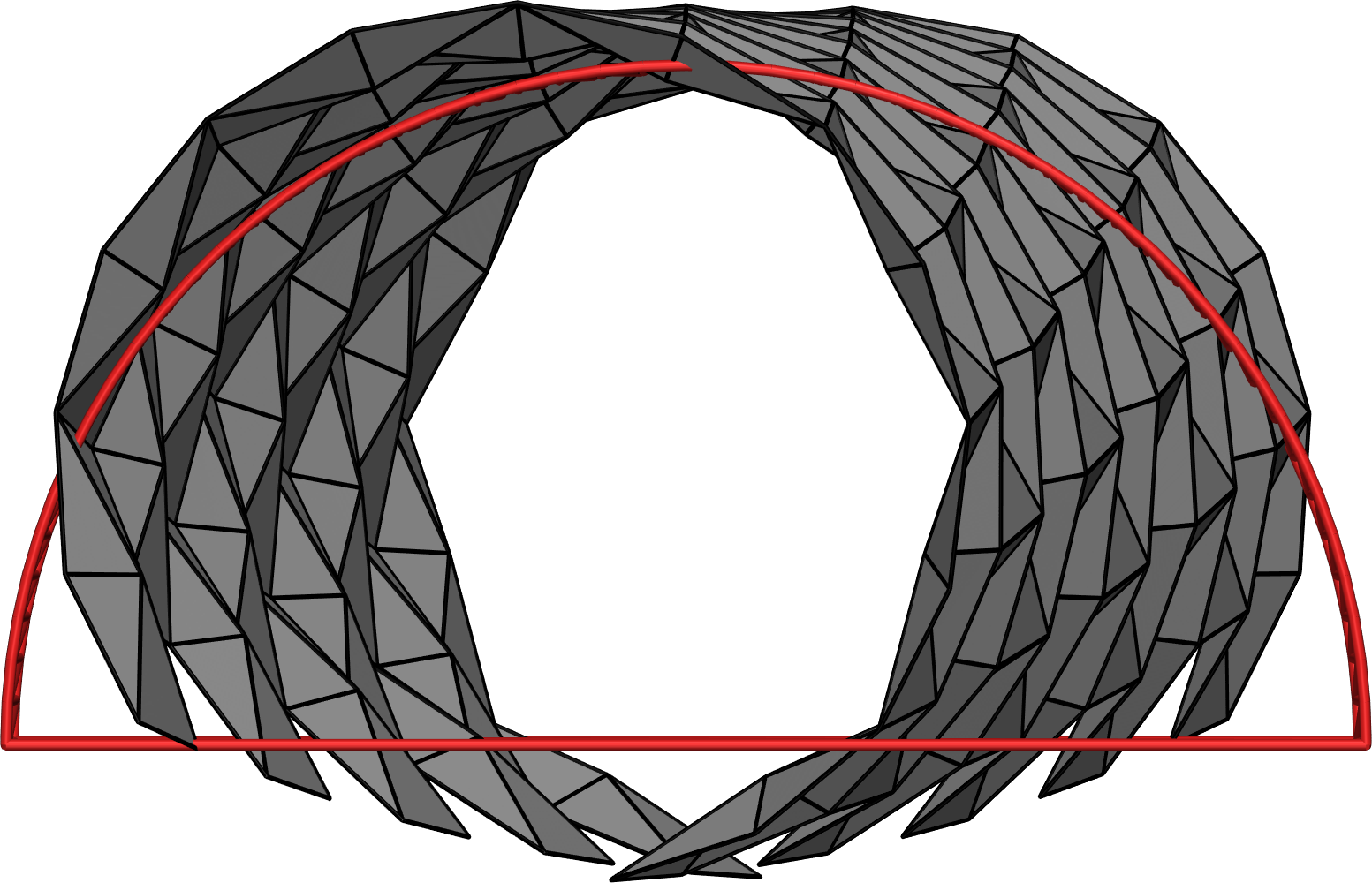} & 
\includegraphics[height=\paramheight, valign=m]{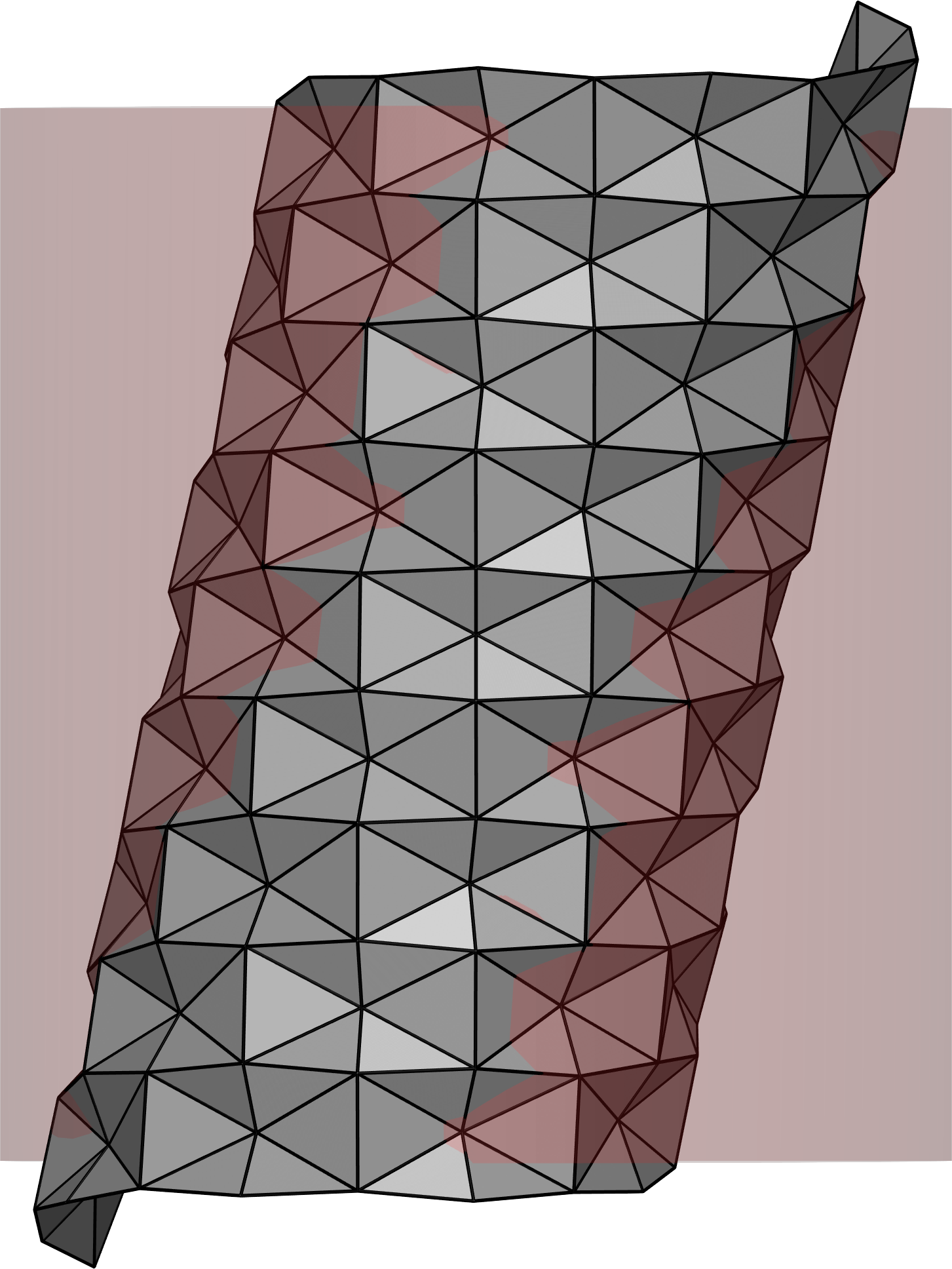} & 
\includegraphics[height=\paramheight, valign=m]{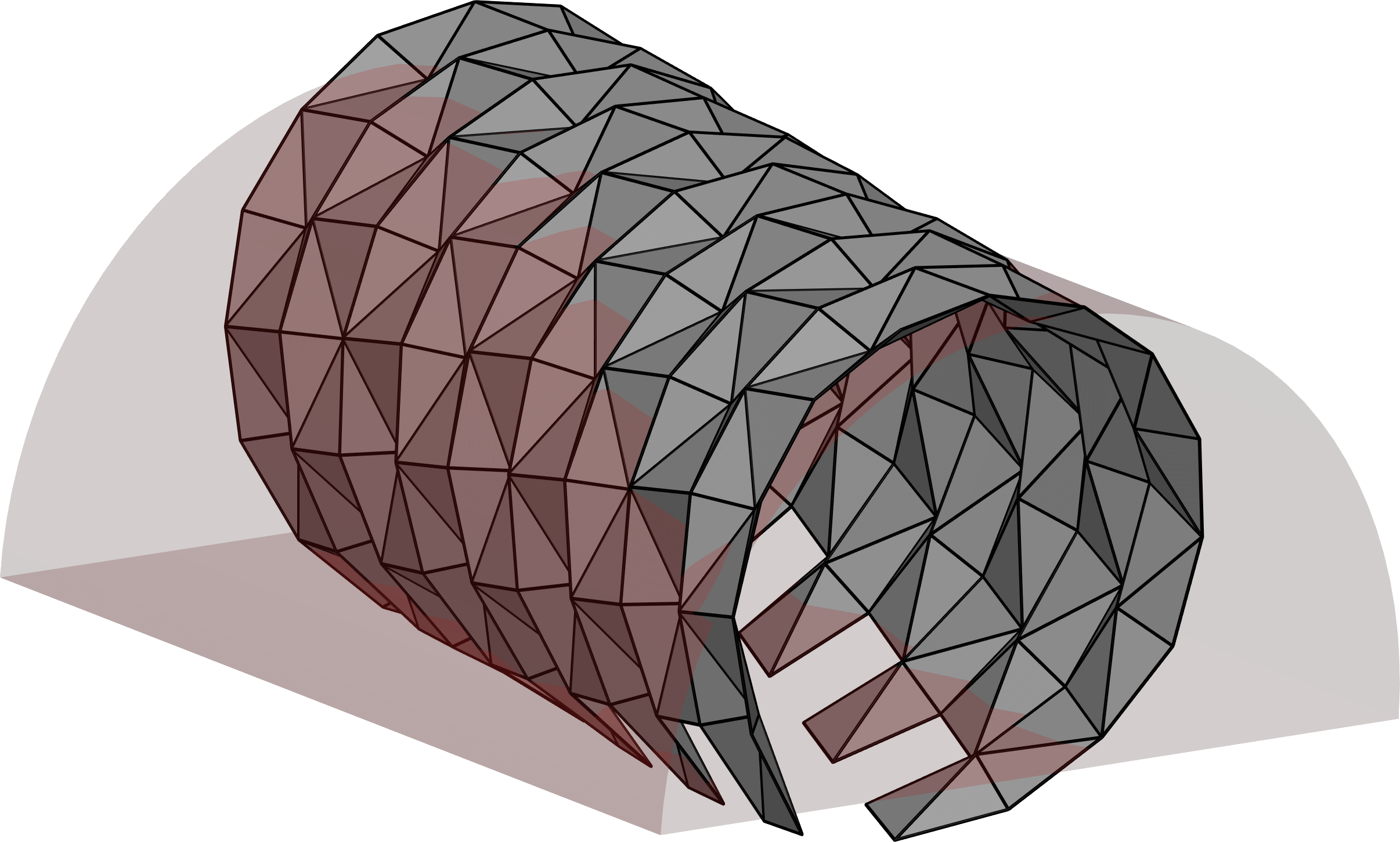} \\
\bottomrule
\end{tblr}
\end{table*}

\subsection{Proof of the curvature theorems}\label{sec:proof_curvature}
 This subsection proves
Theorems~\ref{thm:global_curvature} and~\ref{thm:global-cylinder-I}. Let
\begin{equation*}
    S
    :=
    \begin{pmatrix}
        -1&0&0\\
        0&-1&0\\
        0&0&1
    \end{pmatrix}
\end{equation*}
denote the rotation by \(\pi\) about the \(z\)-axis. Semi-symmetry maps
each neighbor-cell attachment to the attachment in the opposite
tessellation direction. Consequently, conjugation by \(S\) reverses
the corresponding neighbor-cell isometry. In particular,
\begin{equation}
    S\Tro S
    =
    \Tro^{-1}.
    \label{eq:conjugation-Tro}
\end{equation}
Considering only the rotational parts of the two isometries gives
\begin{equation}
    S\Qro S
    =
    \Qro^{\mathsf T}.
    \label{eq:conjugation-Qro}
\end{equation}
Writing
\[
    \Qro=(q_{ij})_{i,j=1,\ldots,3},
\]
equation~\eqref{eq:conjugation-Qro} yields
\begin{equation}
    q_{12}=q_{21},
    \qquad
    q_{13}=-q_{31},
    \qquad
    q_{23}=-q_{32}.
    \label{eq:Qro-symmetry-relations}
\end{equation}

The assumption \(q_{32}^2+q_{13}^2>0\) implies that
\(\Qro-\Qro^{\mathsf T}\neq 0\). Thus, the rotation angle of \(\Qro\)
is neither \(0\) nor \(\pi\). For such a rotation,
the direction of its rotation axis is parallel to the axial vector of
\(\Qro-\Qro^{\mathsf T}\), namely
\begin{equation*}
    \begin{pmatrix}
        q_{32}-q_{23}\\
        q_{13}-q_{31}\\
        q_{21}-q_{12}
    \end{pmatrix}.
\end{equation*}
Using~\eqref{eq:Qro-symmetry-relations}, this vector becomes
\begin{equation*}
    2
    \begin{pmatrix}
        q_{32}\\
        q_{13}\\
        0
    \end{pmatrix}.
\end{equation*}
Thus the screw axis of \(\Tro\) is horizontal. With
\(\tau\) defined by~\eqref{eq:def-tau-global-cylinder}, its unit
direction vector is therefore
\begin{equation*}
    a_\tau
    =
    \frac{1}{
        \sqrt{q_{32}^2+q_{13}^2}
    }
    \begin{pmatrix}
        q_{32}\\
        q_{13}\\
        0
    \end{pmatrix}
    =
    \begin{pmatrix}
        \cos\tau\\
        \sin\tau\\
        0
    \end{pmatrix}.
\end{equation*}

Let \(A\) denote the screw axis of \(\Tro\). Since
\(\Tro^{-1}\) has the same screw axis as \(\Tro\), while conjugation
of an isometry maps its screw axis to the image of that axis,
equation~\eqref{eq:conjugation-Tro} implies
\begin{equation*}
    S(A)
    =
    A.
\end{equation*}
Because \(A\) is horizontal and invariant under the
half-turn \(S\), it intersects the \(z\)-axis. Denote this intersection
point by
\begin{equation*}
    C
    =
    z_C e_z
    =
    \begin{pmatrix}
        0\\
        0\\
        z_C
    \end{pmatrix}.
\end{equation*}
Hence
\begin{equation*}
    A
    =
    \left\{
        C+s\,a_\tau
        :
        s\in\mathbb{R}
    \right\}.
\end{equation*}

Since \(C\) lies on the screw axis of \(\Tro\), there exists a scalar
\(h_{\urcorner}\) such that
\begin{equation}
    \Qro C+\tro
    =
    C+h_{\urcorner}a_\tau.
    \label{eq:screw-axis-point}
\end{equation}
Taking the scalar product with \(a_\tau\), and using
\(\Qro a_\tau=a_\tau\), gives
\begin{equation*}
    h_{\urcorner}
    =
    a_\tau^{\mathsf T}\tro.
\end{equation*}
Substituting \(C=z_Ce_z\) into
equation~\eqref{eq:screw-axis-point} yields
\begin{equation*}
    z_C
    \left(
        e_z-\Qro e_z
    \right)
    =
    \tro-h_{\urcorner}a_\tau.
\end{equation*}
Taking the scalar product of both sides with
\(e_z-\Qro e_z\) gives
\begin{equation*}
    z_C
    =
    \frac{
        \bigl(
            \tro-h_{\urcorner}a_\tau
        \bigr)^{\mathsf T}
        \bigl(
            e_z-\Qro e_z
        \bigr)
    }{
        \left\|
            e_z-\Qro e_z
        \right\|^2
    },
\end{equation*}
which is~\eqref{eq:def-zC-global-cylinder}.

It remains to show that the entire family of points \(I_{m,n}\) has
constant distance from \(A\). Since
\(\Tro\) preserves its own screw axis,
\begin{equation*}
    \Tro(A)
    =
    A.
\end{equation*}
Moreover, by the commutativity
\eqref{eq:commuting-tessellation-generators},
\begin{equation*}
    \Tru\Tro\Tru^{-1}
    =
    \Tro.
\end{equation*}
The screw axis of the left-hand side is
\(\Tru(A)\), whereas the screw axis of the right-hand side
is \(A\). By uniqueness of the screw axis of the
non-trivial rotational isometry \(\Tro\),
\begin{equation*}
    \Tru(A)
    =
    A.
\end{equation*}
Thus both tessellation generators preserve the same line
\(A\).

Rigid motions that preserve a line also preserve the distance to that
line. Therefore, for every \((m,n)\in\mathbb{Z}^2\),
\begin{align*}
    \operatorname{dist}
    \bigl(
        I_{m,n},
        A
    \bigr)
    &=
    \operatorname{dist}
    \bigl(
        \Tro^{\,m}\Tru^{\,n}(I),
        A
    \bigr)
    \notag\\
    &=
    \operatorname{dist}
    \bigl(
        I,
        A
    \bigr).
\end{align*}
Since
\[
    I=c\cos\gamma\,e_z,
    \qquad
    C=z_Ce_z,
\]
and the axis direction \(a_\tau\) is horizontal,
\[
    I-C
    =
    \bigl(
        c\cos\gamma-z_C
    \bigr)e_z
\]
is perpendicular to \(a_\tau\). Hence
\begin{equation*}
    \operatorname{dist}
    \bigl(
        I,
        A
    \bigr)
    =
    \|I-C\|
    =
    \left|
        c\cos\gamma-z_C
    \right|
    =
    r.
\end{equation*}
If \(r>0\), the common constant-distance locus is the circular cylinder
\(C_I\) of radius \(r\) about \(A\). Hence
\begin{equation*}
    I_{m,n}\in C_I
    \qquad
    \text{for all }
    (m,n)\in\mathbb{Z}^2.
\end{equation*}
This cylinder has one principal direction parallel to
its axis and one principal direction tangent to its circular
cross-section. The corresponding principal curvatures are \(0\) and
\(\pm 1/r\), respectively. Hence the magnitude of the non-zero
principal curvature is
\[
    \kappa
    =
    \frac{1}{r}.
\]
This proves Theorem~\ref{thm:global_curvature}.

It remains to verify the explicit representations in
Theorem~\ref{thm:global-cylinder-I}. The vectors

\[
    a_\tau,
    \qquad
    b_\tau,
    \qquad
    e_z
\]
form an orthonormal basis, with \(b_\tau\) and \(e_z\) spanning the
plane perpendicular to the cylinder axis. Consequently,
\begin{equation*}
    X(s,\varphi)
    =
    C
    +
    s\,a_\tau
    +
    r
    \left(
        \cos\varphi\,e_z
        +
        \sin\varphi\,b_\tau
    \right)
\end{equation*}
parametrizes the circular cylinder of radius \(r\) about
\(A\). Expanding this expression gives
\eqref{eq:explicit-global-cylinder-components}.

Furthermore,
\[
    \mathrm{Id}
    -
    a_\tau a_\tau^{\mathsf T}
\]
is the orthogonal projector onto the plane perpendicular to
\(A\). Hence the condition that a point \(X\) have distance
\(r\) from the axis is
\[
    \left\|
        \left(
            \mathrm{Id}
            -
            a_\tau a_\tau^{\mathsf T}
        \right)
        (X-C)
    \right\|^2
    =
    r^2,
\]
which proves~\eqref{eq:implicit-global-cylinder-vector}. Substitution
of
\[
    a_\tau
    =
    (\cos\tau,\sin\tau,0)^{\mathsf T}
\]
gives~\eqref{eq:implicit-global-cylinder}. This proves
Theorem~\ref{thm:global-cylinder-I}.

\subsection{Derivation of realizable planar waterbomb patterns}
Of special interest in the context of civil engineering construction are planar configurations (\(\kappa = 0\)) of the semi-symmetric waterbomb pattern as they could be used as interlayers in floor or ceiling constructions to reduce material consumption. In this section we describe the parameters that lead to planar configurations algebraically. In this case, we say a waterbomb pattern consisting of copies of the same semi-symmetric master cell is planar, if \(\kappa = 0\) or analogously all vertices except for the middle vertices of the waterbomb cells lie in a common plane. For the vertices in the whole pattern to lie in a common plane, in particular the outer vertices of the master cell have to be coplanar. Consider the plane \(E\) spanned by \(\Vl, \; \Vr, \; \Ulo\). The point \(\Uro\) lies in this plane, if and only if the triangle spanned by \(\Vr, \Uro, \Ulo\) is contained in \(E\). 
Because the line \(\overline{\Vl, \Vr}\) is parallel to the \(x\)-axis, the plane \(E\) spanned by \(\Vl, \Vr, \Ulo\) is parallel to the \(xy\)-plane rotated around the \(x\)-axis. Thus, the point \(\Uro\) lies in \(E\) if and only if \(\zro = \zlo\). This is the case in symmetric configurations of the waterbomb cell, i.e. configurations that have \(yz\)-reflection symmetry in addition to the rotational symmetry. Adjusting the angle \(\beta\) from a symmetric configuration leads to a change in the \(z\)-coordinate of the point \(\Uro\). Because \(|\overline{O\Uro}| = |\overline{O\Ulo}|\), the \(z\)-coordinate has to adjust in the opposite direction and therefore the only configurations where \(\Uro\) lies in \(E\) are the symmetric ones. \\
Therefore, we can now assume to be in the symmetric case. Because of the resulting \(xz\)-reflection symmetry all outer vertices of the waterbomb base cell lie in a common plane, if and only if \(\zlo\) equals the \(z\)-coordinate of the point \(\Vl\), i.e. \(\frac{a \left(1-\sin \! \left(\gamma \right)\right)}{\cos \! \left(\gamma \right)}
 = \zlo = c\cos(\gamma)\). This yields the equation 
 \[0 = \frac{a \left(1-\sin \! \left(\mathit{\gamma} \right)\right)}{\cos \! \left(\mathit{\gamma} \right)}-c \cos \! \left(\mathit{\gamma} \right).\]
 Solving for \(a, b, \gamma\) we obtain the equivalent solutions
 \begin{align*}
     a &= \frac{c \cos^{2}\left(\gamma \right)}{1-\sin \! \left(\gamma \right)},\\
     c &= \frac{a \left(1-\sin \! \left(\gamma \right)\right)}{\cos \! \left(\gamma \right)^{2}},\\
    \gamma &= -2 \arctan \! \left(\frac{-c +\sqrt{-a^{2}+2 a c}}{a -c}\right).
 \end{align*}
These parameters result in planar configurations of the global geometry. Note that the condition that the global geometry is planar does not constrain the parameter \(b\).

\section{Conclusions}

This paper has presented a rigorous algebraic derivation of the folding kinematics for the waterbomb tessellation, covering both symmetric and semi-symmetric folding states. By establishing exact functional relationships between the elementary cell design parameters, the folding angles, and the resulting global geometry, we provide a robust alternative to iterative numerical shape-finding methods. The primary use case for this parametrization lies in the domain of civil engineering, specifically in the innovative design and fabrication of thin-walled, textile-reinforced concrete structures. The analytical solution facilitates the "fold-in-fresh" production method, where concrete is cast on a flat formwork and folded into its final spatial configuration before hardening. The algebraic formulation ensures the repeatability and compatibility of precast elements, which is a prerequisite for the industrial manufacturing of modular shells. Furthermore, the direct correlation established between geometric parameters and statically relevant properties—such as global curvature and effective cross-sectional height—allows structural engineers to optimize the load-bearing capacity and material efficiency of the shell during the preliminary design phase.\\
Looking forward, the versatility of the waterbomb pattern opens promising avenues for application in modular floor, wall and shell systems. The ability to tune the depth of the tessellation ribs while maintaining a specific global curvature (or lack thereof, in the case of flat configurations) offers distinct advantages for integrating building services, acoustic dampening, and aesthetic variation within a ceiling system. The algebraic model supports the precise dimensioning required for such modular assemblies. \\
Finally, while this study focused on tessellations composed of uniform cell copies, a significant area for future research is the algebraic description of non-uniform tessellations. Extending the current framework to accommodate cells with varying edge lengths and differing folding parameters throughout the mesh would transcend the cylindrical geometries characteristic of uniform tessellations. Such an advancement would enable the realization of complex, double-curved freeform surfaces using the waterbomb pattern, thereby expanding the morphological vocabulary available to architectural design and structural engineering. In this progression, symmetric folding yields cylinders with a fixed axis direction, semi-symmetric folding adds control of the axis and rib orientation, and abandoning cell uniformity is the natural next step toward two independently non-zero principal curvatures.

\section*{Acknowledgment}
This work was funded by the Deutsche Forschungsgemeinschaft (DFG, German Research Foundation) -- SFB/TRR280, Project-ID 417002380.
The authors acknowledge this support gratefully.

\bibliographystyle{RS} %

\bibliography{bibliography} %

\end{document}